\documentclass[longauth]{aa}

\usepackage{graphicx}
\usepackage{natbib}
\usepackage{scalerel}
\usepackage{comment}
\usepackage{csquotes}
\usepackage{array}
\usepackage[table]{xcolor}

\usepackage{txfonts}
\usepackage[pdfencoding=auto,psdextra]{hyperref}
\hypersetup{
    colorlinks=true,
    linkcolor=blue,
    filecolor=magenta,      
    urlcolor=blue,
    citecolor=blue
}
\makeatletter
\renewcommand*\aa@pageof{, page \thepage{} of \pageref*{LastPage}}
\makeatother

\usepackage[utf8]{inputenc}

\usepackage[switch, modulo]{lineno}
\renewcommand{\linenumbers}[0]{}

\usepackage{euclid}

\begin{document}

\title{\Euclid}
\subtitle{Forecasting the number counts of bright line emitters in the Euclid Wide Survey and Euclid Deep fields}    


\newcommand{\orcid}[1]{} 
\author{Euclid Collaboration: P.~Cassata\orcid{0000-0002-6716-4400}\thanks{\email{paolo.cassata@unipd.it}}\inst{\ref{aff1},\ref{aff2}}
\and S.~Quai\orcid{0000-0002-0449-8163}\inst{\ref{aff3},\ref{aff4}}
\and C.~Mancini\orcid{0000-0002-4297-0561}\inst{\ref{aff5}}
\and L.~Pozzetti\orcid{0000-0001-7085-0412}\inst{\ref{aff4}}
\and L.~Gabarra\orcid{0000-0002-8486-8856}\inst{\ref{aff6}}
\and M.~Talia\orcid{0000-0003-4352-2063}\inst{\ref{aff3},\ref{aff4}}
\and L.~Bisigello\orcid{0000-0003-0492-4924}\inst{\ref{aff2}}
\and M.~Bolzonella\orcid{0000-0003-3278-4607}\inst{\ref{aff4}}
\and G.~De~Lucia\orcid{0000-0002-6220-9104}\inst{\ref{aff7}}
\and A.~Enia\orcid{0000-0002-0200-2857}\inst{\ref{aff8},\ref{aff4}}
\and A.~Feltre\orcid{0000-0001-6865-2871}\inst{\ref{aff9}}
\and F.~Fontanot\orcid{0000-0003-4744-0188}\inst{\ref{aff7},\ref{aff10}}
\and M.~Hirschmann\orcid{0000-0002-3301-3321}\inst{\ref{aff11}}
\and V.~Le~Brun\orcid{0000-0002-5027-1939}\inst{\ref{aff12}}
\and M.~Moresco\orcid{0000-0002-7616-7136}\inst{\ref{aff3},\ref{aff4}}
\and M.~Scodeggio\inst{\ref{aff5}}
\and L.~Scharr\'e\orcid{0000-0003-2551-4430}\inst{\ref{aff11}}
\and M.~Siudek\orcid{0000-0002-2949-2155}\inst{\ref{aff13},\ref{aff14},\ref{aff15}}
\and D.~Vergani\orcid{0000-0003-0898-2216}\inst{\ref{aff4}}
\and G.~Zamorani\orcid{0000-0002-2318-301X}\inst{\ref{aff4}}
\and O.~Cucciati\orcid{0000-0002-9336-7551}\inst{\ref{aff4}}
\and N.~Aghanim\orcid{0000-0002-6688-8992}\inst{\ref{aff16}}
\and B.~Altieri\orcid{0000-0003-3936-0284}\inst{\ref{aff17}}
\and S.~Andreon\orcid{0000-0002-2041-8784}\inst{\ref{aff18}}
\and N.~Auricchio\orcid{0000-0003-4444-8651}\inst{\ref{aff4}}
\and C.~Baccigalupi\orcid{0000-0002-8211-1630}\inst{\ref{aff10},\ref{aff7},\ref{aff19},\ref{aff20}}
\and M.~Baldi\orcid{0000-0003-4145-1943}\inst{\ref{aff21},\ref{aff4},\ref{aff22}}
\and S.~Bardelli\orcid{0000-0002-8900-0298}\inst{\ref{aff4}}
\and P.~Battaglia\orcid{0000-0002-7337-5909}\inst{\ref{aff4}}
\and A.~Biviano\orcid{0000-0002-0857-0732}\inst{\ref{aff7},\ref{aff10}}
\and E.~Branchini\orcid{0000-0002-0808-6908}\inst{\ref{aff23},\ref{aff24},\ref{aff18}}
\and M.~Brescia\orcid{0000-0001-9506-5680}\inst{\ref{aff25},\ref{aff26}}
\and S.~Camera\orcid{0000-0003-3399-3574}\inst{\ref{aff27},\ref{aff28},\ref{aff29}}
\and G.~Ca\~nas-Herrera\orcid{0000-0003-2796-2149}\inst{\ref{aff30}}
\and V.~Capobianco\orcid{0000-0002-3309-7692}\inst{\ref{aff29}}
\and C.~Carbone\orcid{0000-0003-0125-3563}\inst{\ref{aff5}}
\and J.~Carretero\orcid{0000-0002-3130-0204}\inst{\ref{aff31},\ref{aff32}}
\and M.~Castellano\orcid{0000-0001-9875-8263}\inst{\ref{aff33}}
\and G.~Castignani\orcid{0000-0001-6831-0687}\inst{\ref{aff4}}
\and S.~Cavuoti\orcid{0000-0002-3787-4196}\inst{\ref{aff26},\ref{aff34}}
\and A.~Cimatti\inst{\ref{aff35}}
\and C.~Colodro-Conde\inst{\ref{aff13}}
\and G.~Congedo\orcid{0000-0003-2508-0046}\inst{\ref{aff36}}
\and L.~Conversi\orcid{0000-0002-6710-8476}\inst{\ref{aff37},\ref{aff17}}
\and Y.~Copin\orcid{0000-0002-5317-7518}\inst{\ref{aff38}}
\and F.~Courbin\orcid{0000-0003-0758-6510}\inst{\ref{aff39},\ref{aff40},\ref{aff41}}
\and H.~M.~Courtois\orcid{0000-0003-0509-1776}\inst{\ref{aff42}}
\and H.~Degaudenzi\orcid{0000-0002-5887-6799}\inst{\ref{aff43}}
\and S.~de~la~Torre\inst{\ref{aff12}}
\and H.~Dole\orcid{0000-0002-9767-3839}\inst{\ref{aff16}}
\and F.~Dubath\orcid{0000-0002-6533-2810}\inst{\ref{aff43}}
\and X.~Dupac\inst{\ref{aff17}}
\and A.~Ealet\orcid{0000-0003-3070-014X}\inst{\ref{aff38}}
\and S.~Escoffier\orcid{0000-0002-2847-7498}\inst{\ref{aff44}}
\and M.~Farina\orcid{0000-0002-3089-7846}\inst{\ref{aff45}}
\and R.~Farinelli\inst{\ref{aff4}}
\and F.~Faustini\orcid{0000-0001-6274-5145}\inst{\ref{aff33}}
\and S.~Ferriol\inst{\ref{aff38}}
\and P.~Fosalba\orcid{0000-0002-1510-5214}\inst{\ref{aff46},\ref{aff15}}
\and S.~Fotopoulou\orcid{0000-0002-9686-254X}\inst{\ref{aff47}}
\and N.~Fourmanoit\orcid{0009-0005-6816-6925}\inst{\ref{aff44}}
\and M.~Frailis\orcid{0000-0002-7400-2135}\inst{\ref{aff7}}
\and E.~Franceschi\orcid{0000-0002-0585-6591}\inst{\ref{aff4}}
\and M.~Fumana\orcid{0000-0001-6787-5950}\inst{\ref{aff5}}
\and S.~Galeotta\orcid{0000-0002-3748-5115}\inst{\ref{aff7}}
\and K.~George\orcid{0000-0002-1734-8455}\inst{\ref{aff48}}
\and B.~Gillis\orcid{0000-0002-4478-1270}\inst{\ref{aff36}}
\and C.~Giocoli\orcid{0000-0002-9590-7961}\inst{\ref{aff4},\ref{aff22}}
\and J.~Gracia-Carpio\orcid{0000-0003-4689-3134}\inst{\ref{aff49}}
\and A.~Grazian\orcid{0000-0002-5688-0663}\inst{\ref{aff2}}
\and F.~Grupp\inst{\ref{aff49},\ref{aff50}}
\and S.~V.~H.~Haugan\orcid{0000-0001-9648-7260}\inst{\ref{aff51}}
\and S.~Hemmati\orcid{0000-0003-2226-5395}\inst{\ref{aff52}}
\and W.~Holmes\orcid{0009-0007-8554-4646}\inst{\ref{aff53}}
\and F.~Hormuth\inst{\ref{aff54}}
\and A.~Hornstrup\orcid{0000-0002-3363-0936}\inst{\ref{aff55},\ref{aff56}}
\and K.~Jahnke\orcid{0000-0003-3804-2137}\inst{\ref{aff57}}
\and M.~Jhabvala\inst{\ref{aff58}}
\and B.~Joachimi\orcid{0000-0001-7494-1303}\inst{\ref{aff59}}
\and S.~Kermiche\orcid{0000-0002-0302-5735}\inst{\ref{aff44}}
\and A.~Kiessling\orcid{0000-0002-2590-1273}\inst{\ref{aff53}}
\and B.~Kubik\orcid{0009-0006-5823-4880}\inst{\ref{aff38}}
\and M.~K\"ummel\orcid{0000-0003-2791-2117}\inst{\ref{aff50}}
\and M.~Kunz\orcid{0000-0002-3052-7394}\inst{\ref{aff60}}
\and H.~Kurki-Suonio\orcid{0000-0002-4618-3063}\inst{\ref{aff61},\ref{aff62}}
\and A.~M.~C.~Le~Brun\orcid{0000-0002-0936-4594}\inst{\ref{aff63}}
\and S.~Ligori\orcid{0000-0003-4172-4606}\inst{\ref{aff29}}
\and P.~B.~Lilje\orcid{0000-0003-4324-7794}\inst{\ref{aff51}}
\and V.~Lindholm\orcid{0000-0003-2317-5471}\inst{\ref{aff61},\ref{aff62}}
\and I.~Lloro\orcid{0000-0001-5966-1434}\inst{\ref{aff64}}
\and M.~Magliocchetti\orcid{0000-0001-9158-4838}\inst{\ref{aff45}}
\and G.~Mainetti\orcid{0000-0003-2384-2377}\inst{\ref{aff65}}
\and E.~Maiorano\orcid{0000-0003-2593-4355}\inst{\ref{aff4}}
\and O.~Mansutti\orcid{0000-0001-5758-4658}\inst{\ref{aff7}}
\and O.~Marggraf\orcid{0000-0001-7242-3852}\inst{\ref{aff66}}
\and M.~Martinelli\orcid{0000-0002-6943-7732}\inst{\ref{aff33},\ref{aff67}}
\and N.~Martinet\orcid{0000-0003-2786-7790}\inst{\ref{aff12}}
\and F.~Marulli\orcid{0000-0002-8850-0303}\inst{\ref{aff3},\ref{aff4},\ref{aff22}}
\and R.~J.~Massey\orcid{0000-0002-6085-3780}\inst{\ref{aff68}}
\and D.~C.~Masters\orcid{0000-0001-5382-6138}\inst{\ref{aff69}}
\and N.~Mauri\orcid{0000-0001-8196-1548}\inst{\ref{aff35},\ref{aff22}}
\and E.~Medinaceli\orcid{0000-0002-4040-7783}\inst{\ref{aff4}}
\and M.~Melchior\inst{\ref{aff70}}
\and M.~Meneghetti\orcid{0000-0003-1225-7084}\inst{\ref{aff4},\ref{aff22}}
\and E.~Merlin\orcid{0000-0001-6870-8900}\inst{\ref{aff2}}
\and G.~Meylan\orcid{0000-0001-6503-0209}\inst{\ref{aff71}}
\and J.~J.~Mohr\orcid{0000-0002-6875-2087}\inst{\ref{aff48}}
\and P.~Monaco\orcid{0000-0003-2083-7564}\inst{\ref{aff72},\ref{aff7},\ref{aff19},\ref{aff10}}
\and A.~Mora\orcid{0000-0002-1922-8529}\inst{\ref{aff73}}
\and C.~Moretti\orcid{0000-0003-3314-8936}\inst{\ref{aff7},\ref{aff10},\ref{aff19}}
\and L.~Moscardini\orcid{0000-0002-3473-6716}\inst{\ref{aff3},\ref{aff4},\ref{aff22}}
\and C.~Neissner\orcid{0000-0001-8524-4968}\inst{\ref{aff74},\ref{aff32}}
\and S.-M.~Niemi\orcid{0009-0005-0247-0086}\inst{\ref{aff75}}
\and J.~W.~Nightingale\orcid{0000-0002-8987-7401}\inst{\ref{aff76}}
\and C.~Padilla\orcid{0000-0001-7951-0166}\inst{\ref{aff74}}
\and S.~Paltani\orcid{0000-0002-8108-9179}\inst{\ref{aff43}}
\and F.~Pasian\orcid{0000-0002-4869-3227}\inst{\ref{aff7}}
\and V.~Pettorino\orcid{0000-0002-4203-9320}\inst{\ref{aff75}}
\and A.~Pezzotta\orcid{0000-0003-0726-2268}\inst{\ref{aff18}}
\and S.~Pires\orcid{0000-0002-0249-2104}\inst{\ref{aff77}}
\and G.~Polenta\orcid{0000-0003-4067-9196}\inst{\ref{aff78}}
\and M.~Poncet\inst{\ref{aff79}}
\and L.~A.~Popa\inst{\ref{aff80}}
\and F.~Raison\orcid{0000-0002-7819-6918}\inst{\ref{aff49}}
\and A.~Renzi\orcid{0000-0001-9856-1970}\inst{\ref{aff1},\ref{aff81},\ref{aff4}}
\and J.~Rhodes\orcid{0000-0002-4485-8549}\inst{\ref{aff53}}
\and G.~Riccio\inst{\ref{aff26}}
\and I.~Risso\orcid{0000-0003-2525-7761}\inst{\ref{aff23},\ref{aff24},\ref{aff18}}
\and F.~Rizzo\orcid{0000-0002-9407-585X}\inst{\ref{aff7}}
\and E.~Romelli\orcid{0000-0003-3069-9222}\inst{\ref{aff7}}
\and M.~Roncarelli\orcid{0000-0001-9587-7822}\inst{\ref{aff4}}
\and B.~Rusholme\orcid{0000-0001-7648-4142}\inst{\ref{aff52}}
\and R.~Saglia\orcid{0000-0003-0378-7032}\inst{\ref{aff50},\ref{aff49}}
\and Z.~Sakr\orcid{0000-0002-4823-3757}\inst{\ref{aff82},\ref{aff83},\ref{aff84}}
\and A.~G.~S\'anchez\orcid{0000-0003-1198-831X}\inst{\ref{aff49}}
\and D.~Sapone\orcid{0000-0001-7089-4503}\inst{\ref{aff85}}
\and B.~Sartoris\orcid{0000-0003-1337-5269}\inst{\ref{aff49},\ref{aff7}}
\and P.~Schneider\orcid{0000-0001-8561-2679}\inst{\ref{aff66}}
\and T.~Schrabback\orcid{0000-0002-6987-7834}\inst{\ref{aff86}}
\and A.~Secroun\orcid{0000-0003-0505-3710}\inst{\ref{aff44}}
\and E.~Sihvola\orcid{0000-0003-1804-7715}\inst{\ref{aff87}}
\and C.~Sirignano\orcid{0000-0002-0995-7146}\inst{\ref{aff1},\ref{aff81}}
\and G.~Sirri\orcid{0000-0003-2626-2853}\inst{\ref{aff22}}
\and L.~Stanco\orcid{0000-0002-9706-5104}\inst{\ref{aff81}}
\and C.~Surace\orcid{0000-0003-2592-0113}\inst{\ref{aff12}}
\and P.~Tallada-Cresp\'{i}\orcid{0000-0002-1336-8328}\inst{\ref{aff31},\ref{aff32}}
\and A.~N.~Taylor\inst{\ref{aff36}}
\and H.~I.~Teplitz\orcid{0000-0002-7064-5424}\inst{\ref{aff69}}
\and I.~Tereno\orcid{0000-0002-4537-6218}\inst{\ref{aff88},\ref{aff89}}
\and N.~Tessore\orcid{0000-0002-9696-7931}\inst{\ref{aff90}}
\and S.~Toft\orcid{0000-0003-3631-7176}\inst{\ref{aff91},\ref{aff92}}
\and R.~Toledo-Moreo\orcid{0000-0002-2997-4859}\inst{\ref{aff93},\ref{aff94}}
\and F.~Torradeflot\orcid{0000-0003-1160-1517}\inst{\ref{aff32},\ref{aff31}}
\and I.~Tutusaus\orcid{0000-0002-3199-0399}\inst{\ref{aff15},\ref{aff46},\ref{aff83}}
\and J.~Valiviita\orcid{0000-0001-6225-3693}\inst{\ref{aff61},\ref{aff62}}
\and T.~Vassallo\orcid{0000-0001-6512-6358}\inst{\ref{aff7},\ref{aff48}}
\and D.~Vibert\orcid{0009-0008-0607-631X}\inst{\ref{aff12}}
\and Y.~Wang\orcid{0000-0002-4749-2984}\inst{\ref{aff52}}
\and J.~Weller\orcid{0000-0002-8282-2010}\inst{\ref{aff50},\ref{aff49}}
\and I.~A.~Zinchenko\orcid{0000-0002-2944-2449}\inst{\ref{aff95}}
\and E.~Zucca\orcid{0000-0002-5845-8132}\inst{\ref{aff4}}
\and M.~Ballardini\orcid{0000-0003-4481-3559}\inst{\ref{aff96},\ref{aff97},\ref{aff4}}
\and P.~Bergamini\orcid{0000-0003-1383-9414}\inst{\ref{aff4}}
\and S.~Borgani\orcid{0000-0001-6151-6439}\inst{\ref{aff72},\ref{aff10},\ref{aff7},\ref{aff19},\ref{aff98}}
\and A.~Boucaud\orcid{0000-0001-7387-2633}\inst{\ref{aff99}}
\and E.~Bozzo\orcid{0000-0002-8201-1525}\inst{\ref{aff43}}
\and C.~Burigana\orcid{0000-0002-3005-5796}\inst{\ref{aff100},\ref{aff101}}
\and R.~Cabanac\orcid{0000-0001-6679-2600}\inst{\ref{aff83}}
\and A.~Cappi\inst{\ref{aff102},\ref{aff4}}
\and F.~Caro\orcid{0009-0003-1053-0507}\inst{\ref{aff33}}
\and T.~Castro\orcid{0000-0002-6292-3228}\inst{\ref{aff103},\ref{aff7},\ref{aff10}}
\and J.~A.~Escartin~Vigo\inst{\ref{aff49}}
\and J.~Garc\'ia-Bellido\orcid{0000-0002-9370-8360}\inst{\ref{aff82}}
\and T.~Gasparetto\orcid{0000-0002-7913-4866}\inst{\ref{aff33}}
\and V.~Gautard\inst{\ref{aff8}}
\and J.~Macias-Perez\orcid{0000-0002-5385-2763}\inst{\ref{aff104}}
\and R.~Maoli\orcid{0000-0002-6065-3025}\inst{\ref{aff105},\ref{aff33}}
\and R.~B.~Metcalf\orcid{0000-0003-3167-2574}\inst{\ref{aff3},\ref{aff4}}
\and A.~A.~Nucita\inst{\ref{aff106},\ref{aff107},\ref{aff108}}
\and M.~P\"ontinen\orcid{0000-0001-5442-2530}\inst{\ref{aff61}}
\and E.~Sarpa\orcid{0000-0002-1256-655X}\inst{\ref{aff7}}
\and V.~Scottez\orcid{0009-0008-3864-940X}\inst{\ref{aff109},\ref{aff110}}
\and M.~Sereno\orcid{0000-0003-0302-0325}\inst{\ref{aff4},\ref{aff22}}
\and M.~Tenti\orcid{0000-0002-4254-5901}\inst{\ref{aff22}}
\and M.~Tucci\inst{\ref{aff43}}
\and M.~Viel\orcid{0000-0002-2642-5707}\inst{\ref{aff10},\ref{aff7},\ref{aff20},\ref{aff19},\ref{aff98}}
\and M.~Wiesmann\orcid{0009-0000-8199-5860}\inst{\ref{aff51}}
\and J.~A.~Acevedo~Barroso\orcid{0000-0002-9654-1711}\inst{\ref{aff53}}
\and Y.~Akrami\orcid{0000-0002-2407-7956}\inst{\ref{aff82},\ref{aff111}}
\and I.~T.~Andika\orcid{0000-0001-6102-9526}\inst{\ref{aff50}}
\and S.~Anselmi\orcid{0000-0002-3579-9583}\inst{\ref{aff81},\ref{aff1},\ref{aff112}}
\and M.~Archidiacono\orcid{0000-0003-4952-9012}\inst{\ref{aff113},\ref{aff114}}
\and G.~Aric\`o\orcid{0000-0002-2802-2928}\inst{\ref{aff22}}
\and F.~Atrio-Barandela\orcid{0000-0002-2130-2513}\inst{\ref{aff115}}
\and M.~Baes\orcid{0000-0002-3930-2757}\inst{\ref{aff116}}
\and L.~Bazzanini\orcid{0000-0003-0727-0137}\inst{\ref{aff96},\ref{aff4}}
\and D.~Bertacca\orcid{0000-0002-2490-7139}\inst{\ref{aff1},\ref{aff2},\ref{aff81}}
\and M.~Bethermin\orcid{0000-0002-3915-2015}\inst{\ref{aff117}}
\and F.~Beutler\orcid{0000-0003-0467-5438}\inst{\ref{aff36}}
\and L.~Blot\orcid{0000-0002-9622-7167}\inst{\ref{aff118},\ref{aff63}}
\and M.~Bonici\orcid{0000-0002-8430-126X}\inst{\ref{aff119},\ref{aff5}}
\and M.~L.~Brown\orcid{0000-0002-0370-8077}\inst{\ref{aff120}}
\and S.~Bruton\orcid{0000-0002-6503-5218}\inst{\ref{aff121}}
\and A.~Calabro\orcid{0000-0003-2536-1614}\inst{\ref{aff33}}
\and B.~Camacho~Quevedo\orcid{0000-0002-8789-4232}\inst{\ref{aff10},\ref{aff20},\ref{aff7}}
\and P.~Carrilho\orcid{0000-0003-1339-0194}\inst{\ref{aff122}}
\and C.~S.~Carvalho\inst{\ref{aff89}}
\and F.~Cogato\orcid{0000-0003-4632-6113}\inst{\ref{aff3},\ref{aff4}}
\and T.~E.~Collett\orcid{0000-0001-5564-3140}\inst{\ref{aff123}}
\and S.~Conseil\orcid{0000-0002-3657-4191}\inst{\ref{aff38}}
\and A.~R.~Cooray\orcid{0000-0002-3892-0190}\inst{\ref{aff124}}
\and P.~Corcho-Caballero\orcid{0000-0001-6327-7080}\inst{\ref{aff125}}
\and B.~Csizi\orcid{0000-0003-3227-6581}\inst{\ref{aff86}}
\and H.~Dannerbauer\orcid{0000-0001-7147-3575}\inst{\ref{aff13},\ref{aff14}}
\and T.~de~Boer\orcid{0000-0001-5486-2747}\inst{\ref{aff126}}
\and G.~Desprez\orcid{0000-0001-8325-1742}\inst{\ref{aff125}}
\and A.~D\'iaz-S\'anchez\orcid{0000-0003-0748-4768}\inst{\ref{aff127}}
\and S.~Di~Domizio\orcid{0000-0003-2863-5895}\inst{\ref{aff23},\ref{aff24}}
\and J.~M.~Diego\orcid{0000-0001-9065-3926}\inst{\ref{aff128}}
\and V.~Duret\orcid{0009-0009-0383-4960}\inst{\ref{aff44}}
\and M.~Y.~Elkhashab\orcid{0000-0001-9306-2603}\inst{\ref{aff7},\ref{aff19},\ref{aff72},\ref{aff10}}
\and Y.~Fang\orcid{0000-0002-0334-6950}\inst{\ref{aff50}}
\and A.~Farina\orcid{0009-0000-3420-929X}\inst{\ref{aff18},\ref{aff24}}
\and A.~Finoguenov\orcid{0000-0002-4606-5403}\inst{\ref{aff61}}
\and A.~Franco\orcid{0000-0002-4761-366X}\inst{\ref{aff106},\ref{aff107},\ref{aff108}}
\and Y.~Fu\orcid{0000-0002-0759-0504}\inst{\ref{aff30},\ref{aff125}}
\and K.~Ganga\orcid{0000-0001-8159-8208}\inst{\ref{aff99}}
\and R.~Gavazzi\orcid{0000-0002-5540-6935}\inst{\ref{aff12},\ref{aff129}}
\and E.~Gaztanaga\orcid{0000-0001-9632-0815}\inst{\ref{aff15},\ref{aff46},\ref{aff123}}
\and Z.~Ghaffari\orcid{0000-0002-6467-8078}\inst{\ref{aff7},\ref{aff10}}
\and F.~Giacomini\orcid{0000-0002-3129-2814}\inst{\ref{aff22}}
\and F.~Gianotti\orcid{0000-0003-4666-119X}\inst{\ref{aff4}}
\and G.~Gozaliasl\orcid{0000-0002-0236-919X}\inst{\ref{aff130},\ref{aff61}}
\and A.~Gruppuso\orcid{0000-0001-9272-5292}\inst{\ref{aff4},\ref{aff22}}
\and M.~Guidi\orcid{0000-0001-9408-1101}\inst{\ref{aff21},\ref{aff4}}
\and C.~M.~Gutierrez\orcid{0000-0001-7854-783X}\inst{\ref{aff13},\ref{aff14}}
\and A.~Hall\orcid{0000-0002-3139-8651}\inst{\ref{aff36}}
\and C.~Hern\'andez-Monteagudo\orcid{0000-0001-5471-9166}\inst{\ref{aff14},\ref{aff13}}
\and H.~Hildebrandt\orcid{0000-0002-9814-3338}\inst{\ref{aff131}}
\and J.~Hjorth\orcid{0000-0002-4571-2306}\inst{\ref{aff132}}
\and M.~Jauzac\orcid{0000-0003-1974-8732}\inst{\ref{aff133},\ref{aff68},\ref{aff134},\ref{aff135},\ref{aff83}}
\and J.~J.~E.~Kajava\orcid{0000-0002-3010-8333}\inst{\ref{aff136},\ref{aff137},\ref{aff138}}
\and Y.~Kang\orcid{0009-0000-8588-7250}\inst{\ref{aff86}}
\and V.~Kansal\orcid{0000-0002-4008-6078}\inst{\ref{aff139},\ref{aff140}}
\and D.~Karagiannis\orcid{0000-0002-4927-0816}\inst{\ref{aff96},\ref{aff141}}
\and J.~Kim\orcid{0000-0003-2776-2761}\inst{\ref{aff6}}
\and C.~C.~Kirkpatrick\inst{\ref{aff87}}
\and A.~Kov\'acs\orcid{0000-0002-5825-579X}\inst{\ref{aff142},\ref{aff143}}
\and I.~Kova{\v{c}}i{\'{c}}\orcid{0000-0001-6751-3263}\inst{\ref{aff116}}
\and K.~Koyama\orcid{0000-0001-6727-6915}\inst{\ref{aff123}}
\and S.~Kruk\orcid{0000-0001-8010-8879}\inst{\ref{aff17}}
\and M.~C.~Lam\orcid{0000-0002-9347-2298}\inst{\ref{aff36}}
\and M.~Lattanzi\orcid{0000-0003-1059-2532}\inst{\ref{aff97}}
\and F.~Leclercq\orcid{0000-0002-9339-1404}\inst{\ref{aff129}}
\and L.~Legrand\orcid{0000-0003-0610-5252}\inst{\ref{aff144},\ref{aff16}}
\and M.~Lembo\orcid{0000-0002-5271-5070}\inst{\ref{aff129}}
\and G.~F.~Lesci\orcid{0000-0002-4607-2830}\inst{\ref{aff3},\ref{aff4}}
\and J.~Lesgourgues\orcid{0000-0001-7627-353X}\inst{\ref{aff145}}
\and L.~Linke\orcid{0000-0002-2622-8113}\inst{\ref{aff86}}
\and S.~J.~Liu\orcid{0000-0001-7680-2139}\inst{\ref{aff45}}
\and X.~Lopez~Lopez\orcid{0009-0008-5194-5908}\inst{\ref{aff4}}
\and G.~Maggio\orcid{0000-0003-4020-4836}\inst{\ref{aff7}}
\and F.~Mannucci\orcid{0000-0002-4803-2381}\inst{\ref{aff9}}
\and F.~R.~Marleau\orcid{0000-0002-1442-2947}\inst{\ref{aff86}}
\and C.~J.~A.~P.~Martins\orcid{0000-0002-4886-9261}\inst{\ref{aff146},\ref{aff147}}
\and M.~Miluzio\inst{\ref{aff17},\ref{aff148}}
\and G.~Morgante\inst{\ref{aff4}}
\and S.~Nadathur\orcid{0000-0001-9070-3102}\inst{\ref{aff123}}
\and K.~Naidoo\orcid{0000-0002-9182-1802}\inst{\ref{aff123},\ref{aff57}}
\and A.~Navarro-Alsina\orcid{0000-0002-3173-2592}\inst{\ref{aff66}}
\and S.~Nesseris\orcid{0000-0002-0567-0324}\inst{\ref{aff82}}
\and L.~Nicastro\orcid{0000-0001-8534-6788}\inst{\ref{aff4}}
\and F.~Oppizzi\orcid{0000-0003-3904-8370}\inst{\ref{aff24},\ref{aff81},\ref{aff1}}
\and F.~Pace\orcid{0000-0001-8039-0480}\inst{\ref{aff27},\ref{aff28},\ref{aff29},\ref{aff149}}
\and D.~Paoletti\orcid{0000-0003-4761-6147}\inst{\ref{aff4},\ref{aff101}}
\and F.~Passalacqua\orcid{0000-0002-8606-4093}\inst{\ref{aff81}}
\and K.~Paterson\orcid{0000-0001-8340-3486}\inst{\ref{aff57}}
\and L.~Patrizii\inst{\ref{aff22}}
\and C.~Pattison\orcid{0000-0003-3272-2617}\inst{\ref{aff123}}
\and A.~Pisani\orcid{0000-0002-6146-4437}\inst{\ref{aff44}}
\and D.~Potter\orcid{0000-0002-0757-5195}\inst{\ref{aff150}}
\and G.~W.~Pratt\inst{\ref{aff77}}
\and M.~Radovich\orcid{0000-0002-3585-866X}\inst{\ref{aff2}}
\and G.~Rodighiero\orcid{0000-0002-9415-2296}\inst{\ref{aff1},\ref{aff2}}
\and W.~Roster\orcid{0000-0002-9149-6528}\inst{\ref{aff49}}
\and S.~Sacquegna\orcid{0000-0002-8433-6630}\inst{\ref{aff151}}
\and M.~Sahl\'en\orcid{0000-0003-0973-4804}\inst{\ref{aff152}}
\and D.~B.~Sanders\orcid{0000-0002-1233-9998}\inst{\ref{aff126}}
\and A.~Schneider\orcid{0000-0001-7055-8104}\inst{\ref{aff150}}
\and D.~Sciotti\orcid{0009-0008-4519-2620}\inst{\ref{aff33},\ref{aff67}}
\and E.~Sellentin\orcid{0009-0002-2655-3458}\inst{\ref{aff153},\ref{aff30}}
\and S.~Serjeant\orcid{0000-0002-0517-7943}\inst{\ref{aff154}}
\and F.~Shankar\orcid{0000-0001-8973-5051}\inst{\ref{aff155}}
\and L.~C.~Smith\orcid{0000-0002-3259-2771}\inst{\ref{aff156}}
\and J.~G.~Sorce\orcid{0000-0002-2307-2432}\inst{\ref{aff157},\ref{aff16}}
\and I.~Szapudi\orcid{0000-0003-2274-0301}\inst{\ref{aff126}}
\and K.~Tanidis\orcid{0000-0001-9843-5130}\inst{\ref{aff158}}
\and G.~Testera\orcid{0000-0003-2970-766X}\inst{\ref{aff24}}
\and R.~Teyssier\orcid{0000-0001-7689-0933}\inst{\ref{aff159}}
\and S.~Tosi\orcid{0000-0002-7275-9193}\inst{\ref{aff23},\ref{aff18},\ref{aff24}}
\and A.~Troja\orcid{0000-0003-0239-4595}\inst{\ref{aff7}}
\and C.~Uhlemann\orcid{0000-0001-7831-1579}\inst{\ref{aff160},\ref{aff76}}
\and C.~Valieri\inst{\ref{aff22}}
\and A.~Venhola\orcid{0000-0001-6071-4564}\inst{\ref{aff161}}
\and G.~Verza\orcid{0000-0002-1886-8348}\thanks{Deceased}\inst{\ref{aff162},\ref{aff163}}
\and S.~Vinciguerra\orcid{0009-0005-4018-3184}\inst{\ref{aff12}}
\and M.~von~Wietersheim-Kramsta\orcid{0000-0003-4986-5091}\inst{\ref{aff68},\ref{aff133}}
\and N.~A.~Walton\orcid{0000-0003-3983-8778}\inst{\ref{aff156}}
\and L.~Wang\orcid{0000-0002-6736-9158}\inst{\ref{aff164},\ref{aff125}}
\and A.~H.~Wright\orcid{0000-0001-7363-7932}\inst{\ref{aff131}}}
										   
\institute{Dipartimento di Fisica e Astronomia "G. Galilei", Universit\`a di Padova, Via Marzolo 8, 35131 Padova, Italy\label{aff1}
\and
INAF-Osservatorio Astronomico di Padova, Via dell'Osservatorio 5, 35122 Padova, Italy\label{aff2}
\and
Dipartimento di Fisica e Astronomia "Augusto Righi" - Alma Mater Studiorum Universit\`a di Bologna, via Piero Gobetti 93/2, 40129 Bologna, Italy\label{aff3}
\and
INAF-Osservatorio di Astrofisica e Scienza dello Spazio di Bologna, Via Piero Gobetti 93/3, 40129 Bologna, Italy\label{aff4}
\and
INAF-IASF Milano, Via Alfonso Corti 12, 20133 Milano, Italy\label{aff5}
\and
Department of Physics, University of Oxford, Keble Road, Oxford OX1 3RH, UK\label{aff6}
\and
INAF-Osservatorio Astronomico di Trieste, Via G. B. Tiepolo 11, 34143 Trieste, Italy\label{aff7}
\and
CEA Saclay, DFR/IRFU, Service d'Astrophysique, Bat. 709, 91191 Gif-sur-Yvette, France\label{aff8}
\and
INAF-Osservatorio Astrofisico di Arcetri, Largo E. Fermi 5, 50125, Firenze, Italy\label{aff9}
\and
IFPU, Institute for Fundamental Physics of the Universe, via Beirut 2, 34151 Trieste, Italy\label{aff10}
\and
Institute of Physics, Laboratory for Galaxy Evolution, Ecole Polytechnique F\'ed\'erale de Lausanne, Observatoire de Sauverny, CH-1290 Versoix, Switzerland\label{aff11}
\and
Aix-Marseille Universit\'e, CNRS, CNES, LAM, Marseille, France\label{aff12}
\and
Instituto de Astrof\'{\i}sica de Canarias, E-38205 La Laguna, Tenerife, Spain\label{aff13}
\and
Universidad de La Laguna, Dpto. Astrof\'\i sica, E-38206 La Laguna, Tenerife, Spain\label{aff14}
\and
Institute of Space Sciences (ICE, CSIC), Campus UAB, Carrer de Can Magrans, s/n, 08193 Barcelona, Spain\label{aff15}
\and
Universit\'e Paris-Saclay, CNRS, Institut d'astrophysique spatiale, 91405, Orsay, France\label{aff16}
\and
ESAC/ESA, Camino Bajo del Castillo, s/n., Urb. Villafranca del Castillo, 28692 Villanueva de la Ca\~nada, Madrid, Spain\label{aff17}
\and
INAF-Osservatorio Astronomico di Brera, Via Brera 28, 20122 Milano, Italy\label{aff18}
\and
INFN, Sezione di Trieste, Via Valerio 2, 34127 Trieste TS, Italy\label{aff19}
\and
SISSA, International School for Advanced Studies, Via Bonomea 265, 34136 Trieste TS, Italy\label{aff20}
\and
Dipartimento di Fisica e Astronomia, Universit\`a di Bologna, Via Gobetti 93/2, 40129 Bologna, Italy\label{aff21}
\and
INFN-Sezione di Bologna, Viale Berti Pichat 6/2, 40127 Bologna, Italy\label{aff22}
\and
Dipartimento di Fisica, Universit\`a di Genova, Via Dodecaneso 33, 16146, Genova, Italy\label{aff23}
\and
INFN-Sezione di Genova, Via Dodecaneso 33, 16146, Genova, Italy\label{aff24}
\and
Department of Physics "E. Pancini", University Federico II, Via Cinthia 6, 80126, Napoli, Italy\label{aff25}
\and
INAF-Osservatorio Astronomico di Capodimonte, Via Moiariello 16, 80131 Napoli, Italy\label{aff26}
\and
Dipartimento di Fisica, Universit\`a degli Studi di Torino, Via P. Giuria 1, 10125 Torino, Italy\label{aff27}
\and
INFN-Sezione di Torino, Via P. Giuria 1, 10125 Torino, Italy\label{aff28}
\and
INAF-Osservatorio Astrofisico di Torino, Via Osservatorio 20, 10025 Pino Torinese (TO), Italy\label{aff29}
\and
Leiden Observatory, Leiden University, Einsteinweg 55, 2333 CC Leiden, The Netherlands\label{aff30}
\and
Centro de Investigaciones Energ\'eticas, Medioambientales y Tecnol\'ogicas (CIEMAT), Avenida Complutense 40, 28040 Madrid, Spain\label{aff31}
\and
Port d'Informaci\'{o} Cient\'{i}fica, Campus UAB, C. Albareda s/n, 08193 Bellaterra (Barcelona), Spain\label{aff32}
\and
INAF-Osservatorio Astronomico di Roma, Via Frascati 33, 00078 Monteporzio Catone, Italy\label{aff33}
\and
INFN -- Sezione di Napoli, Via Cinthia 6, 80126, Napoli, Italy\label{aff34}
\and
Dipartimento di Fisica e Astronomia "Augusto Righi" - Alma Mater Studiorum Universit\`a di Bologna, Viale Berti Pichat 6/2, 40127 Bologna, Italy\label{aff35}
\and
Institute for Astronomy, University of Edinburgh, Royal Observatory, Blackford Hill, Edinburgh EH9 3HJ, UK\label{aff36}
\and
European Space Agency/ESRIN, Largo Galileo Galilei 1, 00044 Frascati, Roma, Italy\label{aff37}
\and
Universit\'e Claude Bernard Lyon 1, CNRS/IN2P3, IP2I Lyon, UMR 5822, Villeurbanne, F-69100, France\label{aff38}
\and
Institut de Ci\`{e}ncies del Cosmos (ICCUB), Universitat de Barcelona (IEEC-UB), Mart\'{i} i Franqu\`{e}s 1, 08028 Barcelona, Spain\label{aff39}
\and
Instituci\'o Catalana de Recerca i Estudis Avan\c{c}ats (ICREA), Passeig de Llu\'{\i}s Companys 23, 08010 Barcelona, Spain\label{aff40}
\and
Institut de Ciencies de l'Espai (IEEC-CSIC), Campus UAB, Carrer de Can Magrans, s/n Cerdanyola del Vall\'es, 08193 Barcelona, Spain\label{aff41}
\and
UCB Lyon 1, CNRS/IN2P3, IUF, IP2I Lyon, 4 rue Enrico Fermi, 69622 Villeurbanne, France\label{aff42}
\and
Department of Astronomy, University of Geneva, ch. d'Ecogia 16, 1290 Versoix, Switzerland\label{aff43}
\and
Aix-Marseille Universit\'e, CNRS/IN2P3, CPPM, Marseille, France\label{aff44}
\and
INAF-Istituto di Astrofisica e Planetologia Spaziali, via del Fosso del Cavaliere, 100, 00100 Roma, Italy\label{aff45}
\and
Institut d'Estudis Espacials de Catalunya (IEEC),  Edifici RDIT, Campus UPC, 08860 Castelldefels, Barcelona, Spain\label{aff46}
\and
School of Physics, HH Wills Physics Laboratory, University of Bristol, Tyndall Avenue, Bristol, BS8 1TL, UK\label{aff47}
\and
University Observatory, LMU Faculty of Physics, Scheinerstr.~1, 81679 Munich, Germany\label{aff48}
\and
Max Planck Institute for Extraterrestrial Physics, Giessenbachstr. 1, 85748 Garching, Germany\label{aff49}
\and
Universit\"ats-Sternwarte M\"unchen, Fakult\"at f\"ur Physik, Ludwig-Maximilians-Universit\"at M\"unchen, Scheinerstr.~1, 81679 M\"unchen, Germany\label{aff50}
\and
Institute of Theoretical Astrophysics, University of Oslo, P.O. Box 1029 Blindern, 0315 Oslo, Norway\label{aff51}
\and
Caltech/IPAC, 1200 E. California Blvd., Pasadena, CA 91125, USA\label{aff52}
\and
Jet Propulsion Laboratory, California Institute of Technology, 4800 Oak Grove Drive, Pasadena, CA, 91109, USA\label{aff53}
\and
Felix Hormuth Engineering, Goethestr. 17, 69181 Leimen, Germany\label{aff54}
\and
Technical University of Denmark, Elektrovej 327, 2800 Kgs. Lyngby, Denmark\label{aff55}
\and
Cosmic Dawn Center (DAWN), Denmark\label{aff56}
\and
Max-Planck-Institut f\"ur Astronomie, K\"onigstuhl 17, 69117 Heidelberg, Germany\label{aff57}
\and
NASA Goddard Space Flight Center, Greenbelt, MD 20771, USA\label{aff58}
\and
Department of Physics and Astronomy, University College London, Gower Street, London WC1E 6BT, UK\label{aff59}
\and
Universit\'e de Gen\`eve, D\'epartement de Physique Th\'eorique and Centre for Astroparticle Physics, 24 quai Ernest-Ansermet, CH-1211 Gen\`eve 4, Switzerland\label{aff60}
\and
Department of Physics, P.O. Box 64, University of Helsinki, 00014 Helsinki, Finland\label{aff61}
\and
Helsinki Institute of Physics, Gustaf H{\"a}llstr{\"o}min katu 2, University of Helsinki, 00014 Helsinki, Finland\label{aff62}
\and
Laboratoire d'etude de l'Univers et des phenomenes eXtremes, Observatoire de Paris, Universit\'e PSL, Sorbonne Universit\'e, CNRS, 92190 Meudon, France\label{aff63}
\and
SKAO, Jodrell Bank, Lower Withington, Macclesfield SK11 9FT, UK\label{aff64}
\and
Centre de Calcul de l'IN2P3/CNRS, 21 avenue Pierre de Coubertin 69627 Villeurbanne Cedex, France\label{aff65}
\and
Universit\"at Bonn, Argelander-Institut f\"ur Astronomie, Auf dem H\"ugel 71, 53121 Bonn, Germany\label{aff66}
\and
INFN-Sezione di Roma, Piazzale Aldo Moro, 2 - c/o Dipartimento di Fisica, Edificio G. Marconi, 00185 Roma, Italy\label{aff67}
\and
Department of Physics, Institute for Computational Cosmology, Durham University, South Road, Durham, DH1 3LE, UK\label{aff68}
\and
Infrared Processing and Analysis Center, California Institute of Technology, Pasadena, CA 91125, USA\label{aff69}
\and
University of Applied Sciences and Arts of Northwestern Switzerland, School of Engineering, 5210 Windisch, Switzerland\label{aff70}
\and
Institute of Physics, Laboratory of Astrophysics, Ecole Polytechnique F\'ed\'erale de Lausanne (EPFL), Observatoire de Sauverny, 1290 Versoix, Switzerland\label{aff71}
\and
Dipartimento di Fisica - Sezione di Astronomia, Universit\`a di Trieste, Via Tiepolo 11, 34131 Trieste, Italy\label{aff72}
\and
Telespazio UK S.L. for European Space Agency (ESA), Camino bajo del Castillo, s/n, Urbanizacion Villafranca del Castillo, Villanueva de la Ca\~nada, 28692 Madrid, Spain\label{aff73}
\and
Institut de F\'{i}sica d'Altes Energies (IFAE), The Barcelona Institute of Science and Technology, Campus UAB, 08193 Bellaterra (Barcelona), Spain\label{aff74}
\and
European Space Agency/ESTEC, Keplerlaan 1, 2201 AZ Noordwijk, The Netherlands\label{aff75}
\and
School of Mathematics, Statistics and Physics, Newcastle University, Herschel Building, Newcastle-upon-Tyne, NE1 7RU, UK\label{aff76}
\and
Universit\'e Paris-Saclay, Universit\'e Paris Cit\'e, CEA, CNRS, AIM, 91191, Gif-sur-Yvette, France\label{aff77}
\and
Space Science Data Center, Italian Space Agency, via del Politecnico snc, 00133 Roma, Italy\label{aff78}
\and
Centre National d'Etudes Spatiales -- Centre spatial de Toulouse, 18 avenue Edouard Belin, 31401 Toulouse Cedex 9, France\label{aff79}
\and
Institute of Space Science, Str. Atomistilor, nr. 409 M\u{a}gurele, Ilfov, 077125, Romania\label{aff80}
\and
INFN-Padova, Via Marzolo 8, 35131 Padova, Italy\label{aff81}
\and
Instituto de F\'isica Te\'orica UAM-CSIC, Campus de Cantoblanco, 28049 Madrid, Spain\label{aff82}
\and
Institut de Recherche en Astrophysique et Plan\'etologie (IRAP), Universit\'e de Toulouse, CNRS, UPS, CNES, 14 Av. Edouard Belin, 31400 Toulouse, France\label{aff83}
\and
Universit\'e St Joseph; Faculty of Sciences, Beirut, Lebanon\label{aff84}
\and
Departamento de F\'isica, FCFM, Universidad de Chile, Blanco Encalada 2008, Santiago, Chile\label{aff85}
\and
Universit\"at Innsbruck, Institut f\"ur Astro- und Teilchenphysik, Technikerstr. 25/8, 6020 Innsbruck, Austria\label{aff86}
\and
Department of Physics and Helsinki Institute of Physics, Gustaf H\"allstr\"omin katu 2, University of Helsinki, 00014 Helsinki, Finland\label{aff87}
\and
Departamento de F\'isica, Faculdade de Ci\^encias, Universidade de Lisboa, Edif\'icio C8, Campo Grande, PT1749-016 Lisboa, Portugal\label{aff88}
\and
Instituto de Astrof\'isica e Ci\^encias do Espa\c{c}o, Faculdade de Ci\^encias, Universidade de Lisboa, Tapada da Ajuda, 1349-018 Lisboa, Portugal\label{aff89}
\and
Mullard Space Science Laboratory, University College London, Holmbury St Mary, Dorking, Surrey RH5 6NT, UK\label{aff90}
\and
Cosmic Dawn Center (DAWN)\label{aff91}
\and
Niels Bohr Institute, University of Copenhagen, Jagtvej 128, 2200 Copenhagen, Denmark\label{aff92}
\and
Universidad Polit\'ecnica de Cartagena, Departamento de Electr\'onica y Tecnolog\'ia de Computadoras,  Plaza del Hospital 1, 30202 Cartagena, Spain\label{aff93}
\and
European University of Technology EUt+, European Union\label{aff94}
\and
Astronomisches Rechen-Institut, Zentrum f\"ur Astronomie der Universit\"at Heidelberg, M\"onchhofstr. 12-14, 69120 Heidelberg, Germany\label{aff95}
\and
Dipartimento di Fisica e Scienze della Terra, Universit\`a degli Studi di Ferrara, Via Giuseppe Saragat 1, 44122 Ferrara, Italy\label{aff96}
\and
Istituto Nazionale di Fisica Nucleare, Sezione di Ferrara, Via Giuseppe Saragat 1, 44122 Ferrara, Italy\label{aff97}
\and
ICSC - Centro Nazionale di Ricerca in High Performance Computing, Big Data e Quantum Computing, Via Magnanelli 2, Bologna, Italy\label{aff98}
\and
Universit\'e Paris Cit\'e, CNRS, Astroparticule et Cosmologie, 75013 Paris, France\label{aff99}
\and
INAF, Istituto di Radioastronomia, Via Piero Gobetti 101, 40129 Bologna, Italy\label{aff100}
\and
INFN-Bologna, Via Irnerio 46, 40126 Bologna, Italy\label{aff101}
\and
Universit\'e C\^{o}te d'Azur, Observatoire de la C\^{o}te d'Azur, CNRS, Laboratoire Lagrange, Bd de l'Observatoire, CS 34229, 06304 Nice cedex 4, France\label{aff102}
\and
Department of Mathematical Physics, Institute of Physics, University of S\~ao Paulo, R. do Mat\~ao 1371, 05508-090, S\~ao Paulo, SP, Brazil\label{aff103}
\and
Univ. Grenoble Alpes, CNRS, Grenoble INP, LPSC-IN2P3, 53, Avenue des Martyrs, 38000, Grenoble, France\label{aff104}
\and
Dipartimento di Fisica, Sapienza Universit\`a di Roma, Piazzale Aldo Moro 2, 00185 Roma, Italy\label{aff105}
\and
Department of Mathematics and Physics E. De Giorgi, University of Salento, Via per Arnesano, CP-I93, 73100, Lecce, Italy\label{aff106}
\and
INFN, Sezione di Lecce, Via per Arnesano, CP-193, 73100, Lecce, Italy\label{aff107}
\and
INAF-Sezione di Lecce, c/o Dipartimento Matematica e Fisica, Via per Arnesano, 73100, Lecce, Italy\label{aff108}
\and
Institut d'Astrophysique de Paris, 98bis Boulevard Arago, 75014, Paris, France\label{aff109}
\and
ICL, Junia, Universit\'e Catholique de Lille, LITL, 59000 Lille, France\label{aff110}
\and
CERCA/ISO, Department of Physics, Case Western Reserve University, 10900 Euclid Avenue, Cleveland, OH 44106, USA\label{aff111}
\and
Laboratoire Univers et Th\'eorie, Observatoire de Paris, Universit\'e PSL, Universit\'e Paris Cit\'e, CNRS, 92190 Meudon, France\label{aff112}
\and
Dipartimento di Fisica "Aldo Pontremoli", Universit\`a degli Studi di Milano, Via Celoria 16, 20133 Milano, Italy\label{aff113}
\and
INFN-Sezione di Milano, Via Celoria 16, 20133 Milano, Italy\label{aff114}
\and
Departamento de F{\'\i}sica Fundamental. Universidad de Salamanca. Plaza de la Merced s/n. 37008 Salamanca, Spain\label{aff115}
\and
Universiteit Gent, Department of Physics and Astronomy, Proeftuinstraat 86 N3, 9000 Ghent, Belgium
\label{aff116}
\and
Universit\'e de Strasbourg, CNRS, Observatoire astronomique de Strasbourg, UMR 7550, 67000 Strasbourg, France\label{aff117}
\and
Center for Data-Driven Discovery, Kavli IPMU (WPI), UTIAS, The University of Tokyo, Kashiwa, Chiba 277-8583, Japan\label{aff118}
\and
Waterloo Centre for Astrophysics, University of Waterloo, Waterloo, Ontario N2L 3G1, Canada\label{aff119}
\and
Jodrell Bank Centre for Astrophysics, Department of Physics and Astronomy, University of Manchester, Oxford Road, Manchester M13 9PL, UK\label{aff120}
\and
California Institute of Technology, 1200 E California Blvd, Pasadena, CA 91125, USA\label{aff121}
\and
Department of Physics, Astronomy and Mathematics, University of Hertfordshire, College Lane, Hatfield AL10 9AB, UK\label{aff122}
\and
Institute of Cosmology and Gravitation, University of Portsmouth, Portsmouth PO1 3FX, UK\label{aff123}
\and
Department of Physics \& Astronomy, University of California Irvine, Irvine CA 92697, USA\label{aff124}
\and
Kapteyn Astronomical Institute, University of Groningen, PO Box 800, 9700 AV Groningen, The Netherlands\label{aff125}
\and
Institute for Astronomy, University of Hawaii, 2680 Woodlawn Drive, Honolulu, HI 96822, USA\label{aff126}
\and
Departamento F\'isica Aplicada, Universidad Polit\'ecnica de Cartagena, Campus Muralla del Mar, 30202 Cartagena, Murcia, Spain\label{aff127}
\and
Instituto de F\'isica de Cantabria, Edificio Juan Jord\'a, Avenida de los Castros, 39005 Santander, Spain\label{aff128}
\and
Institut d'Astrophysique de Paris, UMR 7095, CNRS, and Sorbonne Universit\'e, 98 bis boulevard Arago, 75014 Paris, France\label{aff129}
\and
Department of Computer Science, Aalto University, PO Box 15400, Espoo, FI-00 076, Finland\label{aff130}
\and
Ruhr University Bochum, Faculty of Physics and Astronomy, Astronomical Institute (AIRUB), German Centre for Cosmological Lensing (GCCL), 44780 Bochum, Germany\label{aff131}
\and
DARK, Niels Bohr Institute, University of Copenhagen, Jagtvej 155, 2200 Copenhagen, Denmark\label{aff132}
\and
Department of Physics, Centre for Extragalactic Astronomy, Durham University, South Road, Durham, DH1 3LE, UK\label{aff133}
\and
Astrophysics Research Centre, University of KwaZulu-Natal, Westville Campus, Durban 4041, South Africa\label{aff134}
\and
School of Mathematics, Statistics \& Computer Science, University of KwaZulu-Natal, Westville Campus, Durban 4041, South Africa\label{aff135}
\and
Department of Physics and Astronomy, Vesilinnantie 5, University of Turku, 20014 Turku, Finland\label{aff136}
\and
Finnish Centre for Astronomy with ESO (FINCA), Quantum, Vesilinnantie 5, University of Turku, 20014 Turku, Finland\label{aff137}
\and
Serco for European Space Agency (ESA), Camino bajo del Castillo, s/n, Urbanizacion Villafranca del Castillo, Villanueva de la Ca\~nada, 28692 Madrid, Spain\label{aff138}
\and
ARC Centre of Excellence for Dark Matter Particle Physics, Melbourne, Australia\label{aff139}
\and
Centre for Astrophysics \& Supercomputing, Swinburne University of Technology,  Hawthorn, Victoria 3122, Australia\label{aff140}
\and
Department of Physics and Astronomy, University of the Western Cape, Bellville, Cape Town, 7535, South Africa\label{aff141}
\and
MTA-CSFK Lend\"ulet Large-Scale Structure Research Group, Konkoly-Thege Mikl\'os \'ut 15-17, H-1121 Budapest, Hungary\label{aff142}
\and
Konkoly Observatory, HUN-REN CSFK, MTA Centre of Excellence, Budapest, Konkoly Thege Mikl\'os {\'u}t 15-17. H-1121, Hungary\label{aff143}
\and
Brazilian Center for Research in Physics (CBPF), Dr. Xavier Sigaud st. 150, zip 22290-180, Rio de Janeiro, RJ, Brazil\label{aff144}
\and
Institute for Theoretical Particle Physics and Cosmology (TTK), RWTH Aachen University, 52056 Aachen, Germany\label{aff145}
\and
Centro de Astrof\'{\i}sica da Universidade do Porto, Rua das Estrelas, 4150-762 Porto, Portugal\label{aff146}
\and
Instituto de Astrof\'isica e Ci\^encias do Espa\c{c}o, Universidade do Porto, CAUP, Rua das Estrelas, PT4150-762 Porto, Portugal\label{aff147}
\and
HE Space for European Space Agency (ESA), Camino bajo del Castillo, s/n, Urbanizacion Villafranca del Castillo, Villanueva de la Ca\~nada, 28692 Madrid, Spain\label{aff148}
\and
Instituto de Astrof\'isica e Ci\^encias do Espa\c{c}o, Faculdade de Ci\^encias, Universidade de Lisboa, Campo Grande, 1749-016 Lisboa, Portugal\label{aff149}
\and
Department of Astrophysics, University of Zurich, Winterthurerstrasse 190, 8057 Zurich, Switzerland\label{aff150}
\and
INAF - Osservatorio Astronomico d'Abruzzo, Via Maggini, 64100, Teramo, Italy\label{aff151}
\and
Theoretical astrophysics, Department of Physics and Astronomy, Uppsala University, Box 516, 751 37 Uppsala, Sweden\label{aff152}
\and
Mathematical Institute, University of Leiden, Einsteinweg 55, 2333 CA Leiden, The Netherlands\label{aff153}
\and
School of Physical Sciences, The Open University, Milton Keynes, MK7 6AA, UK\label{aff154}
\and
School of Physics \& Astronomy, University of Southampton, Highfield Campus, Southampton SO17 1BJ, UK\label{aff155}
\and
Institute of Astronomy, University of Cambridge, Madingley Road, Cambridge CB3 0HA, UK\label{aff156}
\and
Univ. Lille, CNRS, Centrale Lille, UMR 9189 CRIStAL, 59000 Lille, France\label{aff157}
\and
Center for Astrophysics and Cosmology, University of Nova Gorica, Nova Gorica, Slovenia\label{aff158}
\and
Department of Astrophysical Sciences, Peyton Hall, Princeton University, Princeton, NJ 08544, USA\label{aff159}
\and
Fakult\"at f\"ur Physik, Universit\"at Bielefeld, Postfach 100131, 33501 Bielefeld, Germany\label{aff160}
\and
Space physics and astronomy research unit, University of Oulu, Pentti Kaiteran katu 1, FI-90014 Oulu, Finland\label{aff161}
\and
International Centre for Theoretical Physics (ICTP), Strada Costiera 11, 34151 Trieste, Italy\label{aff162}
\and
Center for Computational Astrophysics, Flatiron Institute, 162 5th Avenue, 10010, New York, NY, USA\label{aff163}
\and
SRON Netherlands Institute for Space Research, Landleven 12, 9747 AD, Groningen, The Netherlands\label{aff164}}    

\abstract{
The \Euclid mission will revolutionize our understanding of the visible Universe by collecting images and spectra of millions of galaxies across cosmic time and in diverse environments. In this work, we forecast the number counts of galaxies emitting [\ion{O}{ii}], \hb, [\ion{O}{iii}], \ha, and Pa\,$\beta$ lines in the Euclid Wide Survey (EWS) and the Euclid Deep Survey (EDS). We begin by comparing predictions from different models, including semi-analytic models and empirical scaling relations. We then simulate spectra for galaxies in the \texttt{\texttt{MAMBO}} lightcone using the \texttt{FastSpec} tool, designed to rapidly generate one-dimensional spectra as observed by \Euclid. These simulated spectra are processed with the official \Euclid spectroscopic pipeline to recover redshifts and line fluxes, from which we derive expected number counts. Predictions for \ha{}+[\ion{N}{ii}] emitters are broadly consistent across mocks and with existing observations. For the EWS, the relative root-mean-square scatter among mock predictions for the integrated number counts is approximately $19\%$ for \ha{}+[\ion{N}{ii}], increasing to $44\%$ for [\ion{O}{ii}] and $41\%$ for [\ion{O}{iii}], illustrating the stronger model dependence of the oxygen-line forecasts. After accounting for observational effects such as flux losses and redshift misidentification, our simulations show that corrected flux measurements recover most of the intrinsic line counts, especially in the EDS. From the \texttt{MAMBO}-based \texttt{FastSpec} simulations, we predict that \Euclid will detect about 42 million \ha emitters in the EWS and 1.7 million in the EDS. The same simulations predict the recovery of approximately 0.5 and 0.8 million [\ion{O}{ii}] and [\ion{O}{iii}] emitters, respectively, in the EDS, and about 1.3 and 7.5 million in the EWS, although these non-\ha forecasts remain subject to substantial model-dependent uncertainties, as indicated by the scatter among the different mocks. These samples will be instrumental in pushing large-scale structure and baryon acoustic oscillation studies beyond redshift $z\sim2$, extending the cosmological and galaxy evolution studies with \Euclid beyond what is achievable with \ha alone.
}

%
%
   \titlerunning{\Euclid: forecast of bright line emitters}
   \authorrunning{Euclid Collaboration: P. Cassata et al.}
   
   \maketitle
%
%
%
%
   
\section{\label{sc:Intro}Introduction}
The European Space Agency (ESA) \Euclid mission is designed to investigate dark energy, dark matter, and the evolution of cosmic structures by mapping the geometry and growth of structure over a large fraction of cosmic time \citep{Laureijs11, Racca2016, Scaramella-EP1, EuclidSkyOverview}. \Euclid will survey billions of galaxies over the extragalactic sky, providing optical and near-infrared imaging together with near-infrared slitless spectroscopy. The mission combines the VIS Instrument \citep{EuclidSkyVIS} and the Near-Infrared Spectrometer and Photometer (NISP, \citealt{EuclidSkyNISP}), enabling measurements of galaxy clustering and weak gravitational lensing, which are fundamental probes for cosmology.

In particular, the VIS instrument is equipped with a broad-band filter, $\IE$, covering the $r+i+z$ wavelength range and reaching $5\sigma$ depths of $m_{\rm AB}\leq 26.5$ \citep{EuclidSkyVIS}. The NISP instrument is composed of an imager (NISP-P), equipped with the three filters $\YE$, $\JE$, and $\HE$, reaching $5\sigma$ depths of $m_{\rm AB}\leq 24.5$ each, and a slitless spectrograph (NISP-S), equipped with a blue grism (BGS) and a red grism (RGS), covering the wavelength ranges $0.93<\lambda/\mu{\rm m}<1.37$ and $1.21<\lambda/\mu{\rm m}<1.89$, respectively \citep{EuclidSkyNISP}.

The primary observing strategy of \Euclid is divided between the Euclid Wide Survey \citep[EWS;][]{Scaramella-EP1} and the Euclid Deep Survey (EDS). The EDS consists of three Euclid Deep Fields (EDFs), namely EDF-North, EDF-South, and EDF-Fornax. The EWS covers approximately \num{14\,000} square degrees of the extragalactic sky, focusing on high-precision imaging and spectroscopy at unprecedented depths, and will constitute the main data set for the primary science goals of the mission. In particular, the RGS will be employed to build a complete sample of line emitters with a $3.5\sigma$ detection limit of $2\times
10^{-16}\,{\rm erg\,s^{-1}\,cm^{-2}}$. For the remainder of the paper, we express fluxes $F$ in terms of the unitless quantity
\[
{\cal F} := \frac{F}{10^{-16}\,\mathrm{erg\,s^{-1}\,cm^{-2}}}
= \frac{F}{10^{-19}\,\mathrm{W\,m^{-2}}}.
\] 

This will enable the collection of $\sim 1000$ galaxies per $\mathrm{deg}^2$ at $1\lesssim z\lesssim 2$ through the detection of the \ha emission line \citep{Laureijs11}. The EDS will observe the EDFs with both the BGS and the RGS, reaching a four-fold increase in sensitivity compared to the EWS. It will cover $53\,\mathrm{deg}^2$, reach the faintest depths of ${\cal F}=0.5$ at $3.5\sigma$ \citep{EuclidSkyOverview}, and provide samples that will be used to assess the purity and completeness of the EWS sample.

Galaxy evolution is one of the main scientific objectives of \Euclid. By detecting galaxies across a broad range of redshifts and environments, \Euclid will constrain processes such as star formation, quenching, and morphological transformation, which regulate the transition from star-forming to quiescent galaxy populations. Its data set, covering approximately one third of the sky, will make it possible to map galaxies over a large fraction of cosmic history \citep{MadauDickinson14}, from the local Universe to cosmic noon and to earlier epochs, and across different environments, from groups to proto-clusters and clusters. This will provide constraints on the physical mechanisms through which environment affects galaxy evolution.

A defining feature of \Euclid spectroscopy is its capability to detect bright emission lines from galaxies over wide redshift ranges, including [\ion{O}{ii}]$\lambda3727$ (hereafter [\ion{O}{ii}]), \hb, [\ion{O}{iii}]$\lambda5007$ (hereafter [\ion{O}{iii}]), \ha, and Pa\,$\beta$. These lines provide complementary diagnostics of star formation \citep{Kennicutt_1998,Jansen01}, dust attenuation, gas excitation, metallicity, and AGN activity {\citep{Baldwin81,Baskin05,Kewley06}, and are therefore central to both galaxy evolution and cosmological applications of the \Euclid spectroscopic surveys.

\ha is the primary tracer for the \Euclid spectroscopic cosmology sample at $1\lesssim z\lesssim2$, where it provides a direct probe of star-forming galaxies and enables large-scale structure measurements. 
{The [\ion{N}{ii}]$\lambda\lambda6548,6583$ doublet lies close to \ha and can contribute to the blended \ha{}+[\ion{N}{ii}] complex in low-resolution slitless spectroscopy. In this work, however, \ha is the main line of interest. Unless explicitly stated otherwise, \ha refers to the deblended \ha component. We use \ha{}+[\ion{N}{ii}] only when comparing with models or observations that report the blended complex.

[\ion{O}{ii}] extends the accessible redshift range to $z\gtrsim2$, where \ha is shifted beyond the NISP wavelength coverage, although its interpretation as a star-formation tracer is more sensitive to dust, metallicity, and excitation conditions. [\ion{O}{iii}] is often one of the strongest optical lines in star-forming galaxies and AGN, and is particularly useful, together with [\ion{O}{ii}] and \hb, for probing ionisation conditions and gas-phase metallicity \citep{Kewley02,Maiolino08,Nakajima2014,Curti17}.

The hydrogen recombination lines \hb, \ha, and Pa\,$\beta$ also provide constraints on dust attenuation through line ratios such as the Balmer decrement. For standard Case B recombination, the intrinsic \ha/\hb ratio is 2.86 \citep{Osterbrock06}, and deviations from this value can be used to estimate the colour excess and dust content \citep{Battisti16}. Pa\,$\beta$, being less affected by dust attenuation than the optical lines, provides a complementary tracer of star formation in obscured systems \citep{Calabro18,Calabro19,Cleri22}. In the \Euclid wavelength range, Pa\,$\beta$ probes relatively low-redshift galaxies, where \ha is not accessible to NISP spectroscopy. 

Forecasting galaxy number counts for different spectral lines in the EWS and EDS is crucial for assessing the expected spectroscopic samples and maximising the scientific return of the mission. The expected distribution of line emitters across redshift and survey depth is needed to evaluate the ability of \Euclid to probe galaxy evolution and large-scale structure, and to quantify the impact of observational effects on line detection, redshift recovery, and flux measurement.

{The availability of Euclid Q1 data \citep{Q1-TP001}, and the forthcoming DR1 products, provides the natural next step for testing these forecasts. We do not include Q1 or preliminary DR1 number counts here, because a robust comparison requires validated spectroscopic selections, survey masks, and calibrated estimates of redshift reliability, purity, and completeness. These ingredients are particularly critical for the \ha clustering sample and for high-redshift [\ion{O}{ii}] emitters, and are being addressed in dedicated \Euclid analyses. We therefore restrict the present work to forecasts based on mock catalogues and controlled spectroscopic simulations, leaving comparison with \Euclid data to dedicated Q1/DR1 studies.

We first compare the expected number counts of [\ion{O}{ii}], \hb, [\ion{O}{iii}]$\lambda4959+\lambda5007$, \ha, and Pa\,$\beta$ emitters in the EWS and EDFs using different mock samples. We then use the predictions from the Mocks with Abundance Matching in Bologna (\texttt{MAMBO}; \citealt{Girelli21}), a tool that creates mock samples matching cosmological simulations and observed scaling relations, to construct simulated spectra of galaxies as \Euclid is expected to observe them in the two surveys. From these simulations, processed with the \Euclid spectroscopic pipeline, we derive a second set of number counts that is more directly representative of what \Euclid is expected to recover for the same emission lines within the specific \texttt{MAMBO}+\texttt{FastSpec} framework adopted here.

Although emission-line luminosity functions would provide complementary information, we deliberately restrict the present analysis to catalogue-level number counts and redshift distributions. \Euclid will nevertheless provide strong constraints on emission-line luminosity functions: the large area of the EWS will provide leverage on the bright end, while the greater depth of the EDS will extend the measurements to lower luminosities. A dedicated luminosity-function analysis, including both the luminosity distributions encoded by the different mocks and the impact of the finite simulated volume at the bright end, will require a detailed treatment of the spectroscopic selection function, flux completeness, redshift reliability, effective survey volume, and line-identification systematics, and is therefore left to future work.

The structure of the paper is as follows. In Sect.\,\ref{sc:Data} we present the mock data samples that we will use throughout the paper. In Sect.\,\ref{sc:obs_biases} we present the \Euclid-like simulated spectra that we use to account for observational biases in the number counts of line emitters. In Sect.\,\ref{sc:Results} we present the results. In Sect.\,\ref{sc:Conclusions} we draw the main conclusions of our work.

Throughout the paper we adopt a $\Lambda$CDM cosmology with $\Omega_{\rm m} = 0.315$, $\Omega_{\Lambda} = 0.685$, and $H_0 = 67.3\,\kmsMpc$. All magnitudes are expressed in the AB system \citep{Oke83}.

\section{Mock data}\label{sc:Data} 

In this section we present the mock catalogues that we used in the paper to build our forecasts for line emitters in the EWS and EDS. In the following, we focus only on the aspects of each mock catalogue that are relevant for the emission-line number-count forecasts presented in this work: the survey area and redshift range, the prescription used to assign emission-line fluxes, and the observational selections applied to construct the EWS and EDF samples. When comparing number counts among the different mocks, we apply the same survey-motivated cuts, namely \HE$<24$ and ${\cal F}_{\rm line}>2$ for the EWS, and \HE$<26$ and ${\cal F}_{\rm line}>0.5$ for the EDS. The \HE magnitude and \ha flux are available in all four mock catalogues, while the availability of the other emission lines considered in this work differs among the mocks and is specified in the corresponding subsections below. We do not impose a stellar-mass-complete selection, since the samples considered here are selected by observed near-infrared magnitude and emission-line flux.
\subsection{\texttt{MAMBO}}\label{sc:MAMBO}

\texttt{MAMBO} \citep{Girelli21,lopezlopez24} is an empirical workflow designed to build galaxy mock catalogues by populating dark-matter lightcones through abundance-matching techniques. In this work we use a \texttt{MAMBO} catalogue based on the Millennium Simulation lightcone of \citet{Henriques_2015}, covering a circular area of radius one degree ($\pi\,{\rm deg}^2$). Galaxy properties are assigned empirically through subhalo abundance matching, rather than being evolved through a semi-analytic model, following the approach described in \citet{Girelli_2020} and \citet{Girelli21}. We restrict the catalogue to the redshift range $0<z<4.1$, over which the emission lines considered in this work can fall within the BGS and RGS wavelength coverage.

The workflow statistically classifies galaxies as either star-forming (SF) or quiescent (Q) based on the relative fraction of blue and red populations in observed stellar mass functions (SMFs) at different redshifts, including the SDSS survey \citep{Peng10}, the COSMOS field \citep{Ilbert_2013}, and CANDELS \citep{Grogin11}.

Galaxy physical properties are assigned using the Empirical Galaxy Generator code (EGG; \citealt{Schreiber_2017}), which provides photometry, spectral energy distributions, star-formation rates, dust attenuation, sizes, and emission-line fluxes. The \ha luminosity is inferred from the star-formation rate using the \citet{Kennicutt_1998} calibration, hydrogen recombination lines are obtained from Case B ratios \citep{Osterbrock06}, and the remaining nebular lines are assigned through empirical relations with stellar mass, star-formation rate, and metallicity, calibrated on SDSS and extrapolated in redshift following \citet{Shapley15}. This makes \texttt{MAMBO} suitable for predicting galaxy and emission-line properties over the redshift range relevant for this work. The \texttt{MAMBO} catalogue used here provides fluxes for all five emission lines considered in this work: [\ion{O}{ii}], \hb, [\ion{O}{iii}], \ha, and Pa\,$\beta$.

For the number-count forecasts, we select star-forming galaxies with at least one emission line brighter than ${\cal F}>0.2$, allowing us to probe below the nominal EDF flux limit. We then define two survey-specific parent catalogues by applying the \Euclid \HE-band cuts, namely \HE$<24$ for the EWS and \HE$<26$ for the EDS. The resulting \HE$<26$ parent catalogue contains 366 927 galaxies and constitutes the input sample from which the \texttt{FastSpec} simulations described in Sect. \ref {sc:obs_biases} are constructed. The parent \texttt{MAMBO} catalogue extends well below the magnitude limits adopted here, being approximately complete to \HE$\simeq28$ and to stellar masses of order $10^7\,M_\odot$; therefore, its intrinsic depth does not limit either the EWS or EDS selections.

\subsection{\texttt{SPRITZ}}
\texttt{SPRITZ} \citep{Bisigello21} is a spectro-photometric mock catalogue designed to reproduce the observed galaxy population selected in the infrared. It is based on the observed infrared luminosity functions derived from \textit{Herschel} observations \citep{Gruppioni13}, and includes several galaxy populations: spiral galaxies, starbursts, type~1 and type~2 AGN, composite star-forming/AGN systems, elliptical galaxies, and irregular galaxies. Each population is assigned specific SEDs and observed LFs. The physical properties of simulated galaxies, including stellar mass, star formation rate, and AGN contribution, are estimated through spectral analysis software such as \texttt{Magphys} \citep{daCunha08} or \texttt{sed3fit} \citep{berta13}.

As an empirical mock constructed from observed luminosity and stellar-mass functions, \texttt{SPRITZ} does not have a numerical stellar-mass resolution limit analogous to that of an $N$-body-based simulation; its low-mass population is instead determined by the luminosity and stellar-mass functions adopted as input.

\texttt{SPRITZ} also includes the main emission lines needed for near-infrared spectroscopic forecasts. Hydrogen recombination lines are derived from the star-formation rate using the classical SFR--\ha calibration of \citet{Kennicutt_1998}, assuming Case~B recombination and accounting for dust attenuation. The [\ion{O}{ii}] luminosity is also tied to the star-formation rate, while [\ion{N}{ii}] is computed from \ha as a function of gas-phase metallicity following \citet{Pettini04}. The [\ion{O}{iii}] $\lambda4959$ and $\lambda5007$ lines are assigned by requiring star-forming galaxies to follow the sequence described by \citet{Kewley13}. Among the lines considered in this work, the \texttt{SPRITZ} catalogue used here provides [\ion{O}{ii}], [\ion{O}{iii}], and \ha; \hb and Pa\,$\beta$ are not included in our \texttt{SPRITZ} number-count comparison.

In this work, we use \texttt{SPRITZ} as an independent empirical prediction for the emission-line number counts. We apply the same survey-motivated selections adopted for the other mocks, namely \HE$<24$ and ${\cal F}_{\rm line}>2$ for the EWS, and \HE$<26$ and ${\cal F}_{\rm line}>0.5$ for the EDS.

\subsection{\texttt{GAEA}}

The GAlaxy Evolution and Assembly (\texttt{GAEA}) is a model of galaxy formation built on Millennium Simulation merger trees  \citep{Springel_2005} and including prescriptions for gas cooling, star formation, chemical enrichment, and stellar and AGN feedback \citep{delucia07}. The model has been progressively updated to improve the treatment of chemical enrichment, black hole growth, and feedback processes \citep{delucia14,Hirschmann16,Fontanot20}.  In this work we use the \texttt{GAEA} lightcone described in \cite{EP-Scharre}.

Emission-line luminosities in \texttt{GAEA} are computed using photoionisation models based on \texttt{CLOUDY} calculations  \citep{Ferland13,Gutkin16,Feltre16,Hirschmann17,Hirschmann19}, including ionisation from young stars, AGN narrow-line regions, and evolved stellar populations. Among the five emission lines considered in the number-count analysis, the \texttt{GAEA} catalogue provides [\ion{O}{ii}], \hb, [\ion{O}{iii}], and \ha, but not Pa\,$\beta$. This provides a physically motivated prediction for nebular emission that is independent of the empirical prescriptions adopted in \texttt{MAMBO} and \texttt{SPRITZ}.


The \texttt{GAEA} lightcone used here extends up to $z=2.5$. Consequently, for emission lines whose \Euclid-accessible redshift range extends beyond this limit, the \texttt{GAEA} predictions cover only part of the relevant interval. In particular, we do not include \texttt{GAEA} in the [\ion{O}{ii}] number-count comparisons, because the [\ion{O}{ii}] redshift range probed by \Euclid extends well beyond $z=2.5$, making a direct comparison with the other mocks incomplete. The \texttt{GAEA} lightcone also includes galaxies down to $H<25$, rather than the $H<26$ depth adopted for the EDS. This limit is related to the intrinsic mass resolution of the underlying Millennium Simulation, corresponding approximately to $M_{\star}\sim10^9\,M_{\odot}$ in \texttt{GAEA}. \cite{EP-Scharre} show, however, that most galaxies below this stellar-mass limit would not satisfy the \Euclid emission-line flux limits, reducing the impact of the mass-resolution limit on the spectroscopic samples considered here.

\subsection{Flagship}
The Flagship Simulation \citep{EuclidSkyFlagship}  is a large-volume, high-resolution $N$-body simulation developed to support the scientific preparation of the \Euclid mission. The latest realisation, commonly referred to as Flagship~2 (FS2), follows the evolution of four trillion particles in a box of $3600\,h^{-1}\,{\rm Mpc}$ on a side, resolving dark matter haloes down to masses of $\sim10^{11}\,h^{-1}M_\odot$ \citep{Potter17}. The simulation produces a lightcone covering the full sky up to $z\simeq3$, providing the large cosmological volume needed for \Euclid weak-lensing, galaxy-clustering, and spectroscopic survey forecasts.

Dark matter haloes are identified with the \texttt{ROCKSTAR} halo finder \citep{Behroozi13} and populated with galaxies using halo-occupation and abundance-matching techniques. The resulting galaxy catalogue contains several billion galaxies, with properties assigned to reproduce key observational constraints relevant for \Euclid. These include photometry, spectral energy distributions, sizes, star-formation rates, metallicities, and emission-line fluxes \citep{Blanchard-EP7,EuclidSkyOverview}. This makes Flagship a reference mock for testing survey selections, validating analysis pipelines, and comparing the expected abundance of \Euclid galaxy samples. The Flagship galaxy catalogue is complete to \HE=26, and therefore its intrinsic depth encompasses both the EWS and EDS magnitude selections adopted here.

For the purposes of this work, the most relevant aspect of Flagship is the availability of nebular emission-line fluxes. In particular, \citet{EuclidSkyFlagship} validated the expected \Euclid spectroscopic population by applying EWS-like selections to the \ha and \ha{}+[\ion{N}{ii}] fluxes. For the blended \ha{}+[\ion{N}{ii}] complex, they considered a flux limit ${\cal F}>2$ over the redshift range $0.9<z<1.8$, consistent with the nominal EWS selection. These predictions were compared with empirical models, including those of \citet{Pozzetti2016}, and with observational constraints from slitless spectroscopy \citep{Bagley2020}. This provides a useful validation of the \ha-selected spectroscopic population expected for \Euclid.

The Flagship catalogue also provides fluxes for other emission lines used in this work, including [\ion{O}{ii}], \hb, [\ion{O}{iii}], \ha, and Pa\,$\beta$ enabling tests of line ratios and their evolution with redshift. \citet{EuclidSkyFlagship} showed that the resulting line-ratio distributions broadly reproduce the observed loci of diagnostic diagrams, including comparisons with SDSS at low redshift and with higher-redshift samples such as FMOS-COSMOS \citep{Kashino19}. In addition, the size distribution of \ha-selected emitters was compared with slitless spectroscopic measurements from \citet{Bagley2020}. In this work, we use a Flagship lightcone covering $\sim50\,{\rm deg}^2$, extracted from the CosmoHub database \citep{tallada20,Carretero17}, and apply the same EWS and EDF magnitude and flux selections used for the other mocks.

\section{Accounting for observational biases}\label{sc:obs_biases}
\begin{figure*}[!ht]
    \centering \includegraphics[width=.95\columnwidth]{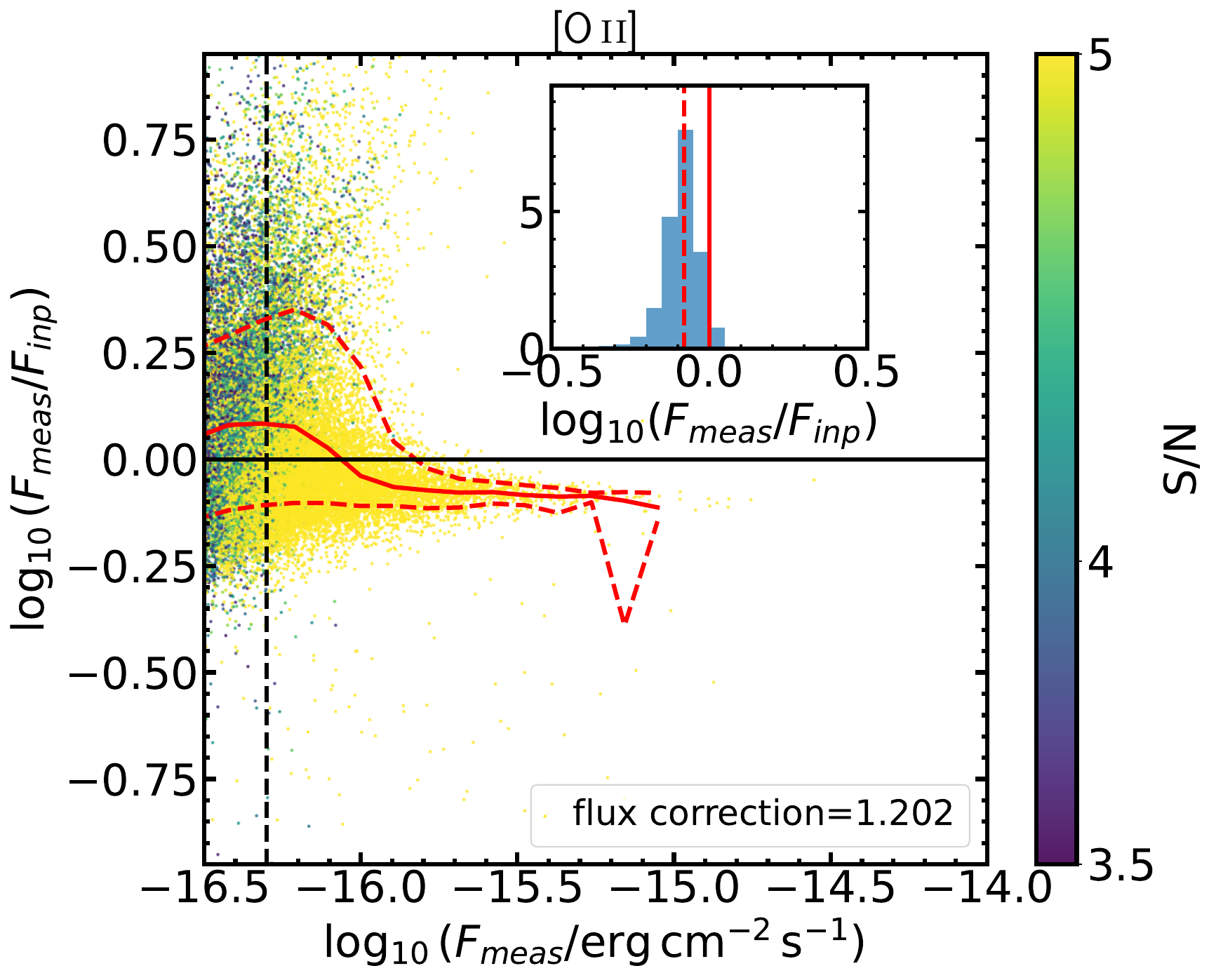}
    \centering \includegraphics[width=.95\columnwidth]{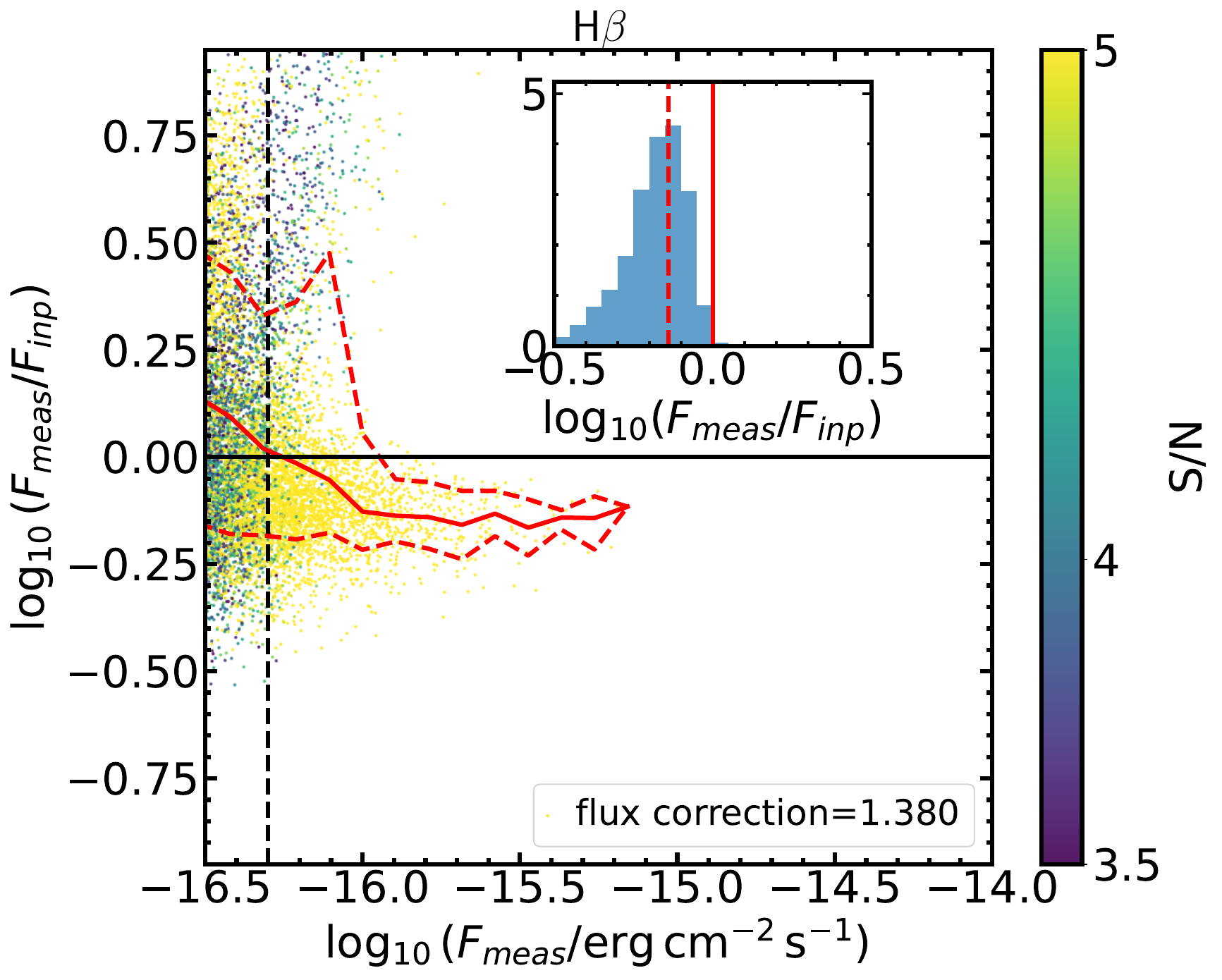}
    \centering \includegraphics[width=.95\columnwidth]{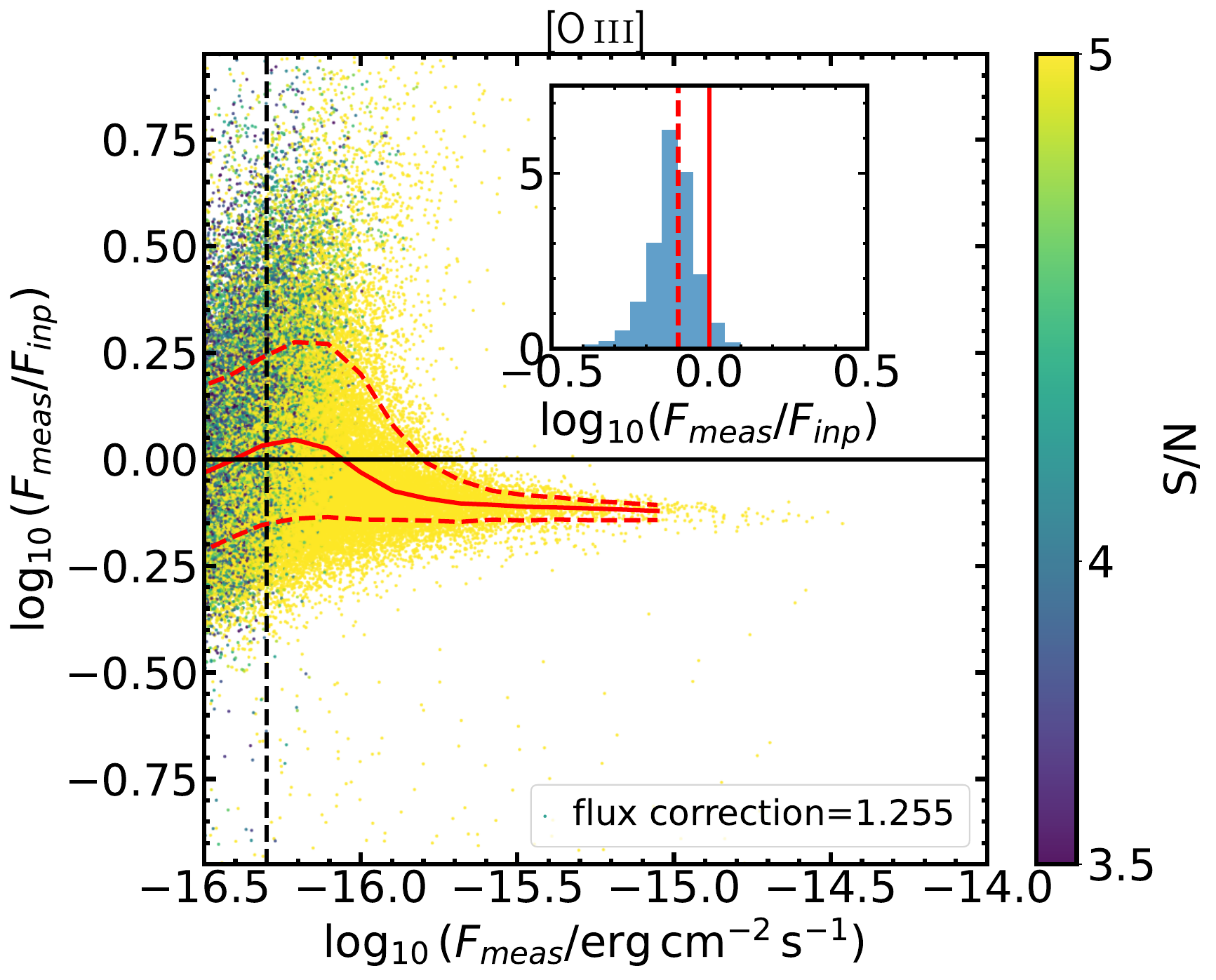}
    \centering \includegraphics[width=.95\columnwidth]{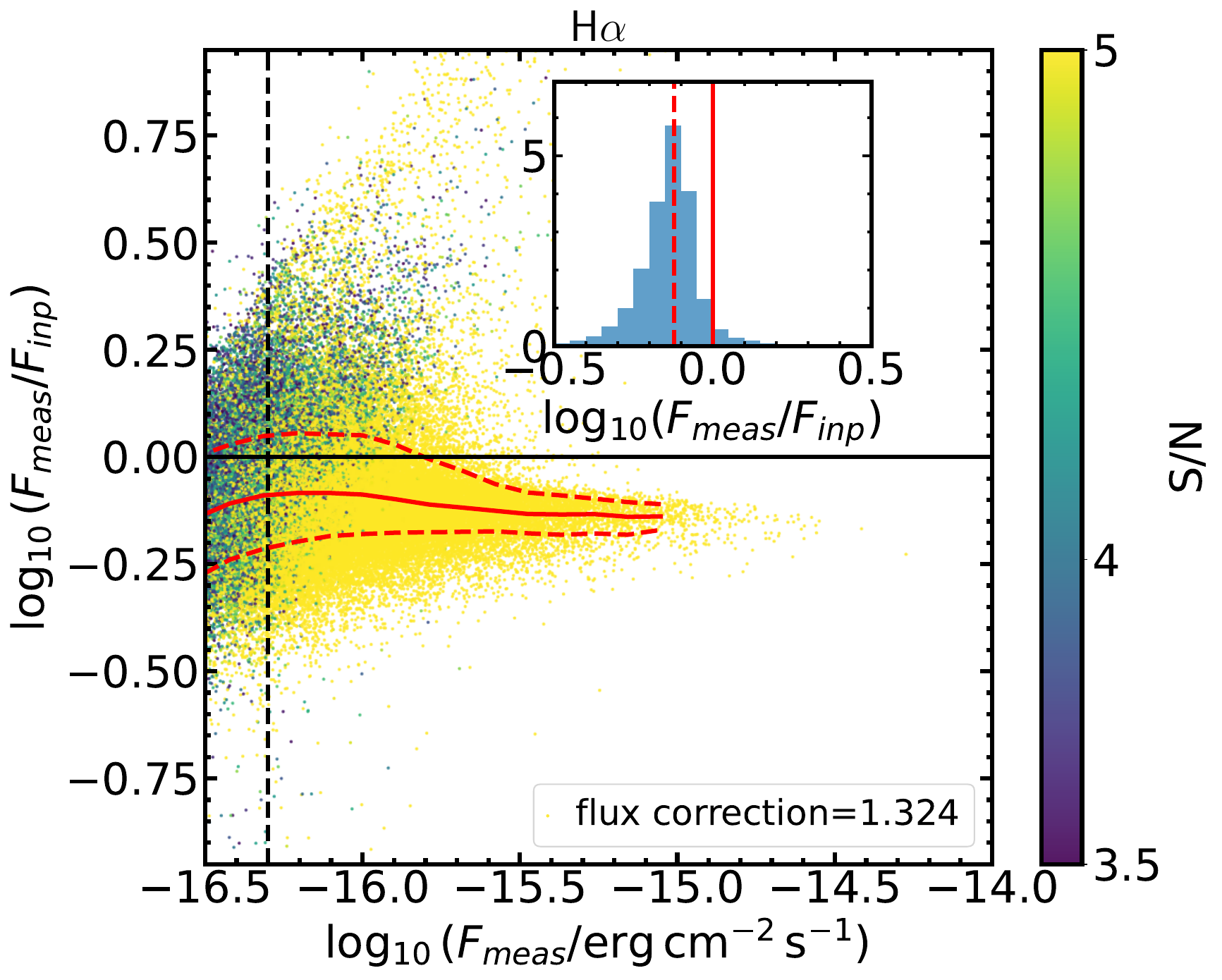}
    \centering \includegraphics[width=.95\columnwidth]{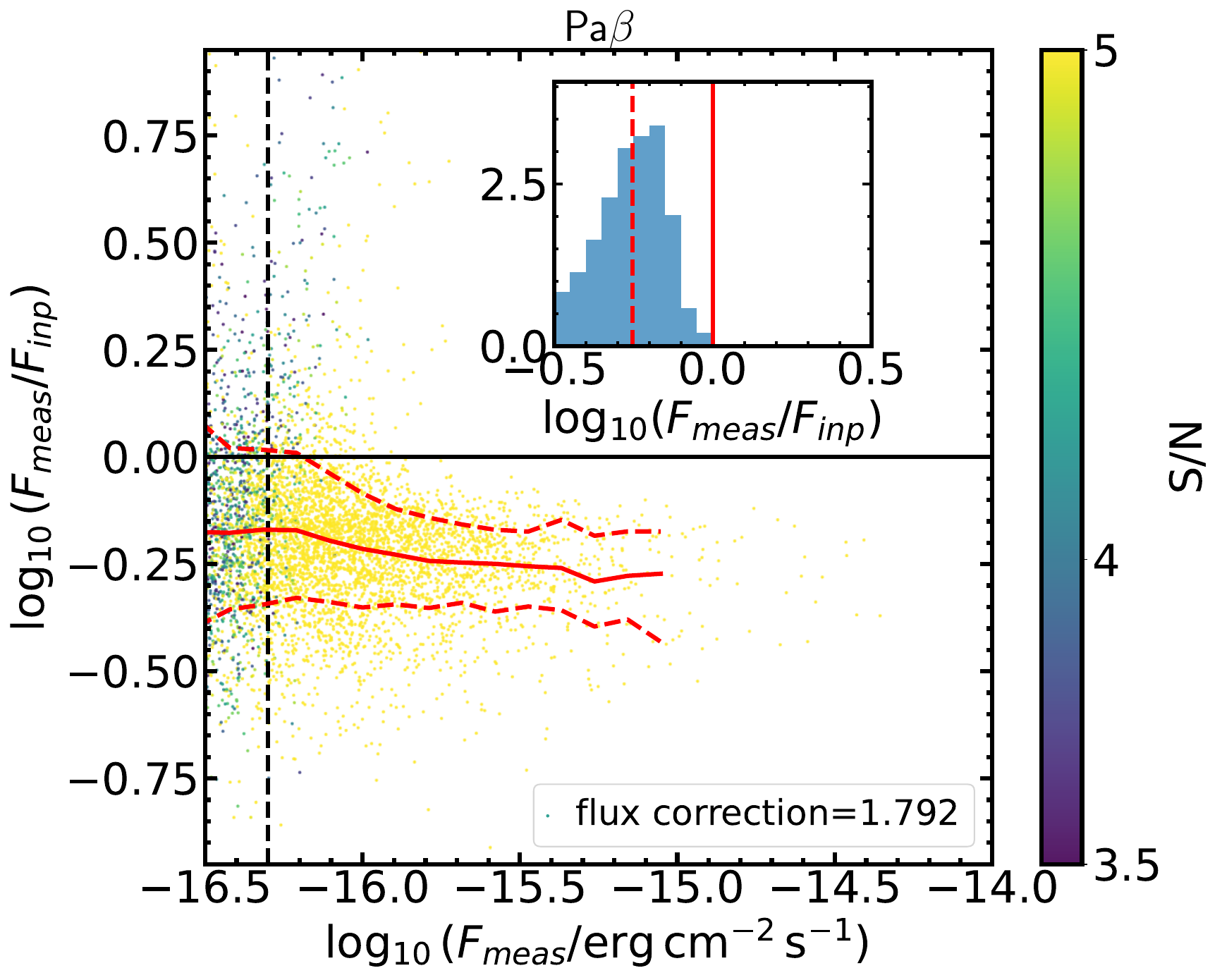}
    \caption{Comparison between the input and measured fluxes for different spectral lines in the \texttt{FastSpec} simulated spectra for the EDS, in the RGS only. Each panel shows a scatter plot of the logarithmic ratio of measured to input flux, as a function of the measured flux, for the five lines considered: [\ion{O}{ii}], \hb, [\ion{O}{iii}]5007+4959, \ha, and Pa\,$\beta$. Colors represent the S/N, with brighter colors indicating a higher S/N. Only lines with S/N $>3.5$ are shown. The colour scale indicates the line S/N and is saturated above S/N=5 for visual clarity. The insets show the distribution of $\logten(F_{\text{meas}}/F_{\text{inp}})$, with red continuous lines marking the zero and the dashed line indicating the median values. The flux correction factors are provided in each panel, indicating the average adjustment necessary for each line to align measured and input fluxes.}
    \label{fig:flux_loss}
\end{figure*}

In order to obtain more realistic forecasts for line emitters, we need to take into account the instrumental effects associated with the NISP spectrograph and quantify the possible biases introduced by the spectroscopic pipeline in measuring redshifts and line fluxes. To this end, we simulated \Euclid-like spectra for the galaxies in the \texttt{MAMBO} parent catalogue described in Sect. \ref{sc:MAMBO} using \texttt{FastSpec}, and then processed them with the official \Euclid spectroscopic pipeline to recover spectroscopic redshifts and line fluxes.


The use of \texttt{MAMBO} as input for the \texttt{FastSpec} simulations should not be interpreted as a preference for this mock over the other catalogues discussed in Sect.~\ref{sc:Data}. At the time this simulation set was produced, \texttt{MAMBO} provided a homogeneous and readily usable catalogue containing all the quantities required to generate the spectra, including continuum SEDs, emission-line fluxes, galaxy sizes, dust attenuation, and \Euclid photometry.


The other mocks are used in Sect.\,\ref{sc:exp_counts} to quantify the model dependence of the intrinsic number-count predictions. Repeating the full spectroscopic simulation and recovery procedure for all mock catalogues would provide a useful assessment of the dependence of the recovered counts on the input galaxy population, but is beyond the scope of the present work. The \texttt{FastSpec}-based results should therefore be interpreted as the outcome of applying the \Euclid spectroscopic measurement process to one representative input catalogue, rather than as a model-independent prediction.

In this section, we will present the results of these simulations and the analysis of the recovered quantities (redshifts, fluxes, ...). We stress that these simulations are intended to quantify the impact of the main instrumental and measurement effects on isolated one-dimensional spectra, and do not include all sources of complexity that will affect real \Euclid slitless observations. In particular, real NISP spectra will be affected by spectral overlap and contamination from neighbouring sources, as expected for slitless spectroscopy, as well as by additional detector- and observation-level systematics. These effects can complicate both line-flux measurements and redshift identification, and are therefore expected to affect the final success rate and contamination of the spectroscopic samples. The estimates derived below should thus be interpreted within the simplified framework of the \texttt{FastSpec} simulations, while a full treatment of spectral overlap requires end-to-end two-dimensional simulations and is beyond the scope of this work.

\subsection{\label{sc:FastSpec Simul} Simulating galaxy spectra with \texttt{FastSpec}} 
We simulated spectra using \texttt{FastSpec} (Euclid Collaboration: Granett et al., in prep.), a tool specifically built to rapidly simulate one-dimensional (1D) \Euclid-like spectra. The simulations include the effects of the point-spread function (PSF), spectral dispersion, and detector noise of the NISP instrument, including dark current, as characterised on the ground \citep{Costille19,Maciaszek22}. They also include diffuse astrophysical foregrounds contributing to the background level, namely zodiacal light and scattered stellar light.


Starting from the \texttt{MAMBO} parent catalogue described in Sect. \ref{sc:MAMBO}, we generated \Euclid-like spectra with \texttt{FastSpec}, following the prescription described in more detail in \cite{EP-Quai}. Throughout this paper, we use the term \texttt{MAMBO}+\texttt{FastSpec} to refer to this spectroscopically simulated version of the \texttt{MAMBO} sample; it is therefore not an independent mock catalogue, but the subset of \texttt{MAMBO} galaxies for which \texttt{FastSpec} spectra were produced and subsequently processed with the \Euclid spectroscopic pipeline.

In the \texttt{FastSpec} simulations used here, the 1D spectra are extracted using a fixed summation aperture of five spatial pixels across the dispersion direction. This choice provides a controlled and efficient setup for testing the recovery of redshifts and line fluxes, but it can introduce aperture flux losses, especially for spatially extended sources, which are preferentially found at lower redshift. More sophisticated extraction schemes, including size-dependent apertures and optimal extraction, are adopted or foreseen for real \Euclid data products.


A noiseless incident spectrum was then created for each galaxy in the \texttt{MAMBO}+\texttt{FastSpec} sample. \texttt{FastSpec} subsequently produced three observed spectra for each incident spectrum: one assuming the \Euclid RGS as disperser and reproducing the observing conditions of the EWS, and two reproducing the observing conditions of the EDS, one with the BGS and one with the RGS. For the EWS simulation, we further restricted the sample to \HE$<24$, since fainter objects would in any case be discarded at the spectroscopic-redshift measurement stage. The resulting EWS simulation contains 255 743 objects.


\subsection{\label{sc:SPE redshift and flux} SPE redshift and flux measurement}


These simulated spectra were then fed to the official \Euclid pipelines designed by the organisation unit spectral extraction and redshift data \citep[hereafter, OU-SPE,][]{Q1-TP007} to measure spectroscopic redshifts and line fluxes. In brief, the OU-SPE redshift measurement is based on a \Euclid-tailored version of the \texttt{AMAZED} algorithm, which fits the extracted 1D spectra with different classes of spectral templates. For galaxies, the fit combines continuum templates with emission-line templates over a grid of trial redshifts, producing a redshift probability distribution. The adopted spectroscopic redshift corresponds to the highest-ranked solution, and the pipeline also provides quality parameters describing the reliability of the measurement.


Since the redshift-measurement pipeline used for the EWS makes use of an \ha prior, designed to maximise the detection of \ha emitters (in practice, single-line emitters are preferentially interpreted as \ha), the redshift determination was performed differently for the EWS and EDS. In the EWS case, the \ha prior was applied, whereas in the EDS the redshift was determined blindly. Thus, consistently with the different \texttt{FastSpec} spectra generated for the two surveys, we kept separate catalogues of recovered quantities for the EWS and EDS.

In both catalogues, line fluxes were measured using two different methods: i) a Gaussian fit to the spectral lines, which in principle allows line complexes such as \ha{}+[\ion{N}{ii}] around 6560\,\AA\ to be deblended (hereafter gaussian fitting, GF); and ii) a direct integration of the line flux above the continuum (hereafter direct integration, DI). In this paper, as well as in the companion works by Euclid Collaboration: Mancini et al. (in prep.) and \cite{EP-Quai}, we use the GF fluxes, since this method is better suited to deblend neighbouring lines, in particular the \ha{}+[\ion{N}{ii}] complex. For both methods, the OU-SPE pipeline also provides an estimate of the signal-to-noise ratio (S/N) for all identified lines; see \cite{Q1-TP007} for details. In the \texttt{FastSpec}-based analysis presented below, we use the deblended \ha flux measured by the GF method. The notation \ha{}+[\ion{N}{ii}] is reserved for the cases in which the blended complex is required for comparison with external number-count estimates.



Each output catalogue includes a spectroscopic redshift $z_{\mathrm{meas}}$ together with several redshift-quality estimators produced by the OU-SPE pipeline. In this work, we use the redshift reliability parameter $R_z$, designed to quantify the reliability of the adopted spectroscopic redshift and ranges from 0, for unreliable solutions, to 1, for highly reliable ones. This parameter is distinct from $z_{\mathrm{prob}}$, which is associated with the probability assigned by the redshift-measurement algorithm to the selected redshift solution.

The development of improved redshift-reliability estimators, based on machine-learning algorithms that exploit the full shape of the redshift probability distribution function rather than only a single peak, is ongoing within OU-SPE. Their training requires large samples of secure redshifts, expected to be provided by the EDFs or, alternatively, by sufficiently representative simulated spectra. For this reason, $R_z$ is expected to be available only in the final calibrated \Euclid spectroscopic products. In the present simulations, however, $R_z$ is available, and we verified that using $z_{\mathrm{prob}}$ as an alternative redshift-quality estimator does not affect our results.


While the spectroscopic redshifts are measured for the EDS considering the blue and the red spectra together, the line fluxes are measured independently on the two sets of spectra. Since the blue and red spectra overlap by $\sim 150\,\AA$, we have two flux measurements for all lines that lie in the overlapping region. Here, we did not attempt to sum or average together the fluxes of the blue and red spectra, but we preferred to select among the two the measurement with the highest S/N.

\subsection{\label{sc:input vs output fluxes} Analysis of recovered fluxes}

One of the advantages of our approach, based on simulated \Euclid-like spectra, is that we can quantify observational effects that are expected to affect real data. In particular, when 1D spectra are extracted from the two-dimensional (2D) dispersed images, part of the galaxy flux can be lost outside the extraction aperture. In the \texttt{FastSpec} simulations used here, the extraction is performed with a fixed window of five spatial pixels across the dispersion direction. The resulting flux loss is therefore expected to depend primarily on the angular size and surface-brightness profile of the source, convolved with the PSF.

Figure\,\ref{fig:flux_loss} shows the comparison of measured and input fluxes for the five lines of interest for this paper, [\ion{O}{ii}], \hb, [\ion{O}{iii}], \ha, and Pa\,$\beta$, using the EDS spectra observed with the RGS as a reference. The ratio between the recovered and input flux is plotted as a function of the recovered flux, so that these plots can be used to derive empirical flux corrections for the \texttt{MAMBO}-based \texttt{FastSpec} samples analysed in the next sections. To produce each of the plots, we selected only objects for which the individual line falls within the RGS, with high redshift reliability ($R_z>0.15$; see next section for a justification of this value), and $\mathrm{S/N}>3.5$. These are the best available spectra in our simulation, and we use them as benchmark to assess the quality of the fluxes for both the EWS and EDS.

However, although the flux loss is expected to depend primarily on angular size and surface-brightness profile, this does not necessarily translate into a simple monotonic trend with measured line flux. The observed flux is not a direct proxy for angular size, since it depends on intrinsic luminosity, redshift, dust attenuation, and line strength. Moreover, for compact or marginally resolved sources the fraction of flux enclosed within a fixed aperture is expected to approach an approximately constant value. In our simulations, we also find that the measured-to-input flux ratio depends on redshift, or equivalently on the observed wavelength of the line, indicating that aperture losses are mixed with wavelength-dependent instrumental, noise, and fitting effects. The trends shown in Fig.~\ref{fig:flux_loss} should therefore be interpreted as averages over the size, redshift, wavelength, and luminosity distribution of the selected emitters, rather than as a pure function of line flux.



For all lines, the recovered flux is on average smaller than the input one, with flux losses of the order of 20\% for [\ion{O}{ii}] and [\ion{O}{iii}], around 35--40\% for \ha and \hb, and 70\% for Pa\,$\beta$. Above ${\cal F}\sim1$, the average offset is relatively stable for all lines. At lower fluxes, instead, the scatter increases significantly and the running average becomes more affected by selection effects, especially for [\ion{O}{ii}], \hb, and [\ion{O}{iii}], where a tail toward large $F_{\rm meas}/F_{\rm inp}$ appears close to the detection threshold. 

We also note that the largest positive values of $F_{\rm meas}/F_{\rm inp}$ are found preferentially at the lowest S/N values, close to the adopted detection threshold. This behaviour is expected, since the sample shown in Fig.\,\ref{fig:flux_loss} is selected on measured quantities: intrinsically faint lines can enter the sample when their measured flux is boosted upward by noise or by uncertainties in the Gaussian fit, while sources scattered below the threshold are excluded. This selection effect produces an asymmetric, non-Gaussian tail toward large $F_{\rm meas}/F_{\rm inp}$ at low S/N.

The flux offsets and trends shown in Fig.~\ref{fig:flux_loss} should therefore be interpreted in the context of the controlled \texttt{MAMBO}+\texttt{FastSpec} simulation framework adopted in this work. They mainly reflect the combination of the input galaxy population and the simplified 1D extraction setup, rather than a limitation of the OU-SPE Gaussian-fitting flux measurements, which have been extensively tested on simplified spectra. In particular, the average correction factors depend on the angular-size, surface-brightness, redshift, wavelength, and luminosity distribution of the selected galaxies. They should therefore not be interpreted as universal flux-loss corrections for the full \Euclid galaxy population.

This caveat is especially relevant for the EWS. The \texttt{MAMBO} lightcone used here covers a much smaller area than the full Euclid Wide Survey, and therefore may not sample the rarest and most luminous galaxies expected in the real survey. If the input mock had a different luminosity, redshift, or angular-size distribution, different average flux-loss corrections could be obtained, even with the same \texttt{FastSpec}+OU-SPE analysis setup. A full calibration of the flux loss as a function of galaxy size, redshift, wavelength, and line luminosity would require a dedicated analysis and is beyond the scope of the present work.

For the purposes of this paper, we therefore adopt average empirical line-by-line corrections. We estimate them above ${\cal F}\sim1$, where the ratio $F_{\rm meas}/F_{\rm inp}$ is more stable and less affected by threshold-selection effects, and report the resulting multiplicative factors in Table~\ref{tab:flux_corr}. These corrections allow us to assess how the \Euclid spectroscopic measurement process affects the recovery of line-emitter number counts within the specific \texttt{MAMBO}+\texttt{FastSpec} framework used here.

\begin{table}[t]
\centering
\caption{Multiplicative flux corrections to be applied to line fluxes, for the five lines analysed in this paper.}
\begin{tabular}{ccccc}
\hline
\hline
\noalign{\vskip 1pt}
[\ion{O}{ii}] &\hb & [\ion{O}{iii}] & \ha & Pa\,$\beta$ \\
1.203 & 1.376 & 1.255 & 1.324 & 1.792\\
\hline
\end{tabular}
\label{tab:flux_corr}
\end{table}

\section{Results and discussion}\label{sc:Results}
In this section we report our estimates for the number counts of [\ion{O}{ii}], \hb, [\ion{O}{iii}], \ha, and Pa\,$\beta$ lines. First, we introduce the predictions of the different mock samples presented in Sect.\,\ref{sc:Data}, then we discuss the predictions based on our \texttt{FastSpec} simulations, using the corrected fluxes as presented in Sect.\,\ref{sc:input vs output fluxes}.

\subsection{Expected number counts from mock samples}\label{sc:exp_counts}
\subsubsection{EWS}

\begin{figure}
    \centering \includegraphics[width=\columnwidth]{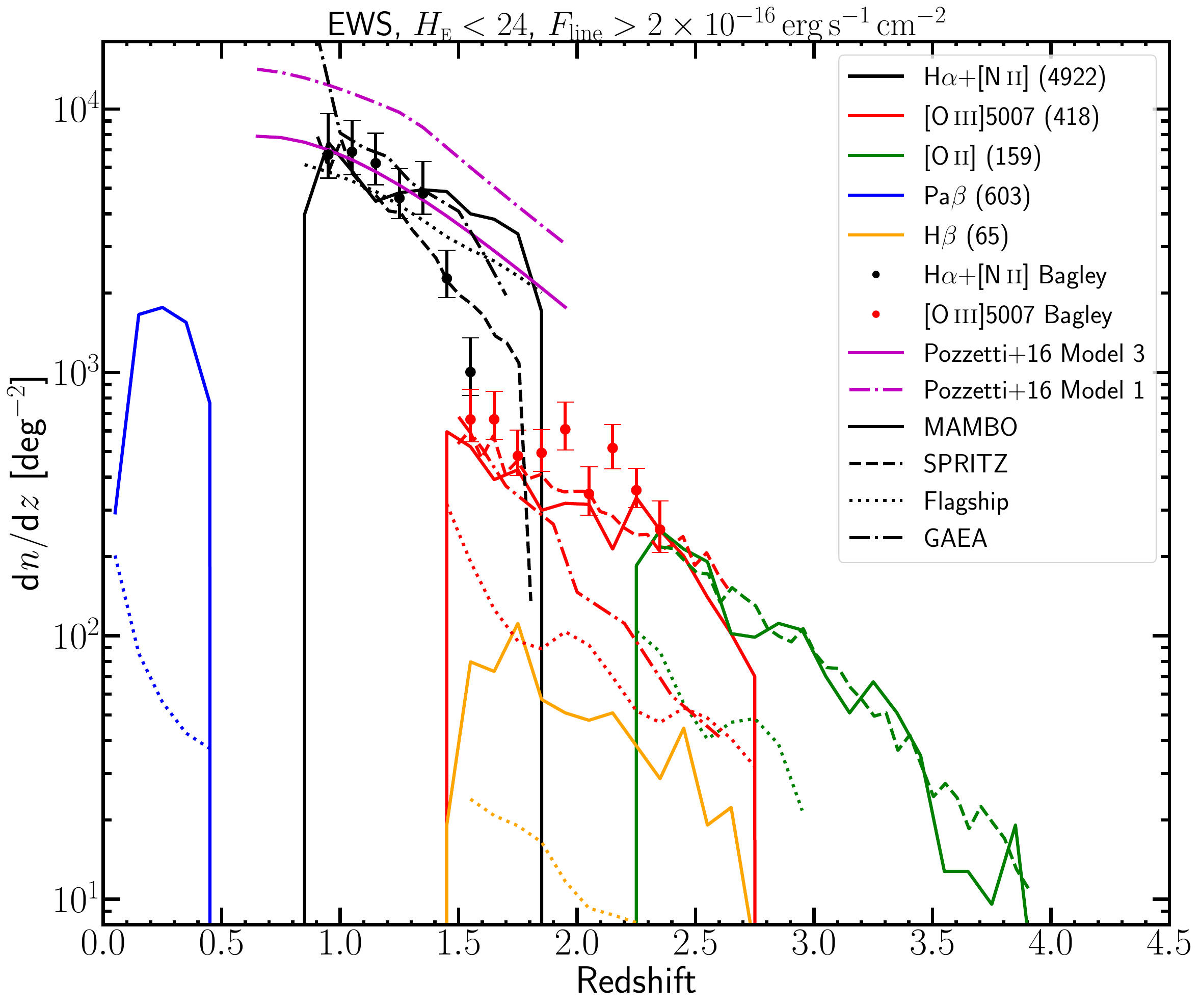}
    \caption{Number counts of emission line galaxies as a function of redshift for the EWS, for lightcones produced by different models. The solid lines represent predictions from the \texttt{MAMBO} lightcone simulation for [\ion{O}{ii}] (green), \hb (orange), [\ion{O}{iii}] (red), \ha{}+[\ion{N}{ii}] (black), and Pa\,$\beta$ (blue). Dashed and dotted lines show predictions from the \texttt{SPRITZ} and Flagship simulations, respectively. The dash-dotted line indicates predictions from the \texttt{GAEA} model. Black and red points with error bars show observed \ha+[NII] and [\ion{O}{iii}] counts from the WISP survey \citep{Bagley2020}, respectively. The magenta line represents the Model 3 prediction for \ha+[NII] in \cite{Pozzetti2016}. The numbers in parenthesis in the legend indicate the total number counts per square degree predicted according to the \texttt{MAMBO} catalogue.
    }
    \label{fig:counts_wide}
\end{figure}

\begin{table}
\centering
\caption{Line emitter number counts per square degree predicted by different simulations for various emission lines, in the EWS. For \texttt{GAEA}, the \hb count is computed only over the redshift interval covered by the lightcone ($z\leq2.5$); therefore, the reported $N=0$ should not be interpreted as a prediction over the full \Euclid RGS-accessible \hb redshift range.} 
\begin{tabular}{lcccccc}
\hline
\hline
\noalign{\vskip 1pt}
Mock lightcone & [\ion{O}{ii}] &\hb   & [\ion{O}{iii}] & \ha{}+[\ion{N}{ii}] & Pa\,$\beta$ \\
\hline
\noalign{\vskip 1pt}
\texttt{MAMBO}  & $159$             & $65$                 & $418$  & $4922$              & $603$ \\ 
Flagship        & $\phantom{0}44$   & $14$                 & $136$  & $4338$               & $\phantom{0}42$ \\ 
\texttt{SPRITZ} & $134$          & $\cdots$             & $399$  & $3282$               & $\cdots$ \\ 
\texttt{GAEA}   & $\cdots$   & $\phantom{0}0$       & $218$  & $5570$               & $\cdots$  \\ 
\hline
\end{tabular}
\label{tab:emitter_counts_wide}
\end{table}

\begin{figure}
    \centering \includegraphics[width=\columnwidth]{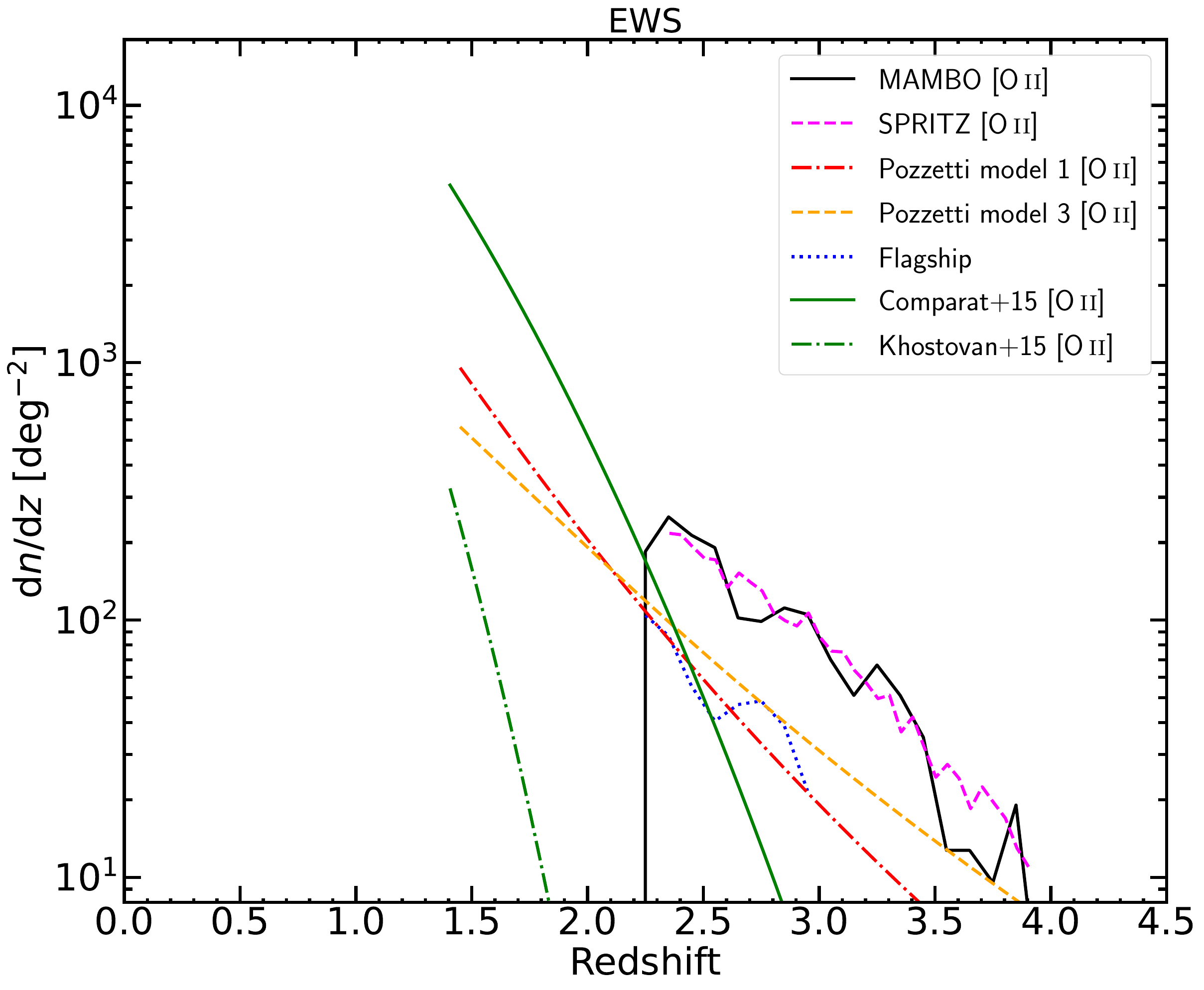}
    \caption{Predicted number counts of [\ion{O}{ii}] emitters in the \texttt{MAMBO} (continuous black), \texttt{SPRITZ} (dashed magenta), and Flagship (dotted blue) simulations, together with the Model 1 (dot-dashed red) and Model 3 (dashed orange) predictions of \cite{Pozzetti2016}, in comparison with number counts obtained from observed luminosity functions in \cite{Comparat15} (continuous green) and \cite{Khostovan15} (dot-dashed  green).}
    \label{fig:counts_wide_o2}
\end{figure}

Figure\,\ref{fig:counts_wide} shows the predicted number counts as a function of redshift, $\mathrm{d}N/\mathrm{d}z$, of emission-line galaxies for various lines, namely [\ion{O}{ii}], \hb, [\ion{O}{iii}]5007, \ha{}+[\ion{N}{ii}], and Pa\,$\beta$, from different mock catalogues (\texttt{MAMBO}, \texttt{SPRITZ}, Flagship, and \texttt{GAEA}), together with the Model\,1 and Model\,3 predictions of \cite{Pozzetti2016}. The observational constraints adopted for the EWS are applied, namely an \HE$<24$ magnitude cut and a line-flux limit ${\cal F}>2$.

In this comparison, we use the blended \ha{}+[\ion{N}{ii}] complex, because some of the observational constraints used for comparison, in particular the WISP measurements of \cite{Bagley2020}, which report the \ha{}+[\ion{N}{ii}] complex rather than a deblended \ha flux. This convention is therefore adopted only when comparing with models or observations that report the blended complex. In the \texttt{FastSpec}-based recovery analysis presented in Sect.\,\ref{sect:fastspec_sim}, instead, we use the deblended \ha flux measured by the Gaussian-fitting method. The same number counts, integrated over the whole redshift interval spanned by each emission line, are also reported in Table\,\ref{tab:emitter_counts_wide}.

The redshift interval shown for \ha{}+[\ion{N}{ii}] reflects the \Euclid RGS wavelength coverage, which allows this complex to be observed over approximately $0.9<z<1.8$. The predicted number counts for \ha{}+[\ion{N}{ii}] are broadly consistent across the different mock catalogues, within a $\pm 45\%$ range. 
Comparison with the models of \cite{Pozzetti2016} suggests a preference for their Model 3, which is in better agreement with both the various mocks and the observational estimates by \cite{Bagley2020}. Also, the mock \ha{}+[\ion{N}{ii}] counts align well with observations by \cite{Bagley2020}, which used slitless spectroscopic data from the WISP survey to estimate the number counts of \ha{}+[\ion{N}{ii}] and [\ion{O}{iii}] emitters in the redshift range probed by \Euclid. It is interesting that the \texttt{SPRITZ} number counts decrease more rapidly than the other models (\texttt{GAEA}, Flagship, \texttt{MAMBO}), in quite good agreement with \cite{Bagley2020}.

For the [\ion{O}{iii}] line, predictions from \texttt{MAMBO} and \texttt{SPRITZ} are in good agreement with the counts observed by \cite{Bagley2020} using WISP data. \texttt{GAEA} matches observations only at lower redshifts (around $z \sim 1.5$) but underpredicts counts by a factor of five at $z \sim 2.5$ compared to \texttt{MAMBO} and \texttt{SPRITZ}. The Flagship mock catalogue predicts systematically two to five times fewer [\ion{O}{iii}] emitters than \texttt{MAMBO} and \texttt{SPRITZ} over the entire redshift range. For [\ion{O}{ii}], we compare predictions from \texttt{MAMBO}, \texttt{SPRITZ}, and Flagship. We do not include \texttt{GAEA} in this comparison because the \texttt{GAEA} lightcone used here extends only to \(z=2.5\), and therefore does not cover the full \Euclid-accessible [\ion{O}{ii}] redshift range. \texttt{MAMBO} and \texttt{SPRITZ} yield broadly consistent number counts over most of the EWS redshift range, and both generally predict more [\ion{O}{ii}] emitters than Flagship, by factors of a few depending on redshift. 


For \hb, the trend is similar to [\ion{O}{ii}]: \texttt{MAMBO} predicts five to eight times more emitters than the Flagship, while \texttt{GAEA} predicts only a few, which are not visible in the figure. Similarly, for Pa\,$\beta$, \texttt{MAMBO} forecasts five to eight times more emitters than the Flagship, whereas \texttt{SPRITZ} and \texttt{GAEA} provide no predictions for this line.


Figure\,\ref{fig:counts_wide_o2} shows the redshift distribution of [\ion{O}{ii}] emitters expected in the EWS, comparing predictions from different theoretical models with observationally motivated estimates. 

Among the mock catalogues, \texttt{MAMBO} and \texttt{SPRITZ} predict broadly similar [\ion{O}{ii}] redshift distributions and normalisations over most of the redshift range probed by the EWS. Flagship generally predicts lower counts and lies closer to the empirical Model 1 and Model 3 prescriptions of \cite{Pozzetti2016}. \texttt{MAMBO} and \texttt{SPRITZ} therefore tend to define the upper envelope of the mock-catalogue predictions, especially at $z \gtrsim 2$. A larger scatter emerges when comparing model predictions with number counts inferred from observed luminosity functions. The estimates by \cite{Comparat15} and \cite{Khostovan15} differ by nearly an order of magnitude around $z \sim 2$, and both decline more rapidly with redshift than the model predictions. While the \cite{Comparat15} counts remain within a factor of $\sim 3 \mbox{--} 5$ of \texttt{MAMBO} at $z \sim 2 \mbox{--} 2.5$ and are close to the Flagship predictions, the \cite{Khostovan15} estimates lie systematically below all models across the entire redshift range.

These discrepancies should be interpreted as a warning regarding the predictive power of current models for [\ion{O}{ii}] emitters at high redshift. While the predictions for \ha{}+[\ion{N}{ii}] are relatively well constrained by existing luminosity-function measurements and by the calibration of the \Euclid spectroscopic sample, the [\ion{O}{ii}] counts are more sensitive to model-dependent assumptions. In particular, they depend on the adopted prescriptions for star formation, dust attenuation, gas metallicity, ionisation conditions, and the redshift evolution of emission-line ratios. These ingredients are treated differently in the various mocks: \texttt{MAMBO} assigns line luminosities through empirical relations tied to galaxy properties, \texttt{SPRITZ} is based on infrared luminosity functions and galaxy spectral templates, and Flagship is calibrated to reproduce the main properties of the \Euclid photometric and spectroscopic samples. The scatter among the model predictions therefore reflects the current uncertainty in modelling non-\ha emitters, especially at the bright flux limits and high redshifts relevant for the EWS.

The comparison with number counts inferred from observed [\ion{O}{ii}] luminosity functions also highlights the uncertainties in empirical extrapolations. The difference between the estimates based on \citet{Comparat15} and \citet{Khostovan15} is mainly driven by the different Schechter-function parameters adopted in the two works. In particular, the \citet{Khostovan15} luminosity functions have a flatter faint-end slope than those adopted by \citet{Comparat15} ($\alpha\simeq -1.3$ instead of $\alpha\simeq -1.6$), reducing the number of galaxies close to the \Euclid flux limit. In addition, the different normalisation and redshift evolution of the Schechter parameters, especially $\phi_\star$, leads to a faster decline of the Khostovan-based counts with redshift. These effects explain why the Khostovan et al. estimates lie systematically below the Comparat et al. predictions in Figs.\,\ref{fig:counts_wide_o2} and \ref{fig:counts_deep_o2}.

The large, homogeneous, and well-controlled spectroscopic samples that will be provided by \Euclid will play a crucial role in resolving these discrepancies, enabling a robust calibration of both semi-analytic models and empirical prescriptions, and significantly improving our understanding of star-forming galaxy populations at $z \gtrsim 2$.

\subsubsection{EDS}

\begin{figure}
    \centering \includegraphics[width=\columnwidth]{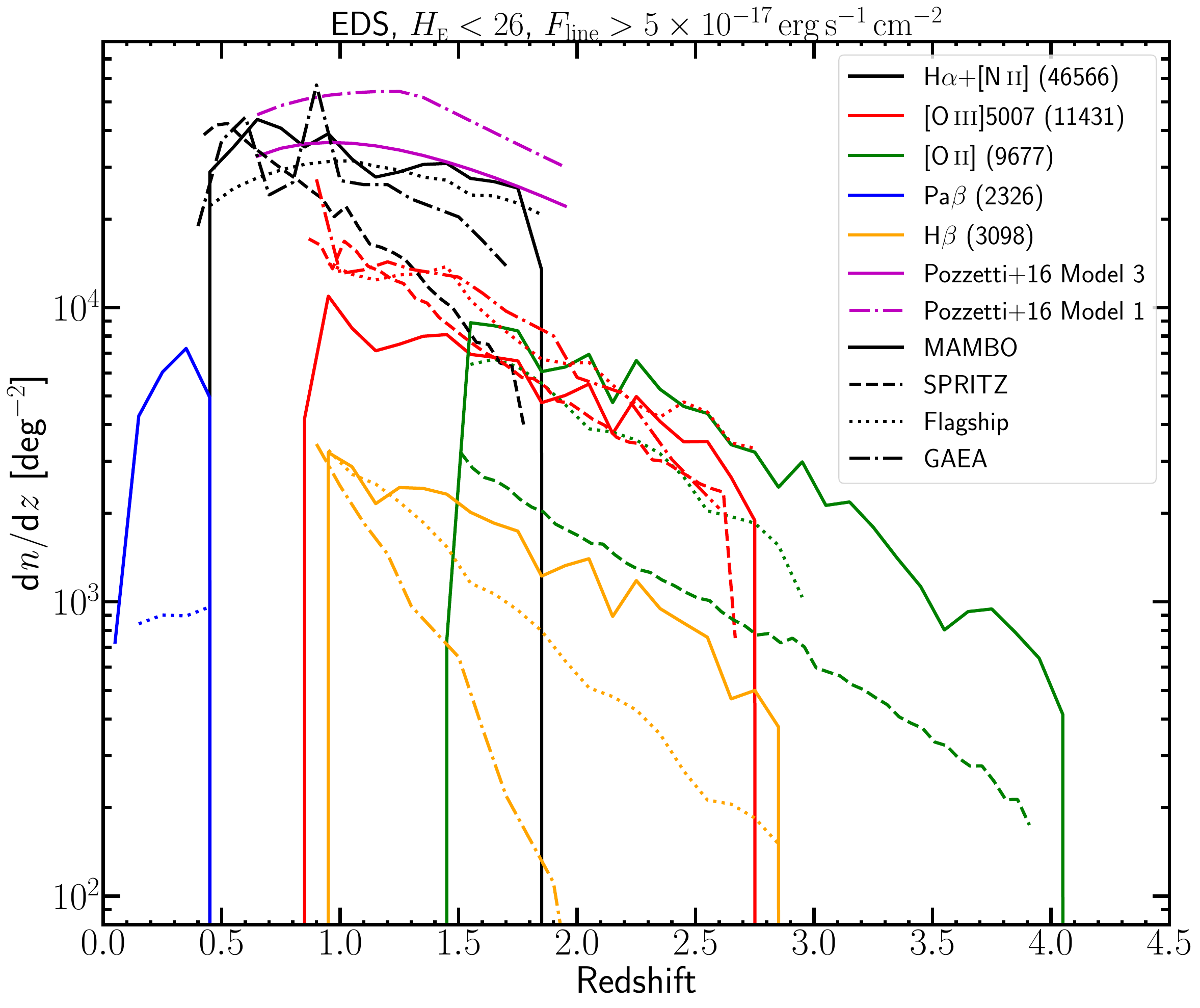}
    \caption{Same as Fig.\,\ref{fig:counts_wide} for the EDS. }
    \label{fig:counts_deep}
\end{figure}

Figure\,\ref{fig:counts_deep} reports the predicted number counts as a function of redshift ($\mathrm{d}N/\mathrm{d}z$) of emission-line galaxies for various lines ([\ion{O}{ii}], \hb, [\ion{O}{iii}]5007, \ha{}+[\ion{N}{ii}],  and Pa\,$\beta$) from different mock catalogues (\texttt{MAMBO}, \texttt{SPRITZ}, Flagship, and \texttt{GAEA}) applying the observational constraints for the EDS, which will reach a flux depth of ${\cal F}=0.5$. As in Fig.~\ref{fig:counts_wide}, we show the blended \ha{}+[\ion{N}{ii}] complex rather than \ha alone, in order to keep the comparison between the EWS and EDS predictions homogeneous.

Similarly to the EWS, the agreement between different mocks is good for \ha{}+[\ion{N}{ii}]. This is likely because \ha is the best observationally constrained emission line for \Euclid-like forecasts, with luminosity functions and number counts that have been extensively studied over the redshift range probed by the RGS. The convergence among the mocks for \ha{}+[\ion{N}{ii}] should therefore be interpreted as a consistency check, rather than as the result of a common calibration. Only the \texttt{SPRITZ} mock catalogue underpredicts \ha{}+[\ion{N}{ii}] emitters: while at the bottom of the accessible redshift interval ($z\sim0.5$) it predicts almost the same number of emitters as the other models, as the redshift increases it predicts fewer and fewer emitters ($\sim5$ times fewer than the others at $z\sim1.8$).

For [\ion{O}{iii}], the predictions among the different models agree within $\pm 50\%$, with \texttt{GAEA} and Flagship predicting more emitters than \texttt{MAMBO} and \texttt{SPRITZ} across the whole accessible redshift range ($0.9\lesssim z\lesssim2.8$). 

For [\ion{O}{ii}], we compare predictions from \texttt{MAMBO}, \texttt{SPRITZ}, and Flagship; \texttt{GAEA} is not included because its \(z\leq2.5\) lightcone does not cover the full [\ion{O}{ii}] redshift interval accessible to the EDS. For \hb, predictions are available from \texttt{MAMBO}, Flagship, and \texttt{GAEA}, although the \texttt{GAEA} prediction is limited to \(z\leq2.5\).
\texttt{MAMBO} and Flagship predict broadly comparable [\ion{O}{ii}] counts, while \texttt{SPRITZ} lies systematically lower, typically by a factor of a few over much of the accessible redshift range. 


Figure\,\ref{fig:counts_deep_o2} shows the same comparison as Fig.\,\ref{fig:counts_wide_o2}, but for the EDS. In this case, the overall level of agreement among the different predictions is noticeably improved, particularly around $z \simeq 1.5$, where most observational constraints are available and where the models are typically calibrated. In this redshift range, \texttt{MAMBO} and Flagship yield broadly consistent number counts, while \texttt{SPRITZ} predicts a lower abundance of [\ion{O}{ii}] emitters by a factor of a few. The estimates inferred from published luminosity functions span an even broader range of normalisations, with the different mock predictions generally lying between or close to the empirical extrapolations depending on redshift.

At higher redshifts, the discrepancies progressively increase. Although model predictions maintain relatively shallow declines with redshift, estimates derived from observed luminosity functions drop more rapidly, leading to growing differences beyond $z \gtrsim 2 \mbox{--} 2.5$. As in the EWS case, the prescriptions by \cite{Comparat15} and \cite{Khostovan15} bracket a wide range of normalisations, with \cite{Comparat15} remaining closer to the Flagship and \texttt{MAMBO} predictions, and \cite{Khostovan15} lying systematically below all models.

The improved agreement found for the EDS compared to the EWS (Fig.\,\ref{fig:counts_wide_o2}) is likely driven by selection effects. The EDS probes a larger interval of the luminosity function, sampling more typical star-forming galaxies rather than being dominated by the rare, extremely luminous emitters that define the bright end. As a result, number counts in the EDS are less affected by small-number statistics and are better anchored to the luminosity range where current models and observational luminosity functions are most robust.

\begin{figure}[t]
    \centering \includegraphics[width=\columnwidth]{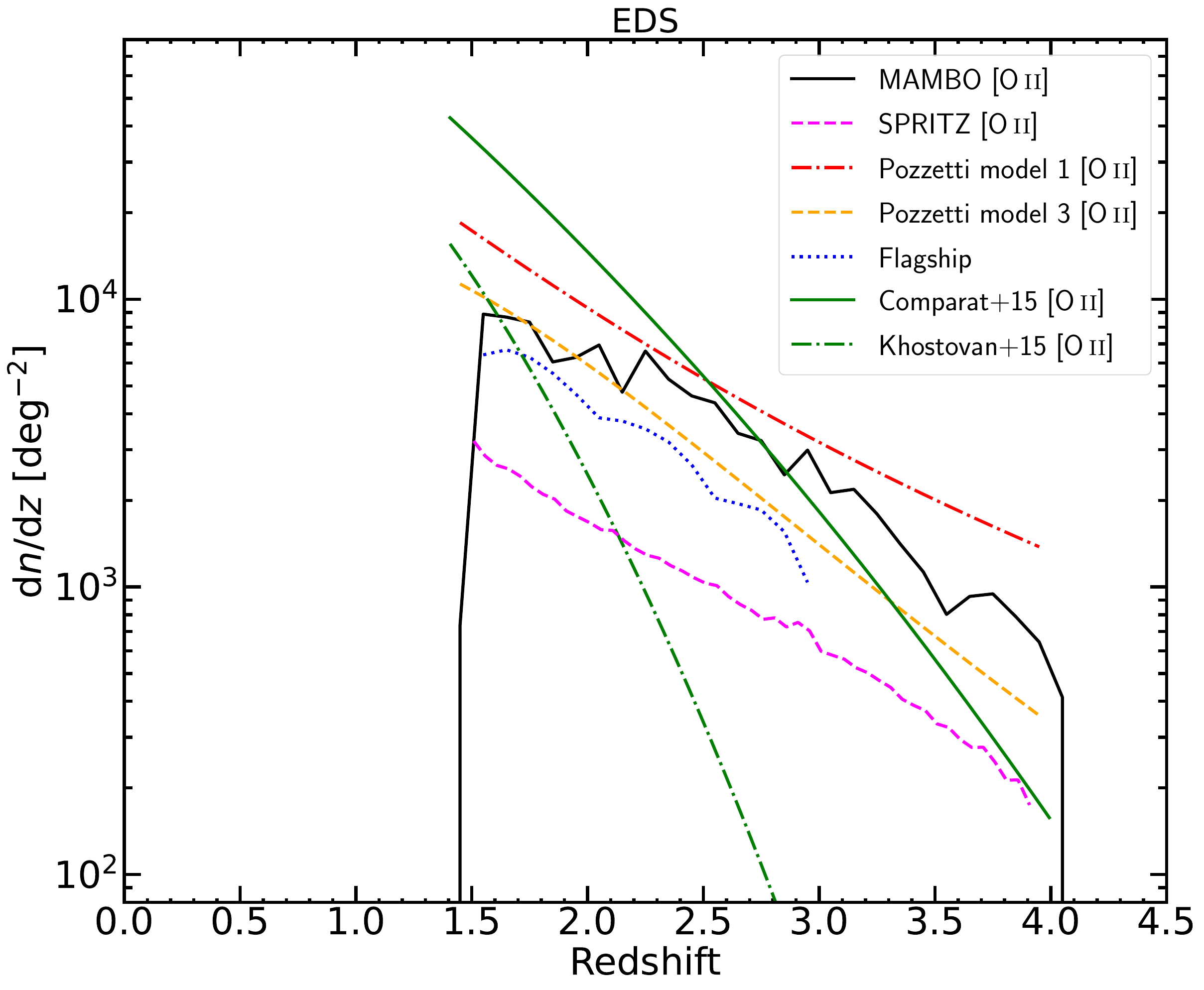}
    \caption{Same as Fig.\,\ref{fig:counts_wide_o2} for the EDS. }
    \label{fig:counts_deep_o2}
\end{figure}

\subsection{Counts from \texttt{FastSpec} simulations}\label{sect:fastspec_sim}
The purpose of this section is to provide an estimate of the expected capacity of \Euclid to recover the number counts of line emitters, as presented in the previous sections and derived from various mock catalogues. In practice, several factors can affect the observed number counts. First, as shown in Fig.\,\ref{fig:flux_loss}, in the \texttt{MAMBO}+\texttt{FastSpec} simulations analysed here the recovered line fluxes are, on average, lower than the input values. The size of this offset depends on the line and on the adopted fixed-aperture extraction setup, and should not be interpreted as a universal mission-level flux bias. Additionally, the estimated redshifts may be inaccurate, with some noise peaks misidentified as emission lines and some real lines possibly confused with others.

To assess the impact of these effects, we computed the number counts for [\ion{O}{ii}], \hb, [\ion{O}{iii}], \ha, and Pa\,$\beta$ using our \texttt{FastSpec} simulations, after applying the flux corrections reported in Table\,\ref{tab:flux_corr}. We define the measured sample as the set of galaxies selected from the simulated spectra using the recovered quantities, namely the corrected measured line flux, the line S/N, and the measured spectroscopic redshift. For each line, galaxies are included in the measured sample if they satisfy the following criteria:

\begin{align}\label{eq:sel}
    &F_{\rm corr} > F_{\rm lim}, \nonumber\\
    &\mathrm{S/N}_{\rm line} > 3.5, \nonumber\\
    &R_z > R_{z,{\rm min}}, \nonumber\\
    &z_{\rm min} < z_{\rm spec} < z_{\rm max}.
\end{align}
Here $F_{\mathrm{corr}}$ is the line flux measured and corrected for aperture loss (Table\,\ref{tab:flux_corr}), $F_{\rm lim}$ is the flux limit of the considered survey ($F_{\rm lim}$=2 and $0.5$ for the EWS and EDS, respectively), S/N$_{\mathrm{line}}$ is the S/N of the line, $R_z$ is the redshift reliability associated with $z_{\mathrm{spec}}$ (estimated by OU-SPE from the simulated spectra), and $R_{z,{\rm min}}$ is the threshold value for $R_z$ (0.2 and 0.15 for EWS and EDS, respectively). The parameters $z_{\mathrm{min}}$ and $z_{\mathrm{max}}$ define the redshift interval over which each line can be observed. 

We stress that this selection is applied to recovered quantities, and therefore the measured sample is not required to be a subset of the intrinsic flux-limited sample selected from the input catalogue. In particular, the measured sample can include both sources with intrinsic flux below $F_{\mathrm{lim}}$ whose recovered, flux-corrected flux is scattered above the adopted threshold, and contaminants due to redshift or line misidentifications.

This effect is analogous to an Eddington-like bias at the flux-selection boundary. Since the selection is applied to the measured, flux-corrected quantity $F_{\rm corr}$, sources with intrinsic fluxes just below $F_{\rm lim}$ can enter the measured sample if their measured flux is scattered upward, or if the average flux-loss correction moves them above the threshold. Conversely, sources intrinsically above the threshold can be scattered below it and missed. If the luminosity function decreases with increasing luminosity near the flux limit, the former effect can dominate, because there are more sources just below the threshold than just above it. In this case, the measured number counts can exceed the intrinsic flux-limited counts even when the redshift contamination remains low. These additional sources should therefore not be interpreted as redshift interlopers, but as a consequence of applying a flux-limited selection to noisy, corrected flux measurements.

We adopt a threshold of S/N$_{\mathrm{line}}>3.5$ for both the EWS and EDS, since S/N$_{\mathrm{line}}=3.5$ corresponds to the average S/N expected for emission lines with flux equal to the $3.5\sigma$ flux sensitivity limits of the \Euclid\ surveys adopted in our simulations (see Appendix A in Euclid Collaboration: Mancini et al., in prep.). Therefore, we require lines to have S/N value above that limit in order to include only reliable and robust lines. In addition, in order to include only objects with a robust $z_{\mathrm{spec}}$, the criterion on $R_z$ is also applied. Euclid Collaboration: Mancini et al. (in prep.) showed in their Appendix A that the threshold values that maximise the success rate while minimizing the contamination are 0.15 for the EDS and 0.2 for the EWS.


Within the measured sample defined by Eq.\,(\ref{eq:sel}), we further define the correct sample as the subset of galaxies for which the recovered spectroscopic redshift is consistent with the true input redshift. Specifically, a measured emitter is included in the correct sample if
\begin{equation}\label{eq:deltaz_EWS}
|z_{\rm spec} - z_{\rm true}| \leq 0.002(1+z_{\rm true})
\end{equation}
for the EWS, and
\begin{equation}\label{eq:deltaz_EDS}
|z_{\rm spec} - z_{\rm true}| \leq 0.001(1+z_{\rm true})
\end{equation}
for the EDS. Here, $z_{\rm spec}$ and $z_{\rm true}$ are the spectroscopic and the simulated redshifts, respectively. The choice of two different thresholds for the EWS and EDS is justified because, in our simulations, the scatter of the spectroscopic redshifts with respect to the simulated ones is larger in the EWS than in the EDS (see Euclid Collaboration: Mancini et al., in prep.).

By construction, the correct sample is therefore a subset of the measured sample, but not necessarily of the intrinsic true flux-limited sample. As a result, the number of correct emitters can exceed the number of true emitters when intrinsically fainter sources are scattered above the measured-flux threshold while retaining a correct redshift identification. Contaminants with incorrect redshifts, instead, contribute to the measured sample but not to the correct sample.

In the following sections, we introduce two parameters that quantify the quality of the recovered samples of line emitters. Specifically, for each line, we define the success rate as the fraction of true emitters that are also found in the measured sample. In other words, we compute the number of emitters in the intersection between the true and measured samples, divided by the total number of true emitters. This metric provides an estimate of how many genuine line emitters are successfully recovered after simulating the observations, redshift estimation, and line flux measurement.

We then define contamination as the fraction of line emitters in the measured sample that have an incorrect spectroscopic redshift (i.e., those not satisfying Eqs.\,\ref{eq:deltaz_EWS} and \ref{eq:deltaz_EDS}). In other words, we compute the complementary value of the number of emitters in the correct sample divided by the number in the measured one. This parameter provides an estimate of the fraction of objects in the measured sample that are not at the correct redshift.

We stress that the success rate and contamination defined here are diagnostic quantities specific to the \texttt{MAMBO}-based \texttt{FastSpec} experiment and are not equivalent to the completeness and redshift purity adopted for the official \Euclid \ha clustering sample. The latter are derived using dedicated survey selections and selection-function modelling, as described in Euclid Collaboration: Granett et al. (in prep.).

As a consistency check, for the EWS \ha sample we find a contamination of 14\% (corresponding to a purity-like value of $\sim$86\%) and a success rate of 63\%. Euclid Collaboration: Granett et al. (in prep.) obtain a redshift purity of $P=0.74$ and a sample completeness of $C=0.45$ for their baseline clustering selection. The values are broadly comparable in scale, although they refer to different input catalogues, selections, and metric definitions.

\subsection{\texttt{FastSpec} simulations: EWS}
We show in Fig.\,\ref{fig:counts_fast_wide} the number counts for the five lines analysed in this paper, for the case of the EWS, for our \texttt{FastSpec} \texttt{MAMBO} spectra. As in the next subsection, for each line we report three different counts: the true number counts, that come from the original \texttt{MAMBO} mock sample (shown already in Fig.\,\ref{fig:counts_wide}); the measured number counts, that comes from counting the emitters in the \texttt{FastSpec} simulated sample that pass the Eq.\,(\ref{eq:sel}) criteria; and the correct number counts, that are those among the measured for which the spectroscopic redshift is correct (i.e., close enough to the true, simulated one, see Eq.\,\ref{eq:deltaz_EWS}). The number counts together with the total number of emitters expected in the whole survey are also reported in Table\,\ref{tab:fastspec_counts}. The table reports as well the success rate and the contamination, as defined above. As discussed in Sect.~\ref{sect:fastspec_sim}, the comparison between true, measured, and correct counts is conditional on the \texttt{MAMBO} input catalogue used for the \texttt{FastSpec} simulations. In particular, the excess of measured or correct counts over the intrinsic true counts depends on the local slope of the underlying number counts around the survey flux limit.

\begin{figure}[t]
    \centering \includegraphics[width=\columnwidth]{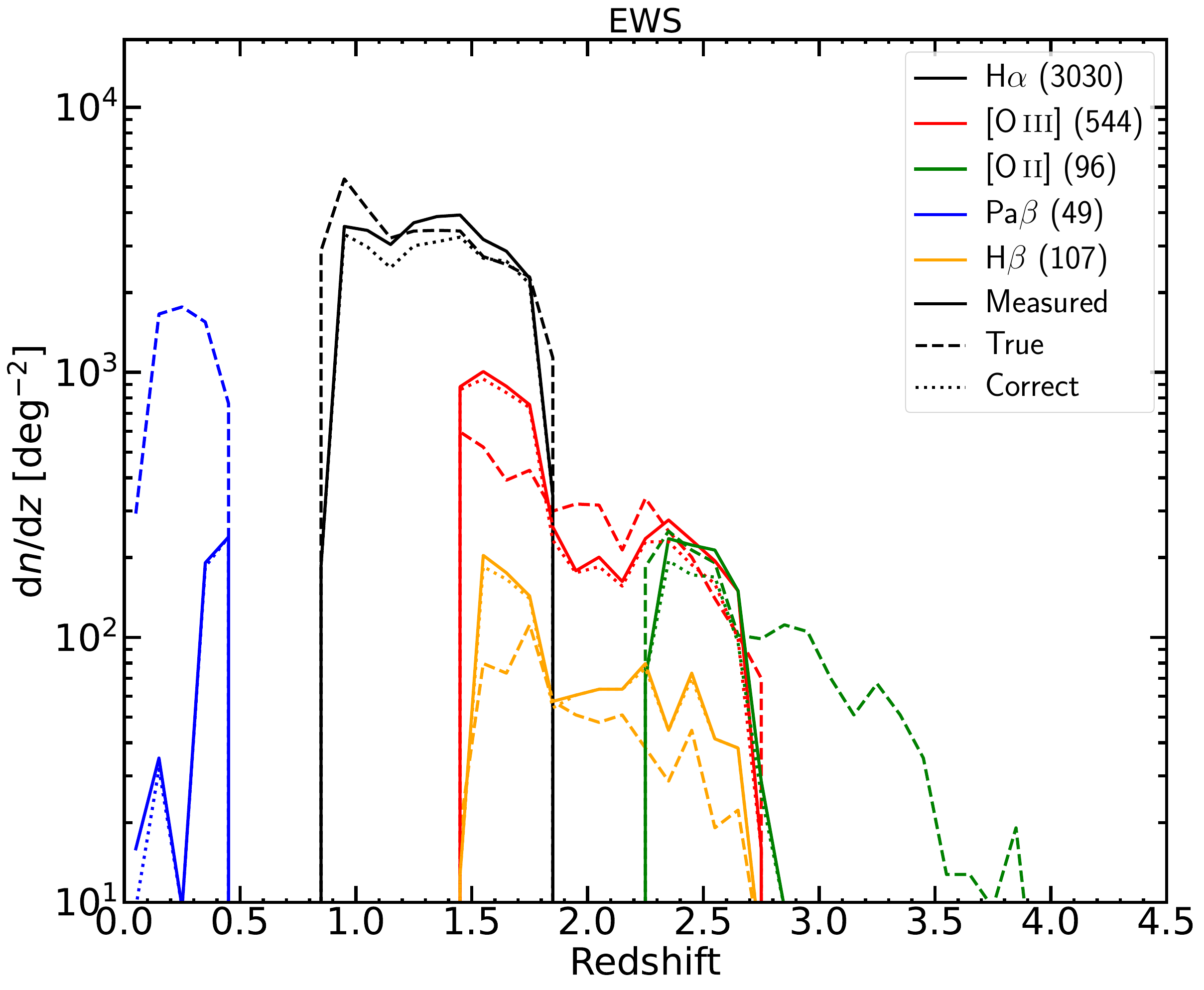}
    \caption{Observed number counts for the EWS \texttt{FastSpec} simulated mock catalogue. The  green, orange,  red, black, and blue lines indicate the number counts for [\ion{O}{ii}], \hb, [\ion{O}{iii}]5007, \ha, and Pa\,$\beta$, respectively. The continuous, dashed, and dotted lines represent number counts for the measured sample (i.e., galaxies from our \texttt{FastSpec} simulated sample), for the true sample (i.e., galaxies from the original \texttt{MAMBO} mock lightcone), and correct sample (i.e., measured lines for which the spectroscopic redshift is correct). In the top right, the number of measured line emitters per square degree are reported.}
    \label{fig:counts_fast_wide}
\end{figure}

From Fig.\,\ref{fig:counts_fast_wide} and Table\,\ref{tab:fastspec_counts}, it is evident that the measured number counts generally differ from the true ones. In particular for [\ion{O}{ii}] the measured number counts closely match the true counts up to $z \sim 2.7$; beyond this redshift, the measured number counts of [\ion{O}{ii}] emitters drop to zero. This effect is clearly due to the \ha prior applied during the spectroscopic redshift estimation by the pipeline: effectively, all single-line emitters (and at $z > 2.7$ [\ion{O}{ii}] is indeed the only line visible in the RGS spectra) are classified as \ha. Consequently, approximately $50\%$ of the [\ion{O}{ii}] emitters at $z > 2.77$ are missed.

For \hb, we find the opposite trend: the measured number counts are approximately 65\% higher than the true ones. This is likely due to the non-Gaussian nature of the flux errors for \hb (see top-right panel of Fig.\,\ref{fig:flux_loss}), which causes the flux correction to scatter a larger number of intrinsically faint sources above the detection threshold than vice versa. For [\ion{O}{iii}], we find a similar trend to \hb, with approximately 30\% more measured counts than the true ones.

Finally, for Pa\,$\beta$, almost $90\%$ of the emitters are expected to be lost in our simulation. This is likely due to the significant flux loss that is suffered by this line in our simulation: Pa\,$\beta$ emitters, in comparison with other line emitters, are at relatively low redshift, therefore cover in general a larger spatial area, and suffer larger flux loss. The flux correction is evidently not enough to bring back up the number of measured emitters. 
The optimal extraction procedure, which was not included in the Q1 spectroscopic processing and is being incorporated into the pipeline for DR1, is expected to mitigate this problem by using source-dependent cross-dispersion profiles during spectral extraction.



It is important to emphasise that the values of success rate and contamination strongly depend on the selection criteria adopted in Eq.\,(\ref{eq:sel}). For example, applying a stricter threshold on the $R_z$ parameter, such as increasing it from 0.2 to 0.3 in the EWS, would reduce contamination by selecting only higher-quality spectroscopic redshifts; however, this would also decrease the success rate. The value we chose provides a good balance between success rate and contamination (see Appendix\,A in Euclid Collaboration: Mancini et al., in prep., for a detailed justification of our threshold). Depending on the specific scientific goals of a given project, the thresholds can be adjusted to favour either a larger recovered sample or a less contaminated one. This flexibility is relevant for galaxy-evolution applications and for non-cosmological selections. The official \ha sample used for \Euclid galaxy-clustering cosmology, instead, will be defined and validated through dedicated selections designed to satisfy the mission requirements on redshift purity and sample completeness; see Euclid Collaboration: Granett et al. (in prep.).

\subsection{\texttt{FastSpec} simulations: EDS}
\begin{figure}[!ht]
    \centering \includegraphics[width=\columnwidth]{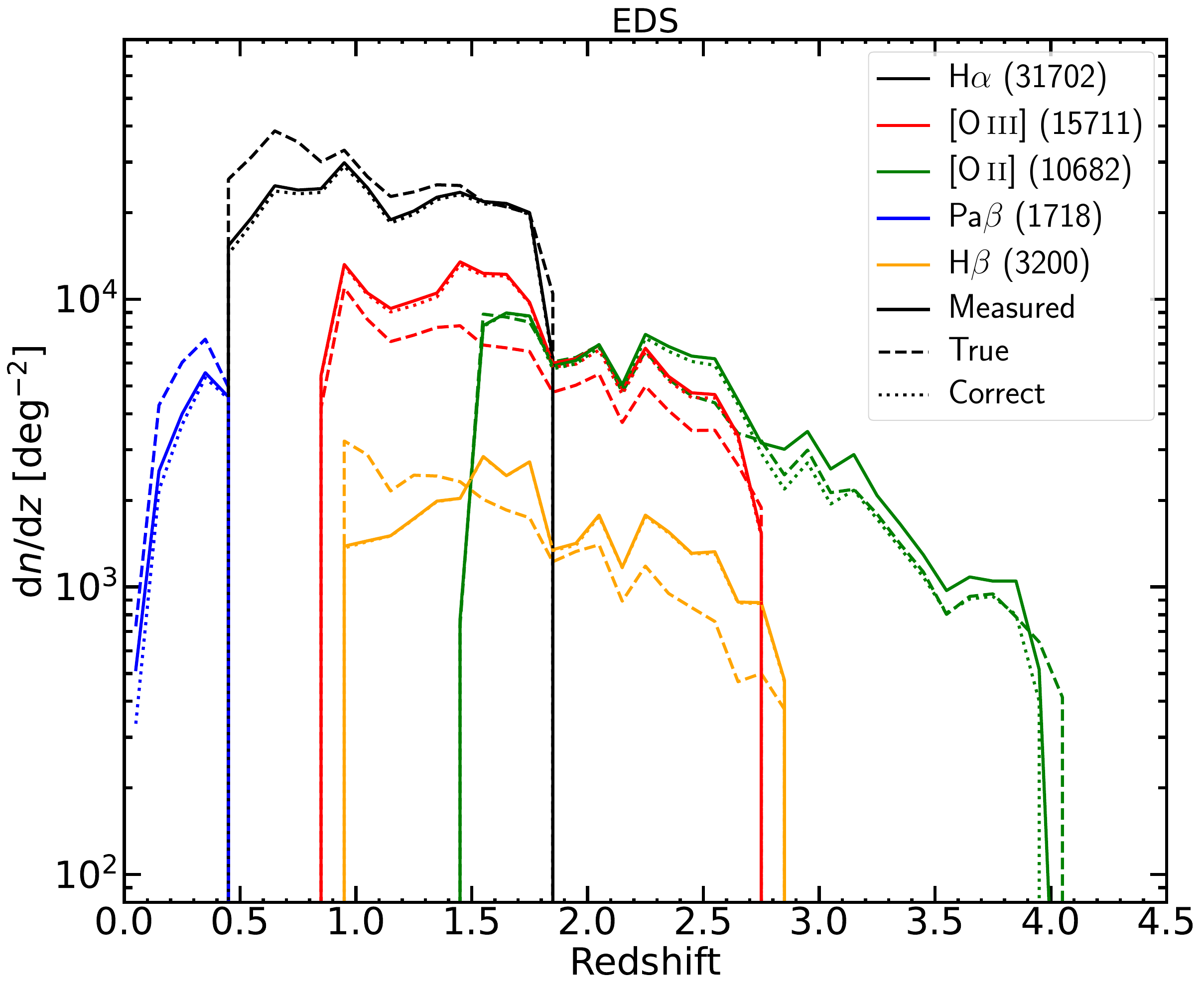}
    \caption{Same as Fig.\,\ref{fig:counts_fast_wide} for the EDS. }
    \label{fig:counts_fast_deep}
\end{figure}

\begin{table*}
\centering
\caption{
Number counts of recovered line emitters in the Euclid Wide Survey (EWS) and Euclid Deep Survey (EDS), based on the \texttt{MAMBO}+\texttt{FastSpec} simulations. For each emission line, we report the intrinsic true counts, the measured counts selected from recovered quantities, and the correct counts, defined as the subset of measured sources with a recovered redshift consistent with the input redshift. We also report the total number of measured emitters expected over the corresponding survey area, the success rate, and the contamination fraction, as defined in Sect.\,\ref{sect:fastspec_sim}.
}
\label{tab:fastspec_counts}
\begin{tabular}{lccccccc}
\hline\hline
Line &
$z$ range &
True &
Measured &
Correct &
Success rate &
Contamination &
Total emitters\\
&
&
deg$^{-2}$ &
deg$^{-2}$ &
deg$^{-2}$ &
&
&
$N_\textrm{obs}$
\\
\hline
\multicolumn{8}{c}{Euclid Wide Survey} \\
\hline
        $[\ion{O}{ii}]$ & $2.25\lesssim z\lesssim 4.07$ & 159  & 96  & 76   & 40\% & 20\% & $1.34\times10^6$\\
       \hb & $1.49\lesssim z\lesssim 2.89$        & 64   & 107  & 102  & 68\% & \phantom{0}4\% & $1.48\times10^6$\\
        $[\ion{O}{iii}]$ & $1.42\lesssim z\lesssim 2.77$& 418  & 544  & 504  & 73\% & \phantom{0}7\% & $7.62\times10^6$\\
        \ha & $0.84\lesssim z\lesssim 1.88$       & 3456 & 3030 & 2611 & 63\% & 14\% & $4.24\times10^7$\\
        Pa\,$\beta$ & $0.00\lesssim z\lesssim 0.47$       &603  & 49   & 47   & \phantom{0}8\%  & \phantom{0}4\% & $6.86\times10^5$\\
\hline
\multicolumn{8}{c}{Euclid Deep Survey} \\
\hline
        $[\ion{O}{ii}]$     & $1.49\lesssim z\lesssim 4.07$ & 9677  & 10682 & 9996  & 80\% & \phantom{0}6\% & $5.66\times10^5$\\
       \hb            & $0.91\lesssim z\lesssim 2.89$ & 3098  & 3200  & 3180  & 58\% & <1\% & $1.70\times10^5$\\
        $$[\ion{O}{iii}]$$  & $0.86\lesssim z\lesssim 2.77$ & 11431 & 15711 & 15331 & 88\% & \phantom{0}2\% & $8.33\times10^5$\\
        \ha           & $0.42\lesssim z\lesssim 1.88$ & 39044 & 31702 & 30810 & 75\% & \phantom{0}2\% & $1.68\times10^6$\\
        Pa\,$\beta$           & $0.00\lesssim z\lesssim 0.47$ & 2326  & 1718  & 1601  & 63\% & \phantom{0}6\% & $8.48\times10^4$\\
        Golden              & $1.49\lesssim z\lesssim 1.88$ & 647   & 653   & 653   & 70\% & \phantom{0}0\% & $3.46\times10^4$\\
\hline
\end{tabular}
\end{table*}

In Fig.\,\ref{fig:counts_fast_deep} and Table\,\ref{tab:fastspec_counts}, we report the number counts for the five emission lines in the case of the EDS, based on our \texttt{FastSpec} simulations. In addition to individual lines, we also provide the number counts of galaxies in what we called the `golden sample', namely galaxies for which [\ion{O}{ii}], \hb, [\ion{O}{iii}], and \ha are all simultaneously detected. This is only possible within the redshift range $1.49\lesssim z\lesssim 1.88$, enabled by the combined coverage of the BGS and RGS.

The measured, true, and correct number counts follow the same definitions as in the previous section. As before, the measured counts are generally close to the true ones, with discrepancies that are smaller than those found in the EWS case. Specifically, for [\ion{O}{ii}], \hb, and [\ion{O}{iii}], we find that the number of measured emitters exceeds the true one by approximately {10\%}, {3\%}, and {37\%}, respectively. Conversely, for \ha and Pa\,$\beta$, the measured counts are approximately {19\%} and {26\%} lower than the corresponding true counts.

The success rates for the EDS are significantly higher than those for the EWS. This result is expected because the longer integration times improve the S/N of the emission lines, while the broader wavelength coverage increases the likelihood of detecting multiple lines from the same galaxy, enabling more precise redshift determinations. In particular, all lines have success rates of at least $58\%$ (the worst case being \hb), with [\ion{O}{ii}], [\ion{O}{iii}], and \ha reaching at least $75\%$. This indicates that for these three lines more than three quarters of the true emitters are indeed recovered in our \texttt{FastSpec} simulated sample. 

We note, however, that these success rates are derived within the framework of our simulations and therefore depend on the adopted flux limits. Since the simulated catalogues are limited to ${\cal F}>0.2$, the contamination from fainter emitters or passive galaxies not included in the simulation is likely underestimated, although the application of quality cuts on $R_z$ (or equivalently $z_{\rm prob}$) is expected to efficiently remove most of these very faint interlopers.

It is remarkable in the EDS that the contamination is also very small for all lines, remaining at or below $6\%$. It is at or below {2\%} for \hb, [\ion{O}{iii}], and \ha, and reaches {6\%} for both [\ion{O}{ii}] and Pa\,$\beta$. This improvement with respect to the EWS case is not unexpected: the broader wavelength coverage offered by the combined BGS and RGS, along with deeper exposures, ensures the detection of at least two lines per spectrum in most cases, thereby reducing line misidentifications.


The so-called `golden sample' further illustrates this point. Since it requires the simultaneous detection of four lines -- possible only in the EDS within the range $1.49\lesssim z\lesssim 1.88$ -- the number of misidentifications is extremely limited. As a consequence, the contamination of this sample is practically zero, with a success rate of $70\%$. 

\section{Conclusions}\label{sc:Conclusions}
In this work, we have presented detailed predictions for the number counts of emission line galaxies detectable by the EWS and EDS, focusing on the key spectral lines [\ion{O}{ii}], \hb, [\ion{O}{iii}], \ha, and Pa\,$\beta$. Our analysis combines different mock galaxy catalogues with realistic \texttt{FastSpec} simulations that incorporate key observational effects such as flux losses, line confusion, and redshift uncertainties.

From the comparison of different catalogues (\texttt{MAMBO}, \texttt{SPRITZ}, Flagship, and \texttt{GAEA}), we find a generally good agreement for \ha emitters, with increasing discrepancies for other lines, especially at high redshift. \texttt{MAMBO} generally lies at the upper end of the model predictions, particularly for [\ion{O}{ii}] and [\ion{O}{iii}], although \texttt{SPRITZ} predicts comparable [\ion{O}{ii}] counts in the EWS and lower counts in the EDS.

Overall, the comparison between model predictions for the [\ion{O}{ii}] line and estimates based on observed [\ion{O}{ii}] luminosity functions shows that discrepancies are larger for the EWS, which is dominated by rare, bright [\ion{O}{ii}] emitters at the luminous end of the luminosity function, while a significantly better agreement is found for the EDS, especially at $z \sim 1.5$ where most models are calibrated; at higher redshifts, differences increase in both surveys, reflecting the growing uncertainties in the extrapolation of current [\ion{O}{ii}] luminosity functions and galaxy evolution models. The large, homogeneous spectroscopic samples of [\ion{O}{ii}] emitters that will be provided by \Euclid will therefore be crucial to discriminate between competing models and to robustly constrain the evolution of the [\ion{O}{ii}] luminosity function over a wide redshift range.

The \texttt{MAMBO}-based \texttt{FastSpec} simulations allow us to quantify both the measured counts and their deviation from the true counts. We find that for the EWS, the success rate ranges from {63\%} to {73\%} for \hb, [\ion{O}{iii}], and \ha, but drops to $\sim 40\%$ for [\ion{O}{ii}] and below $10\%$ for Pa\,$\beta$. This is mainly due to selection effects and observational limitations (e.g., [\ion{O}{ii}] confusion with \ha at high redshift, or flux loss for Pa\,$\beta$). Contamination remains below {10\% } for \hb, [\ion{O}{iii}], and Pa\,$\beta$, and reaches {14\%} for \ha and {20\%} for [\ion{O}{ii}].

The performance improves significantly in the EDS, thanks to longer exposure times and extended wavelength coverage. Here, success rates reach at least $58\%$ for all lines, and contamination remains at or below $6\%$. The combination of depth and spectral coverage enables more reliable identification of emission lines and more precise redshift recovery.


The total numbers of line emitters expected to be recovered in the \texttt{MAMBO}-based \texttt{FastSpec} simulations are also notable. As shown in the last column of Table~\ref{tab:fastspec_counts}, these simulations predict that the EWS will recover of the order of $42\times10^6$ \ha emitters, $7.6\times10^6$ [\ion{O}{iii}] emitters, $1.3\times10^6$ [\ion{O}{ii}] emitters, and $\sim1.5\times10^6$ \hb emitters. For the EDS, despite their much smaller area, the corresponding recovered samples are predicted to contain over $1.7\times10^6$ \ha emitters, $8.3\times10^5$ [\ion{O}{iii}] emitters, $5.7\times10^5$ [\ion{O}{ii}] emitters, and $1.7\times10^5$ \hb emitters.

These numbers should be interpreted as conditional on the \texttt{MAMBO} input catalogue, the adopted empirical flux corrections, and the \texttt{FastSpec}+OU-SPE recovery procedure. In particular, the non-\ha predictions, especially for [\ion{O}{ii}] and [\ion{O}{iii}], remain subject to substantial systematic uncertainty, as indicated by the scatter among the different mock catalogues discussed above.

Even with these caveats, the predicted samples are very large and will enable detailed studies of galaxy evolution and large-scale structure across a broad redshift range. In addition, the samples of [\ion{O}{ii}] and [\ion{O}{iii}] emitters recovered from the EDS and EWS may allow \Euclid to extend large-scale structure studies beyond the redshift range accessible through \ha alone.

In summary, within the \texttt{MAMBO}-based \texttt{FastSpec} framework used here, the recovered samples of emission-line galaxies are expected to be very large, with high success rates and low contamination, particularly in the EDS. The balance between success rate and contamination can be adjusted through the adopted selection thresholds, depending on the specific scientific goals. These simulations therefore confirm the strong potential of \Euclid spectroscopy for galaxy-evolution studies and for exploiting large samples of spectroscopic redshifts, while the assessment of the official \ha clustering sample is left to dedicated cosmology-oriented analyses.

\begin{acknowledgements}
\AckEC
\AckCosmoHub\,
The authors acknowledge support from the ELSA project. "ELSA: Euclid Legacy Science Advanced analysis tools" (Grant Agreement no. 101135203) is funded by the European Union. Views and opinions expressed are however those of the author(s) only and do not necessarily reflect those of the European Union or Innovate UK. Neither the European Union nor the granting authority can be held responsible for them. UK participation is funded through the UK Horizon guarantee scheme under Innovate UK grant 10093177.
\end{acknowledgements}

%
%
\bibliography{Euclid_ECEB, biblio_mia}

@ARTICLE{Q1-TP001,
       author = {{Euclid Collaboration: Aussel}, H. and {Tereno}, I. and {Schirmer}, M. and others},
       title = "{Euclid Quick Data Release (Q1). I. Data release overview}",
      journal = {\aap},
         year = 2026,
        month = jun,
       volume = {711},
          eid = {A1},
        pages = {A1},
          doi = {10.1051/0004-6361/202554610},
       adsurl = {https://ui.adsabs.harvard.edu/abs/2026A&A...711A...1E}
}

@ARTICLE{Q1-TP007,
       author = {{Euclid Collaboration: Le Brun}, V. and {Bethermin}, M. and {Moresco}, M. and others},
       title = "{Euclid Quick Data Release (Q1). VII. Characteristics and limitations of the spectroscopic measurements}",
      journal = {\aap},
         year = 2026,
        month = jun,
       volume = {711},
          eid = {A7},
        pages = {A7},
          doi = {10.1051/0004-6361/202554607},
       adsurl = {https://ui.adsabs.harvard.edu/abs/2026A&A...711A...7E}
}

@ARTICLE{EuclidSkyOverview,
author = {{Euclid Collaboration: Mellier}, Y. and {Abdurro'uf} and {Acevedo~Barroso}, J.A. and others},
	title = {Euclid - I. Overview of the Euclid mission},
	DOI= "10.1051/0004-6361/202450810",
	url= "https://doi.org/10.1051/0004-6361/202450810",
	journal = {A\&A},
	year = 2025,
	volume = 697,
	pages = "A1",
}

@ARTICLE{EuclidSkyVIS,
author = {{Euclid Collaboration: Cropper}, M. and {Al-Bahlawan}, A. and {Amiaux}, J. and others},
	title = {Euclid - II. The VIS instrument},
	DOI= "10.1051/0004-6361/202450996",
	url= "https://doi.org/10.1051/0004-6361/202450996",
	journal = {A\&A},
	year = 2025,
	volume = 697,
	pages = "A2",
}

@ARTICLE{EuclidSkyNISP,
author = {{Euclid Collaboration: Jahnke}, K. and {Gillard}, W. and {Schirmer}, M. and others},
	title = {Euclid - III. The NISP Instrument},
	DOI= "10.1051/0004-6361/202450786",
	url= "https://doi.org/10.1051/0004-6361/202450786",
	journal = {A\&A},
	year = 2025,
	volume = 697,
	pages = "A3",
}

@ARTICLE{EuclidSkyFlagship,
author = {{Euclid Collaboration: Castander}, F. and {Fosalba}, P. and {Stadel}, J. and others},
	title = {Euclid - V. The Flagship galaxy mock catalogue: A comprehensive simulation for the Euclid},
	DOI= "10.1051/0004-6361/202450853",
	url= "https://doi.org/10.1051/0004-6361/202450853",
	journal = {A\&A},
	year = 2025,
	volume = 697,
	pages = "A5",
}

@ARTICLE{EP-Quai,
       author = {{Euclid Collaboration: Quai}, S. and {Pozzetti}, L. and {Talia}, M. and others},
       title = "{Euclid preparation. LXXXII. Predicting star-forming galaxy scaling relations with the spectral stacking code SpectraPyle}",
      journal = {\aap},
         year = 2026,
        month = mar,
       volume = {707},
          eid = {A232},
        pages = {A232},
          doi = {10.1051/0004-6361/202557329},
archivePrefix = {arXiv},
       eprint = {2509.16120},
 primaryClass = {astro-ph.GA},
       adsurl = {https://ui.adsabs.harvard.edu/abs/2026A&A...707A.232E}
}

@ARTICLE{EP-Scharre,
       author = {{Euclid Collaboration: Scharr{\'e}}, L. and {Hirschmann}, M. and {De Lucia}, G. and others},
        title = "{Euclid preparation: XLV. Optical emission-line predictions of intermediate-z galaxy populations in GAEA for the Euclid Deep and Wide Surveys}",
      journal = {\aap},
         year = 2024,
        month = sep,
       volume = {689},
          eid = {A276},
        pages = {A276},
          doi = {10.1051/0004-6361/202449500},
archivePrefix = {arXiv},
       eprint = {2402.03436},
 primaryClass = {astro-ph.GA},
       adsurl = {https://ui.adsabs.harvard.edu/abs/2024A&A...689A.276E}
}

@ARTICLE{Blanchard-EP7,
       author = {{Euclid Collaboration: Blanchard}, A. and {Camera}, S. and {Carbone}, C. and others},
        title = "{Euclid preparation. VII. Forecast validation for Euclid cosmological probes}",
      journal = {\aap},
         year = 2020,
        month = oct,
       volume = {642},
          eid = {A191},
        pages = {A191},
          doi = {10.1051/0004-6361/202038071},
archivePrefix = {arXiv},
       eprint = {1910.09273},
 primaryClass = {astro-ph.CO},
       adsurl = {https://ui.adsabs.harvard.edu/abs/2020A&A...642A.191E}
}

@ARTICLE{Scaramella-EP1,
       author = {{Euclid Collaboration: Scaramella}, R. and {Amiaux}, J. and {Mellier}, Y. and others},
        title = "{Euclid preparation. I. The Euclid Wide Survey}",
      journal = {\aap},
         year = 2022,
        month = jun,
       volume = {662},
          eid = {A112},
        pages = {A112},
          doi = {10.1051/0004-6361/202141938},
archivePrefix = {arXiv},
       eprint = {2108.01201},
 primaryClass = {astro-ph.CO},
       adsurl = {https://ui.adsabs.harvard.edu/abs/2022A&A...662A.112E}
}

@ARTICLE{Laureijs11,
       author = {{Laureijs}, R. and {Amiaux}, J. and {Arduini}, S. and {Augu{\`e}res}, J. -L. and {Brinchmann}, J. and {Cole}, R. and {Cropper}, M. and {Dabin}, C. and {Duvet}, L. and {Ealet}, A. and {Garilli}, B. and {Gondoin}, P. and {Guzzo}, L. and {Hoar}, J. and {Hoekstra}, H. and {Holmes}, R. and {Kitching}, T. and {Maciaszek}, T. and {Mellier}, Y. and {Pasian}, F. and {Percival}, W. and {Rhodes}, J. and {Saavedra Criado}, G. and {Sauvage}, M. and {Scaramella}, R. and {Valenziano}, L. and {Warren}, S. and {Bender}, R. and {Castander}, F. and {Cimatti}, A. and {Le F{\`e}vre}, O. and {Kurki-Suonio}, H. and {Levi}, M. and {Lilje}, P. and {Meylan}, G. and {Nichol}, R. and {Pedersen}, K. and {Popa}, V. and {Rebolo Lopez}, R. and {Rix}, H. -W. and {Rottgering}, H. and {Zeilinger}, W. and {Grupp}, F. and {Hudelot}, P. and {Massey}, R. and {Meneghetti}, M. and {Miller}, L. and {Paltani}, S. and {Paulin-Henriksson}, S. and {Pires}, S. and {Saxton}, C. and {Schrabback}, T. and {Seidel}, G. and {Walsh}, J. and {Aghanim}, N. and {Amendola}, L. and {Bartlett}, J. and {Baccigalupi}, C. and {Beaulieu}, J. -P. and {Benabed}, K. and {Cuby}, J. -G. and {Elbaz}, D. and {Fosalba}, P. and {Gavazzi}, G. and {Helmi}, A. and {Hook}, I. and {Irwin}, M. and {Kneib}, J. -P. and {Kunz}, M. and {Mannucci}, F. and {Moscardini}, L. and {Tao}, C. and {Teyssier}, R. and {Weller}, J. and {Zamorani}, G. and {Zapatero Osorio}, M.~R. and {Boulade}, O. and {Foumond}, J.~J. and {Di Giorgio}, A. and {Guttridge}, P. and {James}, A. and {Kemp}, M. and {Martignac}, J. and {Spencer}, A. and {Walton}, D. and {Bl{\"u}mchen}, T. and {Bonoli}, C. and {Bortoletto}, F. and {Cerna}, C. and {Corcione}, L. and {Fabron}, C. and {Jahnke}, K. and {Ligori}, S. and {Madrid}, F. and {Martin}, L. and {Morgante}, G. and {Pamplona}, T. and {Prieto}, E. and {Riva}, M. and {Toledo}, R. and {Trifoglio}, M. and {Zerbi}, F. and {Abdalla}, F. and {Douspis}, M. and {Grenet}, C. and {Borgani}, S. and {Bouwens}, R. and {Courbin}, F. and {Delouis}, J. -M. and {Dubath}, P. and {Fontana}, A. and {Frailis}, M. and {Grazian}, A. and {Koppenh{\"o}fer}, J. and {Mansutti}, O. and {Melchior}, M. and {Mignoli}, M. and {Mohr}, J. and {Neissner}, C. and {Noddle}, K. and {Poncet}, M. and {Scodeggio}, M. and {Serrano}, S. and {Shane}, N. and {Starck}, J. -L. and {Surace}, C. and {Taylor}, A. and {Verdoes-Kleijn}, G. and {Vuerli}, C. and {Williams}, O.~R. and {Zacchei}, A. and {Altieri}, B. and {Escudero Sanz}, I. and {Kohley}, R. and {Oosterbroek}, T. and {Astier}, P. and {Bacon}, D. and {Bardelli}, S. and {Baugh}, C. and {Bellagamba}, F. and {Benoist}, C. and {Bianchi}, D. and {Biviano}, A. and {Branchini}, E. and {Carbone}, C. and {Cardone}, V. and {Clements}, D. and {Colombi}, S. and {Conselice}, C. and {Cresci}, G. and {Deacon}, N. and {Dunlop}, J. and {Fedeli}, C. and {Fontanot}, F. and {Franzetti}, P. and {Giocoli}, C. and {Garcia-Bellido}, J. and {Gow}, J. and {Heavens}, A. and {Hewett}, P. and {Heymans}, C. and {Holland}, A. and {Huang}, Z. and {Ilbert}, O. and {Joachimi}, B. and {Jennins}, E. and {Kerins}, E. and {Kiessling}, A. and {Kirk}, D. and {Kotak}, R. and {Krause}, O. and {Lahav}, O. and {van Leeuwen}, F. and {Lesgourgues}, J. and {Lombardi}, M. and {Magliocchetti}, M. and {Maguire}, K. and {Majerotto}, E. and {Maoli}, R. and {Marulli}, F. and {Maurogordato}, S. and {McCracken}, H. and {McLure}, R. and {Melchiorri}, A. and {Merson}, A. and {Moresco}, M. and {Nonino}, M. and {Norberg}, P. and {Peacock}, J. and {Pello}, R. and {Penny}, M. and {Pettorino}, V. and {Di Porto}, C. and {Pozzetti}, L. and {Quercellini}, C. and {Radovich}, M. and {Rassat}, A. and {Roche}, N. and {Ronayette}, S. and {Rossetti}, E. and {Sartoris}, B. and {Schneider}, P. and {Semboloni}, E. and {Serjeant}, S. and {Simpson}, F. and {Skordis}, C. and {Smadja}, G. and {Smartt}, S. and {Spano}, P. and {Spiro}, S. and {Sullivan}, M. and {Tilquin}, A. and {Trotta}, R. and {Verde}, L. and {Wang}, Y. and {Williger}, G. and {Zhao}, G. and {Zoubian}, J. and {Zucca}, E.},
        title = "{Euclid Definition Study Report}",
      journal = {ESA/SRE(2011)12},
         year = 2011,
        month = oct,
          eid = {arXiv:1110.3193},
        pages = {arXiv:1110.3193},
          doi = {10.48550/arXiv.1110.3193},
archivePrefix = {arXiv},
       eprint = {1110.3193},
 primaryClass = {astro-ph.CO},
       adsurl = {https://ui.adsabs.harvard.edu/abs/2011arXiv1110.3193L}
}

@INPROCEEDINGS{Maciaszek22,
       author = {{Maciaszek}, T. and {Ealet}, A. and {Gillard}, W. and {Jahnke}, K. and {Barbier}, R. and {Prieto}, E. and {Bon}, W. and {Bonnefoi}, A. and {Caillat}, A. and {Carle}, M. and {Costille}, A. and {Ducret}, F. and {Fabron}, C. and {Foulon}, B. and {Gimenez}, J. -L. and {Grassi}, E. and {Jaquet}, M. and {Le Mignant}, D. and {Martin}, L. and {Pamplona}, T. and {Sanchez}, P. and {Cl{\'e}mens}, J. -C. and {Caillat}, L. and {Niclas}, M. and {Secroun}, A. and {Kubik}, B. and {Ferriol}, S. and {Berthe}, M. and {Barri{\`e}re}, J. -C. and {Fontignie}, J. and {Valenziano}, L. and {Auricchio}, N. and {Battaglia}, P. and {De Rosa}, A. and {Farinelli}, R. and {Franceschi}, E. and {Medinaceli}, E. and {Morgante}, G. and {Sortino}, F. and {Trifoglio}, M. and {Corcione}, L. and {Capobianco}, V. and {Ligori}, S. and {Dusini}, S. and {Borsato}, E. and {Dal Corso}, F. and {Laudisio}, F. and {Sirignano}, C. and {Stanco}, L. and {Ventura}, S. and {Patrizii}, L. and {Chiarusi}, T. and {Fornari}, F. and {Giacomini}, F. and {Margiotta}, A. and {Mauri}, N. and {Pasqualini}, L. and {Sirri}, G. and {Spurio}, M. and {Tenti}, M. and {Travaglini}, R. and {Bonoli}, C. and {Bortoletto}, F. and {Balestra}, A. and {Dalessandro}, M. and {Grupp}, F. and {Penka}, D. and {Steinwagner}, J. and {Hormuth}, F. and {Schirmer}, M. and {Seidel}, G. and {Padilla}, C. and {Casas}, R. and {Lloro}, Y. and {Toledo-Moreo}, R. and {Gomez}, J. and {Colodro-Conde}, C. and {Liz{\'a}n}, D. and {Diaz}, J. -J. and {Lilje}, P. and {Andersen}, M.~I. and {Andersen}, J. and {S{\o}rensen}, A.~N. and {Hornstrup}, A. and {Jessen}, N.~C. and {Thizy}, C. and {Holmes}, W. and {Pniel}, M. and {Jhabvala}, M. and {Pravdo}, S. and {Seiffert}, M. and {Waczynski}, A. and {Laureij}, R.~J. and {Racca}, G. and {Salvignol}, J. -C. and {Boenke}, T. and {Strada}, P. and {Mellier}, Y.},
        title = "{Euclid near infrared spectrometer and photometer instrument flight model presentation, performance, and ground calibration results summary}",
    booktitle = {Space Telescopes and Instrumentation 2022: Optical, Infrared, and Millimeter Wave},
         year = 2022,
       editor = {{Coyle}, Laura E. and {Matsuura}, Shuji and {Perrin}, Marshall D.},
       series = {SPIE Conf. Ser.},
       volume = {12180},
        month = aug,
          eid = {121801K},
        pages = {arXiv:2210.10112},
          doi = {10.1117/12.2630338},
archivePrefix = {arXiv},
       eprint = {2210.10112},
 primaryClass = {astro-ph.IM},
       adsurl = {https://ui.adsabs.harvard.edu/abs/2022SPIE12180E..1KM}
}

@ARTICLE{Bagley2020,
       author = {{Bagley}, Micaela B. and {Scarlata}, Claudia and {Mehta}, Vihang and {Teplitz}, Harry and {Baronchelli}, Ivano and {Eisenstein}, Daniel J. and {Pozzetti}, Lucia and {Cimatti}, Andrea and {Rutkowski}, Michael and {Wang}, Yun and {Merson}, Alexander},
        title = "{HST Grism-derived Forecasts for Future Galaxy Redshift Surveys}",
      journal = {\apj},
         year = 2020,
        month = jul,
       volume = {897},
       number = {1},
          eid = {98},
        pages = {98},
          doi = {10.3847/1538-4357/ab9828},
archivePrefix = {arXiv},
       eprint = {2006.03602},
 primaryClass = {astro-ph.GA},
       adsurl = {https://ui.adsabs.harvard.edu/abs/2020ApJ...897...98B}
}

@ARTICLE{Baskin05,
       author = {{Baskin}, Alexei and {Laor}, Ari},
        title = "{What controls the [OIII]{\ensuremath{\lambda}}5007 line strength in active galactic nuclei?}",
      journal = {\mnras},
         year = 2005,
        month = apr,
       volume = {358},
       number = {3},
        pages = {1043-1054},
          doi = {10.1111/j.1365-2966.2005.08841.x},
archivePrefix = {arXiv},
       eprint = {astro-ph/0501436},
 primaryClass = {astro-ph},
       adsurl = {https://ui.adsabs.harvard.edu/abs/2005MNRAS.358.1043B}
}

@ARTICLE{Jansen01,
       author = {{Jansen}, Rolf A. and {Franx}, Marijn and {Fabricant}, Daniel},
        title = "{[O II] As a Tracer of Current Star Formation}",
      journal = {\apj},
         year = 2001,
        month = apr,
       volume = {551},
       number = {2},
        pages = {825-832},
          doi = {10.1086/320228},
archivePrefix = {arXiv},
       eprint = {astro-ph/0012485},
 primaryClass = {astro-ph},
       adsurl = {https://ui.adsabs.harvard.edu/abs/2001ApJ...551..825J}
}

@article{Ilbert_2013,
       author = {{Ilbert}, O. and {McCracken}, H.~J. and {Le F{\`e}vre}, O. and {Capak}, P. and {Dunlop}, J. and {Karim}, A. and {Renzini}, M.~A. and {Caputi}, K. and {Boissier}, S. and {Arnouts}, S. and {Aussel}, H. and {Comparat}, J. and {Guo}, Q. and {Hudelot}, P. and {Kartaltepe}, J. and {Kneib}, J.~P. and {Krogager}, J.~K. and {Le Floc'h}, E. and {Lilly}, S. and {Mellier}, Y. and {Milvang-Jensen}, B. and {Moutard}, T. and {Onodera}, M. and {Richard}, J. and {Salvato}, M. and {Sanders}, D.~B. and {Scoville}, N. and {Silverman}, J.~D. and {Taniguchi}, Y. and {Tasca}, L. and {Thomas}, R. and {Toft}, S. and {Tresse}, L. and {Vergani}, D. and {Wolk}, M. and {Zirm}, A.},
        title = "{Mass assembly in quiescent and star-forming galaxies since z ≃ 4 from UltraVISTA}",
      journal = {\aap},
         year = 2013,
        month = aug,
       volume = {556},
          eid = {A55},
        pages = {A55},
          doi = {10.1051/0004-6361/201321100},
archivePrefix = {arXiv},
       eprint = {1301.3157},
 primaryClass = {astro-ph.CO},
       adsurl = {https://ui.adsabs.harvard.edu/abs/2013A&A...556A..55I}
}

@ARTICLE{Girelli_2020,
       author = {{Girelli}, G. and {Pozzetti}, L. and {Bolzonella}, M. and {Giocoli}, C. and {Marulli}, F. and {Baldi}, M.},
        title = "{The stellar-to-halo mass relation over the past 12 Gyr. I. Standard {\ensuremath{\Lambda}}CDM model}",
      journal = {\aap},
         year = 2020,
        month = feb,
       volume = {634},
          eid = {A135},
        pages = {A135},
          doi = {10.1051/0004-6361/201936329},
archivePrefix = {arXiv},
       eprint = {2001.02230},
 primaryClass = {astro-ph.CO},
       adsurl = {https://ui.adsabs.harvard.edu/abs/2020A&A...634A.135G}
}

@ARTICLE{Henriques_2015,
       author = {{Henriques}, Bruno M.~B. and {White}, Simon D.~M. and {Thomas}, Peter A. and {Angulo}, Raul and {Guo}, Qi and {Lemson}, Gerard and {Springel}, Volker and {Overzier}, Roderik},
        title = "{Galaxy formation in the Planck cosmology - I. Matching the observed evolution of star formation rates, colours and stellar masses}",
      journal = {\mnras},
         year = 2015,
        month = aug,
       volume = {451},
       number = {3},
        pages = {2663-2680},
          doi = {10.1093/mnras/stv705},
archivePrefix = {arXiv},
       eprint = {1410.0365},
 primaryClass = {astro-ph.GA},
       adsurl = {https://ui.adsabs.harvard.edu/abs/2015MNRAS.451.2663H}
}

@ARTICLE{Kennicutt_1998,
       author = {{Kennicutt}, Robert C., Jr.},
        title = "{Star Formation in Galaxies Along the Hubble Sequence}",
      journal = {\araa},
         year = 1998,
        month = jan,
       volume = {36},
        pages = {189-232},
          doi = {10.1146/annurev.astro.36.1.189},
archivePrefix = {arXiv},
       eprint = {astro-ph/9807187},
 primaryClass = {astro-ph},
       adsurl = {https://ui.adsabs.harvard.edu/abs/1998ARA&A..36..189K}
}

@ARTICLE{Pozzetti2016,
       author = {{Pozzetti}, L. and {Hirata}, C.~M. and {Geach}, J.~E. and {Cimatti}, A. and {Baugh}, C. and {Cucciati}, O. and {Merson}, A. and {Norberg}, P. and {Shi}, D.},
        title = "{Modelling the number density of H{\ensuremath{\alpha}} emitters for future spectroscopic near-IR space missions}",
      journal = {\aap},
         year = 2016,
        month = may,
       volume = {590},
          eid = {A3},
        pages = {A3},
          doi = {10.1051/0004-6361/201527081},
archivePrefix = {arXiv},
       eprint = {1603.01453},
 primaryClass = {astro-ph.GA},
       adsurl = {https://ui.adsabs.harvard.edu/abs/2016A&A...590A...3P}
}

@article{Racca2016,
 title={The Euclid mission design},
   url={http://dx.doi.org/10.1117/12.2230762},
   DOI={10.1117/12.2230762},
   journal={in Space Telescopes
and Instrumentation 2016: Optical, Infrared, and MillimeterWave, ed.
H. A. MacEwen, G. G. Fazio, M. Lystrup, N. Batalha, N. Siegler, E. C.
Tong, Vol. 9904, Society of Photo-Optical
Instrumentation Engineers (SPIE), 0O},
   publisher={SPIE},
   author={Racca, Giuseppe D. and Laureijs, René and Stagnaro, Luca and Salvignol, Jean-Christophe and Lorenzo Alvarez, José and Saavedra Criado, Gonzalo and Gaspar Venancio, Luis and Short, Alex and Strada, Paolo and Bönke, Tobias and et al.},
   editor={MacEwen, Howard A. and Fazio, Giovanni G. and Lystrup, Makenzie and Batalha, Natalie and Siegler, Nicholas and Tong, Edward C.Editors},
   year={2016}
}

@ARTICLE{Schreiber_2017,
       author = {{Schreiber}, C. and {Elbaz}, D. and {Pannella}, M. and {Merlin}, E. and {Castellano}, M. and {Fontana}, A. and {Bourne}, N. and {Boutsia}, K. and {Cullen}, F. and {Dunlop}, J. and {Ferguson}, H.~C. and {Micha{\l}owski}, M.~J. and {Okumura}, K. and {Santini}, P. and {Shu}, X.~W. and {Wang}, T. and {White}, C.},
        title = "{EGG: hatching a mock Universe from empirical prescriptions{\ensuremath{\star}}}",
      journal = {\aap},
         year = 2017,
        month = jun,
       volume = {602},
          eid = {A96},
        pages = {A96},
          doi = {10.1051/0004-6361/201629123},
archivePrefix = {arXiv},
       eprint = {1606.05354},
 primaryClass = {astro-ph.IM},
       adsurl = {https://ui.adsabs.harvard.edu/abs/2017A&A...602A..96S}
}

@ARTICLE{Springel_2005,
       author = {{Springel}, Volker and {White}, Simon D.~M. and {Jenkins}, Adrian and {Frenk}, Carlos S. and {Yoshida}, Naoki and {Gao}, Liang and {Navarro}, Julio and {Thacker}, Robert and {Croton}, Darren and {Helly}, John and {Peacock}, John A. and {Cole}, Shaun and {Thomas}, Peter and {Couchman}, Hugh and {Evrard}, August and {Colberg}, J{\"o}rg and {Pearce}, Frazer},
        title = "{Simulations of the formation, evolution and clustering of galaxies and quasars}",
      journal = {\nat},
         year = 2005,
        month = jun,
       volume = {435},
       number = {7042},
        pages = {629-636},
          doi = {10.1038/nature03597},
archivePrefix = {arXiv},
       eprint = {astro-ph/0504097},
 primaryClass = {astro-ph},
       adsurl = {https://ui.adsabs.harvard.edu/abs/2005Natur.435..629S}
}

@ARTICLE{Fontanot20,
       author = {{Fontanot}, Fabio and {De Lucia}, Gabriella and {Hirschmann}, Michaela and {Xie}, Lizhi and {Monaco}, Pierluigi and {Menci}, Nicola and {Fiore}, Fabrizio and {Feruglio}, Chiara and {Cristiani}, Stefano and {Shankar}, Francesco},
        title = "{The rise of active galactic nuclei in the galaxy evolution and assembly semi-analytic model}",
      journal = {\mnras},
         year = 2020,
        month = aug,
       volume = {496},
       number = {3},
        pages = {3943-3960},
          doi = {10.1093/mnras/staa1716},
archivePrefix = {arXiv},
       eprint = {2002.10576},
 primaryClass = {astro-ph.CO},
       adsurl = {https://ui.adsabs.harvard.edu/abs/2020MNRAS.496.3943F}
}

@ARTICLE{Ferland13,
       author = {{Ferland}, G.~J. and {Porter}, R.~L. and {van Hoof}, P.~A.~M. and {Williams}, R.~J.~R. and {Abel}, N.~P. and {Lykins}, M.~L. and {Shaw}, G. and {Henney}, W.~J. and {Stancil}, P.~C.},
        title = "{The 2013 Release of Cloudy}",
      journal = {\rmxaa},
         year = 2013,
        month = apr,
       volume = {49},
        pages = {137-163},
          doi = {10.48550/arXiv.1302.4485},
archivePrefix = {arXiv},
       eprint = {1302.4485},
 primaryClass = {astro-ph.GA},
       adsurl = {https://ui.adsabs.harvard.edu/abs/2013RMxAA..49..137F}
}

@ARTICLE{Gutkin16,
       author = {{Gutkin}, Julia and {Charlot}, St{\'e}phane and {Bruzual}, Gustavo},
        title = "{Modelling the nebular emission from primeval to present-day star-forming galaxies}",
      journal = {\mnras},
         year = 2016,
        month = oct,
       volume = {462},
       number = {2},
        pages = {1757-1774},
          doi = {10.1093/mnras/stw1716},
archivePrefix = {arXiv},
       eprint = {1607.06086},
 primaryClass = {astro-ph.GA},
       adsurl = {https://ui.adsabs.harvard.edu/abs/2016MNRAS.462.1757G}
}

@ARTICLE{Feltre16,
       author = {{Feltre}, A. and {Charlot}, S. and {Gutkin}, J.},
        title = "{Nuclear activity versus star formation: emission-line diagnostics at ultraviolet and optical wavelengths}",
      journal = {\mnras},
         year = 2016,
        month = mar,
       volume = {456},
       number = {3},
        pages = {3354-3374},
          doi = {10.1093/mnras/stv2794},
archivePrefix = {arXiv},
       eprint = {1511.08217},
 primaryClass = {astro-ph.GA},
       adsurl = {https://ui.adsabs.harvard.edu/abs/2016MNRAS.456.3354F}
}

@ARTICLE{Hirschmann17,
       author = {{Hirschmann}, Michaela and {Charlot}, Stephane and {Feltre}, Anna and {Naab}, Thorsten and {Choi}, Ena and {Ostriker}, Jeremiah P. and {Somerville}, Rachel S.},
        title = "{Synthetic nebular emission from massive galaxies - I: origin of the cosmic evolution of optical emission-line ratios}",
      journal = {\mnras},
         year = 2017,
        month = dec,
       volume = {472},
       number = {2},
        pages = {2468-2495},
          doi = {10.1093/mnras/stx2180},
archivePrefix = {arXiv},
       eprint = {1706.00010},
 primaryClass = {astro-ph.GA},
       adsurl = {https://ui.adsabs.harvard.edu/abs/2017MNRAS.472.2468H}
}

@ARTICLE{Potter17,
       author = {{Potter}, Douglas and {Stadel}, Joachim and {Teyssier}, Romain},
        title = "{PKDGRAV3: beyond trillion particle cosmological simulations for the next era of galaxy surveys}",
      journal = {Computational Astrophysics and Cosmology},
         year = 2017,
        month = may,
       volume = {4},
       number = {1},
          eid = {2},
        pages = {2},
          doi = {10.1186/s40668-017-0021-1},
archivePrefix = {arXiv},
       eprint = {1609.08621},
 primaryClass = {astro-ph.IM},
       adsurl = {https://ui.adsabs.harvard.edu/abs/2017ComAC...4....2P}
}

@ARTICLE{Hirschmann19,
       author = {{Hirschmann}, Michaela and {Charlot}, Stephane and {Feltre}, Anna and {Naab}, Thorsten and {Somerville}, Rachel S. and {Choi}, Ena},
        title = "{Synthetic nebular emission from massive galaxies - II. Ultraviolet-line diagnostics of dominant ionizing sources}",
      journal = {\mnras},
         year = 2019,
        month = jul,
       volume = {487},
       number = {1},
        pages = {333-353},
          doi = {10.1093/mnras/stz1256},
archivePrefix = {arXiv},
       eprint = {1811.07909},
 primaryClass = {astro-ph.GA},
       adsurl = {https://ui.adsabs.harvard.edu/abs/2019MNRAS.487..333H}
}

@ARTICLE{Behroozi13,
       author = {{Behroozi}, Peter S. and {Wechsler}, Risa H. and {Wu}, Hao-Yi},
        title = "{The ROCKSTAR Phase-space Temporal Halo Finder and the Velocity Offsets of Cluster Cores}",
      journal = {\apj},
         year = 2013,
        month = jan,
       volume = {762},
       number = {2},
          eid = {109},
        pages = {109},
          doi = {10.1088/0004-637X/762/2/109},
archivePrefix = {arXiv},
       eprint = {1110.4372},
 primaryClass = {astro-ph.CO},
       adsurl = {https://ui.adsabs.harvard.edu/abs/2013ApJ...762..109B}
}

@ARTICLE{Nakajima2014,
       author = {{Nakajima}, Kimihiko and {Ouchi}, Masami},
        title = "{Ionization state of inter-stellar medium in galaxies: evolution, SFR-M$_{*}$-Z dependence, and ionizing photon escape}",
      journal = {\mnras},
         year = 2014,
        month = jul,
       volume = {442},
       number = {1},
        pages = {900-916},
          doi = {10.1093/mnras/stu902},
archivePrefix = {arXiv},
       eprint = {1309.0207},
 primaryClass = {astro-ph.CO},
       adsurl = {https://ui.adsabs.harvard.edu/abs/2014MNRAS.442..900N}
}

@ARTICLE{Kewley02,
       author = {{Kewley}, L.~J. and {Dopita}, M.~A.},
        title = "{Using Strong Lines to Estimate Abundances in Extragalactic H II Regions and Starburst Galaxies}",
      journal = {\apjs},
         year = 2002,
        month = sep,
       volume = {142},
       number = {1},
        pages = {35-52},
          doi = {10.1086/341326},
archivePrefix = {arXiv},
       eprint = {astro-ph/0206495},
 primaryClass = {astro-ph},
       adsurl = {https://ui.adsabs.harvard.edu/abs/2002ApJS..142...35K}
}

@ARTICLE{Kewley06,
       author = {{Kewley}, Lisa J. and {Groves}, Brent and {Kauffmann}, Guinevere and {Heckman}, Tim},
        title = "{The host galaxies and classification of active galactic nuclei}",
      journal = {\mnras},
         year = 2006,
        month = nov,
       volume = {372},
       number = {3},
        pages = {961-976},
          doi = {10.1111/j.1365-2966.2006.10859.x},
archivePrefix = {arXiv},
       eprint = {astro-ph/0605681},
 primaryClass = {astro-ph},
       adsurl = {https://ui.adsabs.harvard.edu/abs/2006MNRAS.372..961K}
}

@ARTICLE{Maiolino08,
       author = {{Maiolino}, R. and {Nagao}, T. and {Grazian}, A. and {Cocchia}, F. and {Marconi}, A. and {Mannucci}, F. and {Cimatti}, A. and {Pipino}, A. and {Ballero}, S. and {Calura}, F. and {Chiappini}, C. and {Fontana}, A. and {Granato}, G.~L. and {Matteucci}, F. and {Pastorini}, G. and {Pentericci}, L. and {Risaliti}, G. and {Salvati}, M. and {Silva}, L.},
        title = "{AMAZE. I. The evolution of the mass-metallicity relation at z > 3}",
      journal = {\aap},
         year = 2008,
        month = sep,
       volume = {488},
       number = {2},
        pages = {463-479},
          doi = {10.1051/0004-6361:200809678},
archivePrefix = {arXiv},
       eprint = {0806.2410},
 primaryClass = {astro-ph},
       adsurl = {https://ui.adsabs.harvard.edu/abs/2008A&A...488..463M}
}

@ARTICLE{Curti17,
       author = {{Curti}, M. and {Cresci}, G. and {Mannucci}, F. and {Marconi}, A. and {Maiolino}, R. and {Esposito}, S.},
        title = "{New fully empirical calibrations of strong-line metallicity indicators in star-forming galaxies}",
      journal = {\mnras},
         year = 2017,
        month = feb,
       volume = {465},
       number = {2},
        pages = {1384-1400},
          doi = {10.1093/mnras/stw2766},
archivePrefix = {arXiv},
       eprint = {1610.06939},
 primaryClass = {astro-ph.GA},
       adsurl = {https://ui.adsabs.harvard.edu/abs/2017MNRAS.465.1384C}
}

@ARTICLE{Baldwin81,
       author = {{Baldwin}, J.~A. and {Phillips}, M.~M. and {Terlevich}, R.},
        title = "{Classification parameters for the emission-line spectra of extragalactic objects.}",
      journal = {\pasp},
         year = 1981,
        month = feb,
       volume = {93},
        pages = {5-19},
          doi = {10.1086/130766},
       adsurl = {https://ui.adsabs.harvard.edu/abs/1981PASP...93....5B}
}

@ARTICLE{Calabro18,
       author = {{Calabr{\`o}}, A. and {Daddi}, E. and {Cassata}, P. and {Onodera}, M. and {Gobat}, R. and {Puglisi}, A. and {Jin}, S. and {Liu}, D. and {Amor{\'\i}n}, R. and {Arimoto}, N. and {Boquien}, M. and {Carraro}, R. and {Elbaz}, D. and {Ibar}, E. and {Juneau}, S. and {Mannucci}, F. and {M{\'e}ndez Hern{\'a}nez}, H. and {Oliva}, E. and {Rodighiero}, G. and {Valentino}, F. and {Zanella}, A.},
        title = "{Near-infrared Emission Lines in Starburst Galaxies at 0.5 < z < 0.9: Discovery of a Merger Sequence of Extreme Obscurations}",
      journal = {\apjl},
         year = 2018,
        month = aug,
       volume = {862},
       number = {2},
          eid = {L22},
        pages = {L22},
          doi = {10.3847/2041-8213/aad33e},
archivePrefix = {arXiv},
       eprint = {1807.04350},
 primaryClass = {astro-ph.GA},
       adsurl = {https://ui.adsabs.harvard.edu/abs/2018ApJ...862L..22C}
}

@ARTICLE{Oke83,
       author = {{Oke}, J.~B. and {Gunn}, J.~E.},
        title = "{Secondary standard stars for absolute spectrophotometry.}",
      journal = {\apj},
         year = 1983,
        month = mar,
       volume = {266},
        pages = {713-717},
          doi = {10.1086/160817},
       adsurl = {https://ui.adsabs.harvard.edu/abs/1983ApJ...266..713O}
}

@ARTICLE{Pettini04,
       author = {{Pettini}, Max and {Pagel}, Bernard E.~J.},
        title = "{[OIII]/[NII] as an abundance indicator at high redshift}",
      journal = {\mnras},
         year = 2004,
        month = mar,
       volume = {348},
       number = {3},
        pages = {L59-L63},
          doi = {10.1111/j.1365-2966.2004.07591.x},
archivePrefix = {arXiv},
       eprint = {astro-ph/0401128},
 primaryClass = {astro-ph},
       adsurl = {https://ui.adsabs.harvard.edu/abs/2004MNRAS.348L..59P}
}

@ARTICLE{delucia14,
       author = {{De Lucia}, Gabriella and {Tornatore}, Luca and {Frenk}, Carlos S. and {Helmi}, Amina and {Navarro}, Julio F. and {White}, Simon D.~M.},
        title = "{Elemental abundances in Milky Way-like galaxies from a hierarchical galaxy formation model}",
      journal = {\mnras},
         year = 2014,
        month = nov,
       volume = {445},
       number = {1},
        pages = {970-987},
          doi = {10.1093/mnras/stu1752},
archivePrefix = {arXiv},
       eprint = {1407.7867},
 primaryClass = {astro-ph.GA},
       adsurl = {https://ui.adsabs.harvard.edu/abs/2014MNRAS.445..970D}
}

@ARTICLE{Hirschmann16,
       author = {{Hirschmann}, Michaela and {De Lucia}, Gabriella and {Fontanot}, Fabio},
        title = "{Galaxy assembly, stellar feedback and metal enrichment: the view from the GAEA model}",
      journal = {\mnras},
         year = 2016,
        month = sep,
       volume = {461},
       number = {2},
        pages = {1760-1785},
          doi = {10.1093/mnras/stw1318},
archivePrefix = {arXiv},
       eprint = {1512.04531},
 primaryClass = {astro-ph.GA},
       adsurl = {https://ui.adsabs.harvard.edu/abs/2016MNRAS.461.1760H}
}

@ARTICLE{delucia07,
       author = {{De Lucia}, Gabriella and {Blaizot}, J{\'e}r{\'e}my},
        title = "{The hierarchical formation of the brightest cluster galaxies}",
      journal = {\mnras},
         year = 2007,
        month = feb,
       volume = {375},
       number = {1},
        pages = {2-14},
          doi = {10.1111/j.1365-2966.2006.11287.x},
archivePrefix = {arXiv},
       eprint = {astro-ph/0606519},
 primaryClass = {astro-ph},
       adsurl = {https://ui.adsabs.harvard.edu/abs/2007MNRAS.375....2D}
}

@ARTICLE{Grogin11,
       author = {{Grogin}, Norman A. and {Kocevski}, Dale D. and {Faber}, S.~M. and {Ferguson}, Henry C. and {Koekemoer}, Anton M. and {Riess}, Adam G. and {Acquaviva}, Viviana and {Alexander}, David M. and {Almaini}, Omar and {Ashby}, Matthew L.~N. and {Barden}, Marco and {Bell}, Eric F. and {Bournaud}, Fr{\'e}d{\'e}ric and {Brown}, Thomas M. and {Caputi}, Karina I. and {Casertano}, Stefano and {Cassata}, Paolo and {Castellano}, Marco and {Challis}, Peter and {Chary}, Ranga-Ram and {Cheung}, Edmond and {Cirasuolo}, Michele and {Conselice}, Christopher J. and {Roshan Cooray}, Asantha and {Croton}, Darren J. and {Daddi}, Emanuele and {Dahlen}, Tomas and {Dav{\'e}}, Romeel and {de Mello}, Du{\'\i}lia F. and {Dekel}, Avishai and {Dickinson}, Mark and {Dolch}, Timothy and {Donley}, Jennifer L. and {Dunlop}, James S. and {Dutton}, Aaron A. and {Elbaz}, David and {Fazio}, Giovanni G. and {Filippenko}, Alexei V. and {Finkelstein}, Steven L. and {Fontana}, Adriano and {Gardner}, Jonathan P. and {Garnavich}, Peter M. and {Gawiser}, Eric and {Giavalisco}, Mauro and {Grazian}, Andrea and {Guo}, Yicheng and {Hathi}, Nimish P. and {H{\"a}ussler}, Boris and {Hopkins}, Philip F. and {Huang}, Jia-Sheng and {Huang}, Kuang-Han and {Jha}, Saurabh W. and {Kartaltepe}, Jeyhan S. and {Kirshner}, Robert P. and {Koo}, David C. and {Lai}, Kamson and {Lee}, Kyoung-Soo and {Li}, Weidong and {Lotz}, Jennifer M. and {Lucas}, Ray A. and {Madau}, Piero and {McCarthy}, Patrick J. and {McGrath}, Elizabeth J. and {McIntosh}, Daniel H. and {McLure}, Ross J. and {Mobasher}, Bahram and {Moustakas}, Leonidas A. and {Mozena}, Mark and {Nandra}, Kirpal and {Newman}, Jeffrey A. and {Niemi}, Sami-Matias and {Noeske}, Kai G. and {Papovich}, Casey J. and {Pentericci}, Laura and {Pope}, Alexandra and {Primack}, Joel R. and {Rajan}, Abhijith and {Ravindranath}, Swara and {Reddy}, Naveen A. and {Renzini}, Alvio and {Rix}, Hans-Walter and {Robaina}, Aday R. and {Rodney}, Steven A. and {Rosario}, David J. and {Rosati}, Piero and {Salimbeni}, Sara and {Scarlata}, Claudia and {Siana}, Brian and {Simard}, Luc and {Smidt}, Joseph and {Somerville}, Rachel S. and {Spinrad}, Hyron and {Straughn}, Amber N. and {Strolger}, Louis-Gregory and {Telford}, Olivia and {Teplitz}, Harry I. and {Trump}, Jonathan R. and {van der Wel}, Arjen and {Villforth}, Carolin and {Wechsler}, Risa H. and {Weiner}, Benjamin J. and {Wiklind}, Tommy and {Wild}, Vivienne and {Wilson}, Grant and {Wuyts}, Stijn and {Yan}, Hao-Jing and {Yun}, Min S.},
        title = "{CANDELS: The Cosmic Assembly Near-infrared Deep Extragalactic Legacy Survey}",
      journal = {\apjs},
         year = 2011,
        month = dec,
       volume = {197},
       number = {2},
          eid = {35},
        pages = {35},
          doi = {10.1088/0067-0049/197/2/35},
archivePrefix = {arXiv},
       eprint = {1105.3753},
 primaryClass = {astro-ph.CO},
       adsurl = {https://ui.adsabs.harvard.edu/abs/2011ApJS..197...35G}
}

@ARTICLE{Calabro19,
       author = {{Calabr{\`o}}, A. and {Daddi}, E. and {Puglisi}, A. and {Oliva}, E. and {Gobat}, R. and {Cassata}, P. and {Amor{\'\i}n}, R. and {Arimoto}, N. and {Boquien}, M. and {Carraro}, R. and {Delvecchio}, I. and {Ibar}, E. and {Jin}, S. and {Juneau}, S. and {Liu}, D. and {Onodera}, M. and {Mannucci}, F. and {M{\'e}ndez-Hern{\'a}ndez}, H. and {Rodighiero}, G. and {Valentino}, F. and {Zanella}, A.},
        title = "{Deciphering an evolutionary sequence of merger stages in infrared-luminous starburst galaxies at z {\ensuremath{\sim}} 0.7}",
      journal = {\aap},
         year = 2019,
        month = mar,
       volume = {623},
          eid = {A64},
        pages = {A64},
          doi = {10.1051/0004-6361/201834522},
archivePrefix = {arXiv},
       eprint = {1901.05013},
 primaryClass = {astro-ph.GA},
       adsurl = {https://ui.adsabs.harvard.edu/abs/2019A&A...623A..64C}
}

@ARTICLE{berta13,
       author = {{Berta}, S. and {Lutz}, D. and {Santini}, P. and {Wuyts}, S. and {Rosario}, D. and {Brisbin}, D. and {Cooray}, A. and {Franceschini}, A. and {Gruppioni}, C. and {Hatziminaoglou}, E. and {Hwang}, H.~S. and {Le Floc'h}, E. and {Magnelli}, B. and {Nordon}, R. and {Oliver}, S. and {Page}, M.~J. and {Popesso}, P. and {Pozzetti}, L. and {Pozzi}, F. and {Riguccini}, L. and {Rodighiero}, G. and {Roseboom}, I. and {Scott}, D. and {Symeonidis}, M. and {Valtchanov}, I. and {Viero}, M. and {Wang}, L.},
        title = "{Panchromatic spectral energy distributions of Herschel sources}",
      journal = {\aap},
         year = 2013,
        month = mar,
       volume = {551},
          eid = {A100},
        pages = {A100},
          doi = {10.1051/0004-6361/201220859},
archivePrefix = {arXiv},
       eprint = {1301.4496},
 primaryClass = {astro-ph.CO},
       adsurl = {https://ui.adsabs.harvard.edu/abs/2013A&A...551A.100B}
}

@ARTICLE{daCunha08,
       author = {{da Cunha}, Elisabete and {Charlot}, St{\'e}phane and {Elbaz}, David},
        title = "{A simple model to interpret the ultraviolet, optical and infrared emission from galaxies}",
      journal = {\mnras},
         year = 2008,
        month = aug,
       volume = {388},
       number = {4},
        pages = {1595-1617},
          doi = {10.1111/j.1365-2966.2008.13535.x},
archivePrefix = {arXiv},
       eprint = {0806.1020},
 primaryClass = {astro-ph},
       adsurl = {https://ui.adsabs.harvard.edu/abs/2008MNRAS.388.1595D}
}

@ARTICLE{Kashino19,
       author = {{Kashino}, Daichi and {Silverman}, John D. and {Sanders}, David and {Kartaltepe}, Jeyhan and {Daddi}, Emanuele and {Renzini}, Alvio and {Rodighiero}, Giulia and {Puglisi}, Annagrazia and {Valentino}, Francesco and {Juneau}, St{\'e}phanie and {Arimoto}, Nobuo and {Nagao}, Tohru and {Ilbert}, Olivier and {Le F{\`e}vre}, Olivier and {Koekemoer}, Anton M.},
        title = "{The FMOS-COSMOS Survey of Star-forming Galaxies at z {\ensuremath{\sim}} 1.6. VI. Redshift and Emission-line Catalog and Basic Properties of Star-forming Galaxies}",
      journal = {\apjs},
         year = 2019,
        month = mar,
       volume = {241},
       number = {1},
          eid = {10},
        pages = {10},
          doi = {10.3847/1538-4365/ab06c4},
archivePrefix = {arXiv},
       eprint = {1812.01529},
 primaryClass = {astro-ph.GA},
       adsurl = {https://ui.adsabs.harvard.edu/abs/2019ApJS..241...10K}
}

@INPROCEEDINGS{Costille19,
       author = {{Costille}, Anne and {Caillat}, A. and {Rossin}, C. and {Pascal}, S. and {Sanchez}, P. and {Foulon}, B. and {Pari{\`e}s}, C. and {Tisserand}, S. and {Salaun}, Y. and {Weber}, T.},
        title = "{Optical performance of NISP grisms flight models for EUCLID mission}",
    booktitle = {International Conference on Space Optics {\textemdash} ICSO 2018},
         year = 2019,
       editor = {{Sodnik}, Zoran and {Karafolas}, Nikos and {Cugny}, Bruno},
       series = {Society of Photo-Optical Instrumentation Engineers (SPIE) Conference Series},
       volume = {11180},
        month = jul,
          eid = {1118013},
        pages = {1118013},
          doi = {10.1117/12.2535958},
       adsurl = {https://ui.adsabs.harvard.edu/abs/2019SPIE11180E..13C}
}

@ARTICLE{Gruppioni13,
       author = {{Gruppioni}, C. and {Pozzi}, F. and {Rodighiero}, G. and {Delvecchio}, I. and {Berta}, S. and {Pozzetti}, L. and {Zamorani}, G. and {Andreani}, P. and {Cimatti}, A. and {Ilbert}, O. and {Le Floc'h}, E. and {Lutz}, D. and {Magnelli}, B. and {Marchetti}, L. and {Monaco}, P. and {Nordon}, R. and {Oliver}, S. and {Popesso}, P. and {Riguccini}, L. and {Roseboom}, I. and {Rosario}, D.~J. and {Sargent}, M. and {Vaccari}, M. and {Altieri}, B. and {Aussel}, H. and {Bongiovanni}, A. and {Cepa}, J. and {Daddi}, E. and {Dom{\'\i}nguez-S{\'a}nchez}, H. and {Elbaz}, D. and {F{\"o}rster Schreiber}, N. and {Genzel}, R. and {Iribarrem}, A. and {Magliocchetti}, M. and {Maiolino}, R. and {Poglitsch}, A. and {P{\'e}rez Garc{\'\i}a}, A. and {Sanchez-Portal}, M. and {Sturm}, E. and {Tacconi}, L. and {Valtchanov}, I. and {Amblard}, A. and {Arumugam}, V. and {Bethermin}, M. and {Bock}, J. and {Boselli}, A. and {Buat}, V. and {Burgarella}, D. and {Castro-Rodr{\'\i}guez}, N. and {Cava}, A. and {Chanial}, P. and {Clements}, D.~L. and {Conley}, A. and {Cooray}, A. and {Dowell}, C.~D. and {Dwek}, E. and {Eales}, S. and {Franceschini}, A. and {Glenn}, J. and {Griffin}, M. and {Hatziminaoglou}, E. and {Ibar}, E. and {Isaak}, K. and {Ivison}, R.~J. and {Lagache}, G. and {Levenson}, L. and {Lu}, N. and {Madden}, S. and {Maffei}, B. and {Mainetti}, G. and {Nguyen}, H.~T. and {O'Halloran}, B. and {Page}, M.~J. and {Panuzzo}, P. and {Papageorgiou}, A. and {Pearson}, C.~P. and {P{\'e}rez-Fournon}, I. and {Pohlen}, M. and {Rigopoulou}, D. and {Rowan-Robinson}, M. and {Schulz}, B. and {Scott}, D. and {Seymour}, N. and {Shupe}, D.~L. and {Smith}, A.~J. and {Stevens}, J.~A. and {Symeonidis}, M. and {Trichas}, M. and {Tugwell}, K.~E. and {Vigroux}, L. and {Wang}, L. and {Wright}, G. and {Xu}, C.~K. and {Zemcov}, M. and {Bardelli}, S. and {Carollo}, M. and {Contini}, T. and {Le F{\'e}vre}, O. and {Lilly}, S. and {Mainieri}, V. and {Renzini}, A. and {Scodeggio}, M. and {Zucca}, E.},
        title = "{The Herschel PEP/HerMES luminosity function - I. Probing the evolution of PACS selected Galaxies to z ≃ 4}",
      journal = {\mnras},
         year = 2013,
        month = jun,
       volume = {432},
       number = {1},
        pages = {23-52},
          doi = {10.1093/mnras/stt308},
archivePrefix = {arXiv},
       eprint = {1302.5209},
 primaryClass = {astro-ph.CO},
       adsurl = {https://ui.adsabs.harvard.edu/abs/2013MNRAS.432...23G}
}

@ARTICLE{MadauDickinson14,
       author = {{Madau}, Piero and {Dickinson}, Mark},
        title = "{Cosmic Star-Formation History}",
      journal = {\araa},
         year = 2014,
        month = aug,
       volume = {52},
        pages = {415-486},
          doi = {10.1146/annurev-astro-081811-125615},
archivePrefix = {arXiv},
       eprint = {1403.0007},
 primaryClass = {astro-ph.CO},
       adsurl = {https://ui.adsabs.harvard.edu/abs/2014ARA&A..52..415M}
}

@BOOK{Osterbrock06,
       author = {{Osterbrock}, Donald E. and {Ferland}, Gary J.},
        title = "{Astrophysics of gaseous nebulae and active galactic nuclei}",
         year = 2006,
       adsurl = {https://ui.adsabs.harvard.edu/abs/2006agna.book.....O},
      publisher = {University Science Books},
}

@ARTICLE{Battisti16,
       author = {{Battisti}, A.~J. and {Calzetti}, D. and {Chary}, R. -R.},
        title = "{Characterizing Dust Attenuation in Local Star-forming Galaxies: UV and Optical Reddening}",
      journal = {\apj},
         year = 2016,
        month = feb,
       volume = {818},
       number = {1},
          eid = {13},
        pages = {13},
          doi = {10.3847/0004-637X/818/1/13},
archivePrefix = {arXiv},
       eprint = {1601.00208},
 primaryClass = {astro-ph.GA},
       adsurl = {https://ui.adsabs.harvard.edu/abs/2016ApJ...818...13B}
}

@ARTICLE{Cleri22,
       author = {{Cleri}, Nikko J. and {Trump}, Jonathan R. and {Backhaus}, Bren E. and {Momcheva}, Ivelina and {Papovich}, Casey and {Simons}, Raymond and {Weiner}, Benjamin and {Estrada-Carpenter}, Vicente and {Finkelstein}, Steven L. and {Giavalisco}, Mauro and {Ji}, Zhiyuan and {Jung}, Intae and {Matharu}, Jasleen and {Martinez}, Felix and {Sturm}, Megan R.},
        title = "{CLEAR: Paschen-{\ensuremath{\beta}} Star Formation Rates and Dust Attenuation of Low-redshift Galaxies}",
      journal = {\apj},
         year = 2022,
        month = apr,
       volume = {929},
       number = {1},
          eid = {3},
        pages = {3},
          doi = {10.3847/1538-4357/ac5a4c},
archivePrefix = {arXiv},
       eprint = {2009.00617},
 primaryClass = {astro-ph.GA},
       adsurl = {https://ui.adsabs.harvard.edu/abs/2022ApJ...929....3C}
}

@ARTICLE{Comparat15,
       author = {{Comparat}, Johan and {Richard}, Johan and {Kneib}, Jean-Paul and {Ilbert}, Olivier and {Gonzalez-Perez}, Violeta and {Tresse}, Laurence and {Zoubian}, Julien and {Arnouts}, Stephane and {Brownstein}, Joel R. and {Baugh}, Carlton and {Delubac}, Timothee and {Ealet}, Anne and {Escoffier}, Stephanie and {Ge}, Jian and {Jullo}, Eric and {Lacey}, Cedric and {Ross}, Nicholas P. and {Schlegel}, David and {Schneider}, Donald P. and {Steele}, Oliver and {Tasca}, Lidia and {Yeche}, Christophe and {Lesser}, Michael and {Jiang}, Zhaoji and {Jing}, Yipeng and {Fan}, Zhou and {Fan}, Xiaohui and {Ma}, Jun and {Nie}, Jundan and {Wang}, Jiali and {Wu}, Zhenyu and {Zhang}, Tianmeng and {Zhou}, Xu and {Zhou}, Zhimin and {Zou}, Hu},
        title = "{The 0.1 <z < 1.65 evolution of the bright end of the [O ii] luminosity function}",
      journal = {\aap},
         year = 2015,
        month = mar,
       volume = {575},
          eid = {A40},
        pages = {A40},
          doi = {10.1051/0004-6361/201424767},
archivePrefix = {arXiv},
       eprint = {1408.1523},
 primaryClass = {astro-ph.GA},
       adsurl = {https://ui.adsabs.harvard.edu/abs/2015A&A...575A..40C}
}

@ARTICLE{Khostovan15,
       author = {{Khostovan}, A.~A. and {Sobral}, D. and {Mobasher}, B. and {Best}, P.~N. and {Smail}, I. and {Stott}, J.~P. and {Hemmati}, S. and {Nayyeri}, H.},
        title = "{Evolution of the H {\ensuremath{\beta}} + [O III] and [O II] luminosity functions and the [O II] star formation history of the Universe up to z {\ensuremath{\sim}} 5 from HiZELS}",
      journal = {\mnras},
         year = 2015,
        month = oct,
       volume = {452},
       number = {4},
        pages = {3948-3968},
          doi = {10.1093/mnras/stv1474},
archivePrefix = {arXiv},
       eprint = {1503.00004},
 primaryClass = {astro-ph.GA},
       adsurl = {https://ui.adsabs.harvard.edu/abs/2015MNRAS.452.3948K}
}

@PHDTHESIS{Girelli21,
    author = {{Girelli}, G.},
    title = "{Painting dark matter halos with galaxies in mock samples for future surveys}",
    school = {Alma Mater Studiorum Universit\`a di Bologna \url{http://amsdottorato.unibo.it/9820/}},
    year = 2021,
    month = May,
    url = {http://amsdottorato.unibo.it/9820/},
    doi = {10.48676/unibo/amsdottorato/9820}
}

@ARTICLE{Peng10,
       author = {{Peng}, Ying-jie and {Lilly}, Simon J. and {Kova{\v{c}}}, Katarina and
         {Bolzonella}, Micol and {Pozzetti}, Lucia and {Renzini}, Alvio and
         {Zamorani}, Gianni and {Ilbert}, Olivier and {Knobel}, Christian and
         {Iovino}, Angela and {Maier}, Christian and {Cucciati}, Olga and
         {Tasca}, Lidia and {Carollo}, C. Marcella and {Silverman}, John and
         {Kampczyk}, Pawel and {de Ravel}, Loic and {Sanders}, David and
         {Scoville}, Nicholas and {Contini}, Thierry and {Mainieri}, Vincenzo and
         {Scodeggio}, Marco and {Kneib}, Jean-Paul and {Le F{\`e}vre}, Olivier and
         {Bardelli}, Sandro and {Bongiorno}, Angela and {Caputi}, Karina and
         {Coppa}, Graziano and {de la Torre}, Sylvain and {Franzetti}, Paolo and
         {Garilli}, Bianca and {Lamareille}, Fabrice and
         {Le Borgne}, Jean-Francois and {Le Brun}, Vincent and {Mignoli}, Marco and
         {Perez Montero}, Enrique and {Pello}, Roser and {Ricciardelli}, Elena and
         {Tanaka}, Masayuki and {Tresse}, Laurence and {Vergani}, Daniela and
         {Welikala}, Niraj and {Zucca}, Elena and {Oesch}, Pascal and
         {Abbas}, Ummi and {Barnes}, Luke and {Bordoloi}, Rongmon and
         {Bottini}, Dario and {Cappi}, Alberto and {Cassata}, Paolo and
         {Cimatti}, Andrea and {Fumana}, Marco and {Hasinger}, Gunther and
         {Koekemoer}, Anton and {Leauthaud}, Alexei and {Maccagni}, Dario and
         {Marinoni}, Christian and {McCracken}, Henry and {Memeo}, Pierdomenico and
         {Meneux}, Baptiste and {Nair}, Preethi and {Porciani}, Cristiano and
         {Presotto}, Valentina and {Scaramella}, Roberto},
        title = "{Mass and Environment as Drivers of Galaxy Evolution in SDSS and zCOSMOS and the Origin of the Schechter Function}",
      journal = {\apj},
         year = 2010,
        month = sep,
       volume = {721},
       number = {1},
        pages = {193-221},
          doi = {10.1088/0004-637X/721/1/193},
archivePrefix = {arXiv},
       eprint = {1003.4747},
 primaryClass = {astro-ph.CO},
       adsurl = {https://ui.adsabs.harvard.edu/abs/2010ApJ...721..193P}
}

@ARTICLE{Shapley15,
       author = {{Shapley}, Alice E. and {Reddy}, Naveen A. and {Kriek}, Mariska and {Freeman}, William R. and {Sanders}, Ryan L. and {Siana}, Brian and {Coil}, Alison L. and {Mobasher}, Bahram and {Shivaei}, Irene and {Price}, Sedona H. and {de Groot}, Laura},
        title = "{The MOSDEF Survey: Excitation Properties of z {\ensuremath{\sim}} 2.3 Star-forming Galaxies}",
      journal = {\apj},
         year = 2015,
        month = mar,
       volume = {801},
       number = {2},
          eid = {88},
        pages = {88},
          doi = {10.1088/0004-637X/801/2/88},
archivePrefix = {arXiv},
       eprint = {1409.7071},
 primaryClass = {astro-ph.GA},
       adsurl = {https://ui.adsabs.harvard.edu/abs/2015ApJ...801...88S}
}

@ARTICLE{Kewley13,
       author = {{Kewley}, Lisa J. and {Maier}, Christian and {Yabe}, Kiyoto and {Ohta}, Kouji and {Akiyama}, Masayuki and {Dopita}, Michael A. and {Yuan}, Tiantian},
        title = "{The Cosmic BPT Diagram: Confronting Theory with Observations}",
      journal = {\apjl},
         year = 2013,
        month = sep,
       volume = {774},
       number = {1},
          eid = {L10},
        pages = {L10},
          doi = {10.1088/2041-8205/774/1/L10},
archivePrefix = {arXiv},
       eprint = {1307.0514},
 primaryClass = {astro-ph.CO},
       adsurl = {https://ui.adsabs.harvard.edu/abs/2013ApJ...774L..10K}
}

@ARTICLE{Bisigello21,
       author = {{Bisigello}, L. and {Gruppioni}, C. and {Feltre}, A. and {Calura}, F. and {Pozzi}, F. and {Vignali}, C. and {Barchiesi}, L. and {Rodighiero}, G. and {Negrello}, M.},
        title = "{Simulating the infrared sky with a SPRITZ}",
      journal = {\aap},
         year = 2021,
        month = jul,
       volume = {651},
          eid = {A52},
        pages = {A52},
          doi = {10.1051/0004-6361/202039909},
archivePrefix = {arXiv},
       eprint = {2011.07074},
 primaryClass = {astro-ph.GA},
       adsurl = {https://ui.adsabs.harvard.edu/abs/2021A&A...651A..52B}
}

@ARTICLE{tallada20,
       author = {{Tallada}, P. and {Carretero}, J. and {Casals}, J. and {Acosta-Silva}, C. and {Serrano}, S. and {Caubet}, M. and {Castander}, F.~J. and {C{\'e}sar}, E. and {Crocce}, M. and {Delfino}, M. and {Eriksen}, M. and {Fosalba}, P. and {Gazta{\~n}aga}, E. and {Merino}, G. and {Neissner}, C. and {Tonello}, N.},
        title = "{CosmoHub: Interactive exploration and distribution of astronomical data on Hadoop}",
      journal = {Astronomy and Computing},
         year = 2020,
        month = jul,
       volume = {32},
          eid = {100391},
        pages = {100391},
          doi = {10.1016/j.ascom.2020.100391},
archivePrefix = {arXiv},
       eprint = {2003.03217},
 primaryClass = {astro-ph.IM},
       adsurl = {https://ui.adsabs.harvard.edu/abs/2020A&C....3200391T}
}

@INPROCEEDINGS{carretero17,
       author = {{Carretero}, J. and {Tallada}, P. and {Casals}, J. and {Caubet}, M. and {Castander}, F. and {Blot}, L. and {Alarc{\'o}n}, A. and {Serrano}, S. and {Fosalba}, P. and {Acosta-Silva}, C. and {Tonello}, N. and {Torradeflot}, F. n. and {Eriksen}, M. and {Neissner}, C. and {Delfino}, M.},
        title = "{CosmoHub and SciPIC: Massive cosmological data analysis, distribution and generation using a Big Data platform}",
    booktitle = {Proceedings of the European Physical Society Conference on High Energy Physics. 5-12 July},
         year = 2017,
        month = jul,
          eid = {488},
        pages = {488},
          doi = {10.22323/1.314.0488},
       adsurl = {https://ui.adsabs.harvard.edu/abs/2017ehep.confE.488C}
}

@ARTICLE{lopezlopez24,
       author = {{L{\'o}pez-L{\'o}pez}, X. and {Bolzonella}, M. and {Pozzetti}, L. and {Salvato}, M. and {Bisigello}, L. and {Feltre}, A. and {L{\'o}pez}, I.~E. and {Viitanen}, A. and {Allevato}, V. and {Bongiorno}, A. and {Girelli}, G. and {Buchner}, J. and {Charlot}, S. and {Ricci}, F. and {Schreiber}, C. and {Zamorani}, G.},
        title = "{MAMBO: An empirical galaxy and AGN mock catalogue for the exploitation of future surveys}",
      journal = {\aap},
         year = 2024,
        month = nov,
       volume = {691},
          eid = {A136},
        pages = {A136},
          doi = {10.1051/0004-6361/202451509},
archivePrefix = {arXiv},
       eprint = {2409.06700},
 primaryClass = {astro-ph.GA},
       adsurl = {https://ui.adsabs.harvard.edu/abs/2024A&A...691A.136L}
}

%

\begin{appendix}
  

%
\end{appendix}

\end{document}